\documentclass[10pt]{article}
\usepackage[utf8]{inputenc}

\usepackage{setspace}
\usepackage[colorlinks=true, allcolors=blue]{hyperref}

\usepackage[title]{appendix}
\usepackage{float}
\usepackage{booktabs}
\usepackage{subcaption}
\usepackage{longtable}
\usepackage{array}
\usepackage{makecell}
\usepackage{pifont}
\newcommand{\cmark}{\ding{51}}
\newcommand{\xmark}{\ding{55}}
\usepackage{titlesec}
\titleformat{\subsubsection}
  {\normalfont\itshape}{\thesubsubsection}{1em}{}

\makeatletter
\renewcommand{\fps@figure}{H}      
\renewcommand{\fps@table}{H}  
\makeatother 

\usepackage{tikz}
\usetikzlibrary{shapes,arrows}
\usetikzlibrary{positioning}
\usetikzlibrary{decorations.pathreplacing}
\usetikzlibrary{calligraphy}
\usepackage{pgfplots}
\pgfplotsset{compat=1.9}
\usepgfplotslibrary{statistics}

\usepackage[left]{lineno}

\renewcommand*{\thesubfigure}{\textsf{\textbf{\Alph{subfigure}}}}

\newcolumntype{x}[1]{>{\centering\arraybackslash\hspace{0pt}}p{#1}}
\usepackage{multirow}

\usepackage{bm}
\usepackage{ntheorem}
\theoremseparator{:}
\newtheorem{hyp}{Hypothesis}
\usepackage{cleveref}
\usepackage{csquotes}
\usepackage{siunitx}

\title{\singlespacing Explainable AI improves task performance \\ in human-AI collaboration}

\author{Julian Senoner$^{1\ast}$, Simon Schallmoser$^{2,3\ast}$, Bernhard Kratzwald$^{1}$,\\ Stefan Feuerriegel$^{2,3}$, Torbjørn Netland$^{1\dagger}$\\
\normalsize{$^{1}$ETH Zurich, Zurich, Switzerland}\\
\normalsize{$^{2}$LMU Munich, Munich, Germany}\\
\normalsize{$^{3}$Munich Center for Machine Learning (MCML), Munich, Germany}\\
\normalsize{$^\ast$Contributed equally}\\
\normalsize{$^\dagger$Correspondence:  Torbjørn Netland (tnetland@ethz.ch)}}

\date{}

\begin{document}

\maketitle

\newpage

\begin{abstract}
\noindent
Artificial intelligence (AI) provides considerable opportunities to assist human work. However, one crucial challenge of human-AI collaboration is that many AI algorithms operate in a black-box manner where the way how the AI makes predictions remains opaque. This makes it difficult for humans to validate a prediction made by AI against their own domain knowledge. For this reason, we hypothesize that augmenting humans with explainable AI as a decision aid improves task performance in human-AI collaboration. To test this hypothesis, we analyze the effect of augmenting domain experts with explainable AI in the form of visual heatmaps. We then compare participants that were either supported by (a)~black-box AI or (b)~explainable AI, where the latter supports them to follow AI predictions when the AI is accurate or overrule the AI when the AI predictions are wrong. We conducted two preregistered experiments with representative, real-world visual inspection tasks from manufacturing and medicine. The first experiment was conducted with factory workers from an electronics factory, who performed $N=9,600$ assessments of whether electronic products have defects. The second experiment was conducted with radiologists, who performed $N=5,650$ assessments of chest X-ray images to identify lung lesions. The results of our experiments with domain experts performing real-world tasks show that task performance improves when participants are supported by explainable AI instead of black-box AI. For example, in the manufacturing setting, we find that augmenting participants with explainable AI (as opposed to black-box AI) leads to a five-fold decrease in the median error rate of human decisions, which gives a significant improvement in task performance. 
\end{abstract}

\begin{center}
\begin{tabular}{p{14.5cm}}
\end{tabular}
\end{center}

\newpage
\sloppy
\raggedbottom
\clubpenalty = 10000
\widowpenalty = 10000
\displaywidowpenalty = 10000

\section*{Introduction}
\label{sec:introduction}


Artificial intelligence (AI) provides considerable opportunities to assist human work in various domains \cite{Brynjolfsson.2017, AIIndex.2024}. For example, in manufacturing, AI is widely used to support humans when inspecting the quality of produced products to identify defects \cite{Bertolini.2021}. Similarly, in medicine, disease diagnosis now makes increasing use of AI systems. For instance, a recent survey found that  AI is used by about 15\% of radiologists at least weekly \cite{Scheetz.2021}. More broadly, an analysis showed that about 30\% of all jobs in the United States are at high exposure to be assisted by AI \cite{IMF.2024}. Hence, the importance of human-AI collaboration is expected to grow in the near future.


However, many questions regarding the effective design of human-AI collaborations remain open. One particular challenge in the use of AI for human work is that state-of-the-art AI algorithms, which frequently involve millions of trainable parameters \cite{Simonyan.2014, He.2016}, operate as \textquote{black-box} algorithms. The term \textquote{black-box} refers to the opacity of these systems, meaning that the internal workings and decision-making processes of these algorithms are not transparent or easily understandable by humans \cite{Rudin.2019}. This can have crucial implications in practice as the lack of transparency makes it difficult -- or even impossible -- for humans to validate the predictions made by an AI against their domain knowledge. Hence, without being able to assess whether a prediction generated by an AI is accurate, humans will not be able to correct predictions of the AI, because of which the unique expertise of workers is essentially lost, which will make the collaboration between domain experts and AI largely ineffective. 


Increasing efforts have been made to overcome the black-box nature of AI by developing methods that generate \emph{explanations} for how AI algorithms reach their decisions \cite{Guidotti.2018, Murdoch.2019, Gunning.2019, Cheng.2019, Liao.2020, Bodria.2023, Dwivedi.2023}. Explainable AI refers to a set of methods that support humans in understanding how AI algorithms map certain inputs (e.g., lung X-rays, patient characteristics) to certain outputs (e.g., probability estimates for pneumonia) \cite{Samek.2019, Samek.2021}. Explainable AI can be broadly categorized into inherently interpretable models and post-hoc explanation techniques (see Supplement~\ref{sec:literature_review} for an extended literature review on explainable AI). For inherently interpretable algorithms, the decision-making of the algorithm can be inspected by humans, e.g., by inspecting the coefficients in linear regression or the splitting rules in decision trees \cite{Molnar.2019}. Post-hoc explanation techniques are required when the inner workings of an AI algorithm become too complex to be understood by humans such as in deep neural networks. For example, one approach is to approximate the behavior of a black-box AI with a simpler model (e.g., a linear model) that can be interpreted \cite{Ribeiro.2016}. Other methods rely on game theory to estimate the contribution of each model input to the model output while considering possible interaction effects \cite{Lundberg.2017}. Common methods for explaining AI algorithms in computer vision include the use of heatmaps. These heatmaps visually highlight the areas that are most relevant to the predictions made by the AI \cite{Simonyan.2013, Selvaraju.2017}. Such explanation techniques are commonly used by AI engineers in the development of AI algorithms. Hence, this literature stream is orthogonal to the use of explainable AI in our work, where we use post-hoc explanation techniques to improve decisions by domain experts in real-world job tasks. 


Several works have studied behavioral dimensions of human-AI collaboration. For example, it has been examined whether humans are willing to delegate work to AI \cite{Fuegener.2021b, Fuegener.2021c, Bauer.2023}. Another common dimension is algorithm aversion, where humans are averse to following decisions by algorithms and instead rely on their own (mis)judgment \cite{Dietvorst.2015, Dietvorst.2016, Dietvorst.2020, Burton.2020, David.2021}. An antecedent to algorithm aversion is \emph{trust in AI}, critically influencing whether humans adopt or reject AI recommendations \cite{Choung.2023, Nourani.2019, Panigutti.2022}. Oppositely to algorithm aversion, overreliance is also a problem negatively impacting the effectiveness of human-AI collaboration \cite{Bansal.2021, Vasconcelos.2023, Chen.2023}. That is, humans risk following AI predictions blindly without attentively performing the task. While all these dimensions are interesting from a behavioral perspective, the main outcome of interest for business and healthcare organizations is \emph{task performance}. However, the impact of explainable AI on task performance in human-AI collaboration in real-world job tasks remains unclear.


We hypothesize that augmenting domain experts with explainable AI, as opposed to black-box AI, improves task performance in human-AI collaboration. Specifically, we treat explainable AI as a form of decision aid that supports domain experts in better understanding algorithmic decisions. Experts can then compare the explanations to their domain knowledge, thereby validating whether the AI is correct or overwriting the AI if is not correct. Here, explainable AI does not provide more information from an AI perspective (i.e., identical predictive performance). However, for domain experts, it gives rich additional information by making the AI predictions more accessible. Thus, we expect that domain experts supported by explainable AI will outperform those supported by black-box AI in two ways: (1)~they are more likely to follow AI predictions when they are accurate, and (2)~they are more likely to overrule AI predictions when they were wrong.


Previous research has studied the effect of explainable AI on task performance in human-AI collaboration (see Supplement~\ref{sec:literature_review} for a detailed overview), yet with key limitations. In particular, existing works are typically restricted by either (i) recruiting laypeople or (ii) overly simplified tasks that are not representative of real job tasks \cite{Bucinca.2020, Chu.2020, Alufaisan.2021, Schemmer.2023}. However, a realistic estimate of the effect of explainable AI on task performance requires a real-world task performed by domain experts. Such works that actually study real-world tasks with domain experts are on the other hand restricted by (i) comparing explainable AI vs humans alone \cite{Lundberg.2020, Das.2023}, (ii) using no real explainable AI \cite{Gaube.2023}, or (iii) research designs that do not isolate the effect of explainable AI on task performance \cite{Jesus.2021, Sivaraman.2023, Jabbour.2023}. In contrast, the strength of our work is that we study the effect of explainable AI on task performance relative to black-box AI in human-AI collaborations with \emph{real-world tasks} and actual \emph{domain experts}.

In this paper, we analyze the effect of augmenting domain experts with explainable AI on task performance in human-AI collaboration. For this, we conducted two preregistered experiments in which domain experts were asked to solve real-world visual inspection tasks in manufacturing (Study~1) and medicine (Study~2). We followed a between-subject design where we randomly assigned participants to two treatments: (a)~black-box AI (i.e., where AI predictions are opaque) and (b)~explainable AI (i.e., where AI predictions are explained). The latter thus offers not only the prediction from the AI but further shows explanations in the form of a visual heatmap as a decision aid. Heatmaps are frequently used and are considered state-of-the-art with respect to their localization performance across various settings \cite{Bergmann.2019a, Bergmann.2019b, Binder.2021, Saporta.2022}. Study~1 was conducted in a manufacturing setting, where participants had to identify quality defects in electronic products. For this, we specifically recruited actual factory workers performing $N=9,600$ assessments of electronic products at \emph{Siemens}. Study~2 was conducted in a medical setting, where participants had to identify lung lesions on chest X-ray images. To that end, medical professionals, i.e., radiologists, were recruited and performed $N=5,650$ assessments of chest X-ray images. In both studies, participants performed better when being supported by explainable AI as a decision aid.

The tasks of both experiments are representative of many real-world human-AI collaborations. The manufacturing task is an identical, one-to-one copy of a real-world job task at \emph{Siemens} and, hence, highly representative of visual inspection tasks in manufacturing \cite{Juran.1979, Hoyle.2007}. Visual inspection tasks are standard in the manufacturing industry. Regardless of how much manufacturers have sought to build quality into products and processes, labor-intensive inspection tasks still abound \cite{Baudin.2023}. In healthcare, visual inspection tasks are common across many different subdisciplines such as dermatology, radiology, pathology, ophthalmology, and dentistry, among many others. As concrete examples, physicians have to inspect, for instance, skin lesions in dermatology, tissues in pathology, and lung lesions in radiology \cite{Tschandl.2020, Pantanowitz.2011, Saporta.2022}. Hence, establishing whether physicians benefit from explainable AI in visual inspection tasks is highly relevant for setting correct disease diagnosis and subsequent treatment.

\section*{Results}
\label{sec:results}

\begin{figure}[H]
\centering
\vspace{-1cm}
\begin{subfigure}[t]{0.56\textwidth}
  \caption{}
  \label{fig:experiment_manufacturing}
  \centering
  \includegraphics[width=1\linewidth]{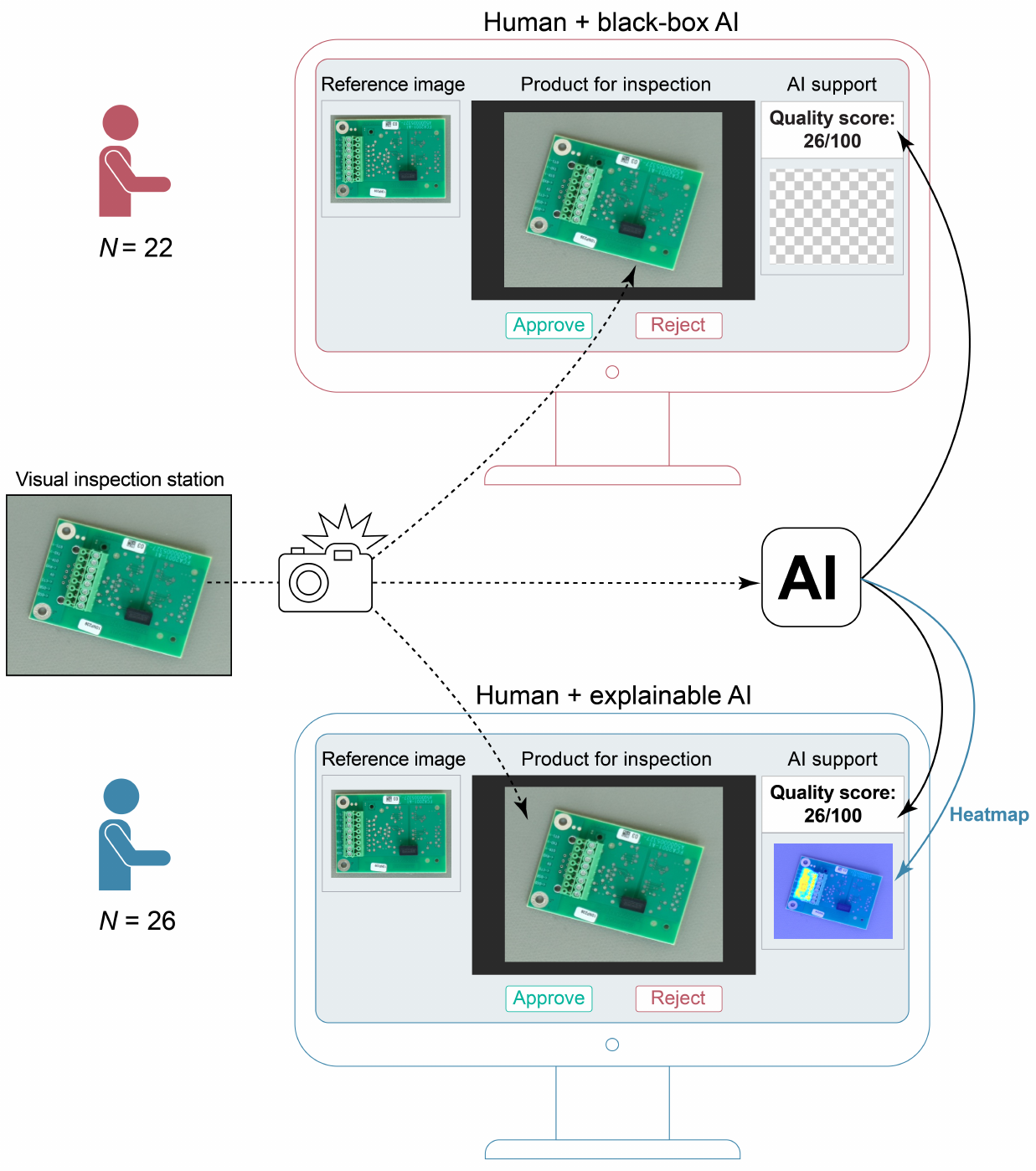}
\end{subfigure}
\begin{subfigure}[t]{0.56\textwidth}
  \caption{}
  \label{fig:experiment_medical}
  \centering
  \includegraphics[width=1\linewidth]{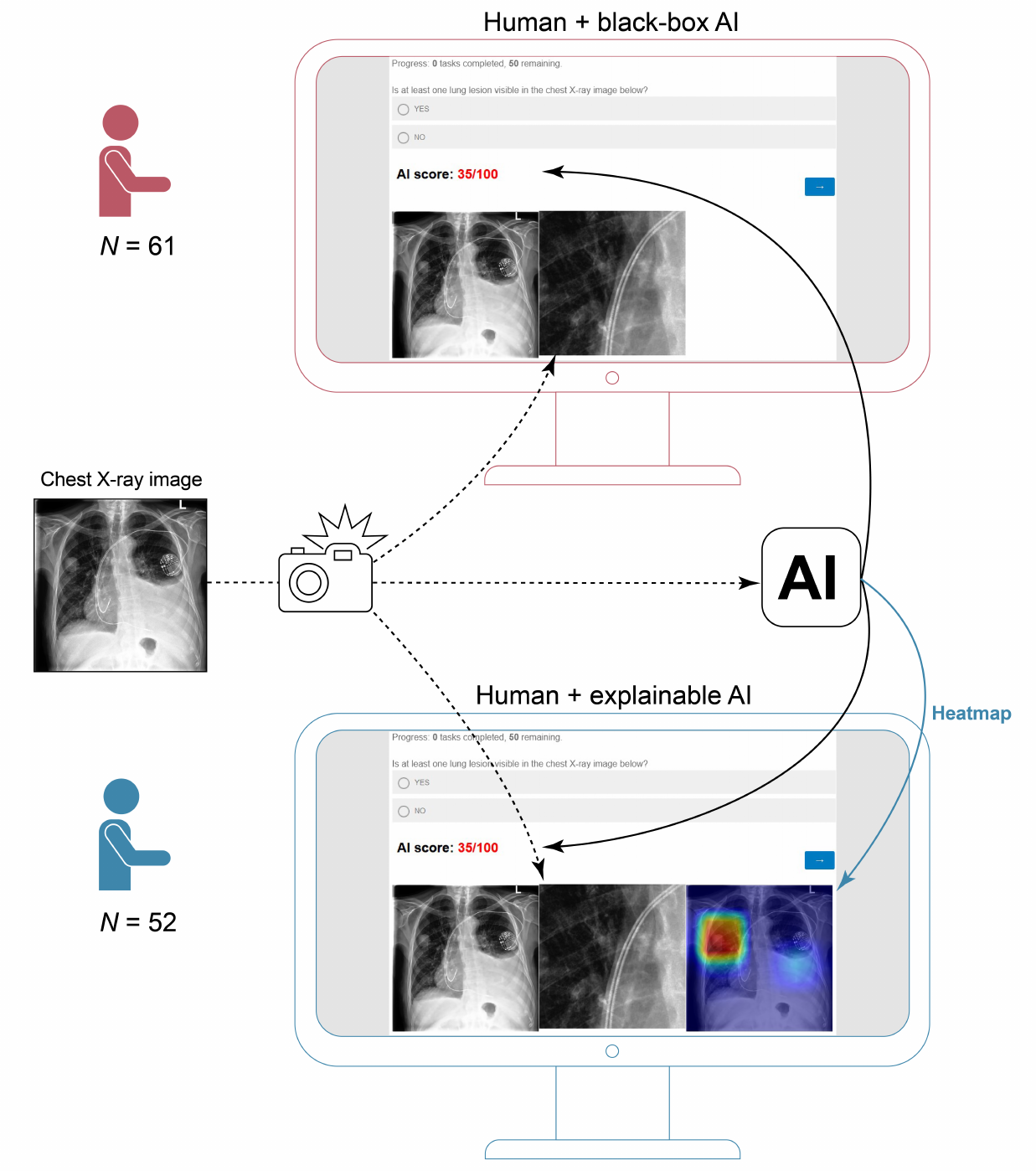}
\end{subfigure}
\caption{\textbf{Overview of the experiments for assessing the effect of explainable AI on task performance.} (\textbf{\textsf{A}})~Experimental design of the manufacturing experiment where factory workers were asked to \textquote{approve} images of faultless products and to \textquote{reject} images of defective products through a computer interface. (\textbf{\textsf{B}})~Experimental design of the medical experiment where radiologists were asked to decide whether lung lesions are visible in the chest X-ray image. In both experiments, participants were randomly assigned to one of the two treatments: (a)~black-box AI or (b)~explainable AI.}
\label{fig:experimental_design}
\end{figure}

To analyze the effect of explainable AI on task performance in human-AI collaboration, we conducted two randomized experiments across two different settings, i.e., in manufacturing (Study~1) and medicine (Study~2). In both experiments, participants had to perform a visual inspection task. In the manufacturing experiment, factory workers were asked to inspect electronic products and to identify defective products. In the medical experiment, radiologists were asked to decide whether lung lesions are visible in chest X-ray images. Participants were randomly assigned to one of two different treatments aiding them in the task: (a)~black-box AI or (b)~explainable AI (\Cref{fig:experimental_design}). Participants with black-box AI received an opaque AI score as a decision aid. Participants with explainable AI received the same score and an additional decision aid: the explanation of the score in the form of a heatmap. The heatmap does not provide more information from an AI perspective (the score is identical) but allows users to verify the prediction made by the AI. However, heatmaps provide a clear and intuitive way of highlighting quality defects/lung lesions \cite{Ibrahim.2023}. We hypothesized that explainable AI as a decision aid improves task performance of domain experts in human-AI collaboration. Details on both experiments are provided in the \hyperref[sec:methods]{Methods} section.

\clearpage
\subsection*{Study 1: Manufacturing experiment}


The manufacturing experiment was conducted at a factory of \emph{Siemens}, a global industrial conglomerate particularly known for its consumer and industrial electronics products. The objective of the task was to identify quality defects in electronic products (e.g., missing components, wrong components, and faulty components) with high accuracy. The task is representative of a real-world inspection task at \emph{Siemens} and is analogous to other inspection tasks in manufacturing \cite{Juran.1979, Hoyle.2007}. Factory workers from \emph{Siemens} were asked to visually inspect 200 images using a computer interface and label them as faultless or defective. The inspection task had to be completed within 35 minutes, which corresponds to realistic field conditions.


Workers were randomly assigned to two treatments where they were either supported by black-box AI or explainable AI (\Cref{fig:experiment_manufacturing}). In both treatment arms, workers received a reference image of a faultless product (note that this is common practice at \emph{Siemens}). In addition, they were provided with AI predictions of the product quality given by a numerical \textquote{quality score} between 0 (most certainly defect) and 100 (most certainly faultless). Workers with explainable AI additionally received a tailored decision aid in the form of visual heatmaps that indicated the assumed location of potential quality defects. The quality scores in both treatment arms were identical, so that differences in task performance could only be attributed to the explanations (i.e., the heatmaps).


The objective of the field experiment was to obtain accurate estimates of the treatment effect under real-world conditions. Hence, we ran the experiment with actual domain experts rather than laypeople. The results are based on the entire available workforce of one shift of factory workers from \emph{Siemens}. The factory workers performed in total $N=9,600$ assessments of electronic products. We then analyzed the effect of explainable AI on task performance using the balanced accuracy and defect detection rate (i.e., proportion of correctly identified defective products among all actual defective products) based on the quality assessments in the visual inspection task.

\begin{figure}[H]
\centering
\includegraphics[width=0.95\linewidth]{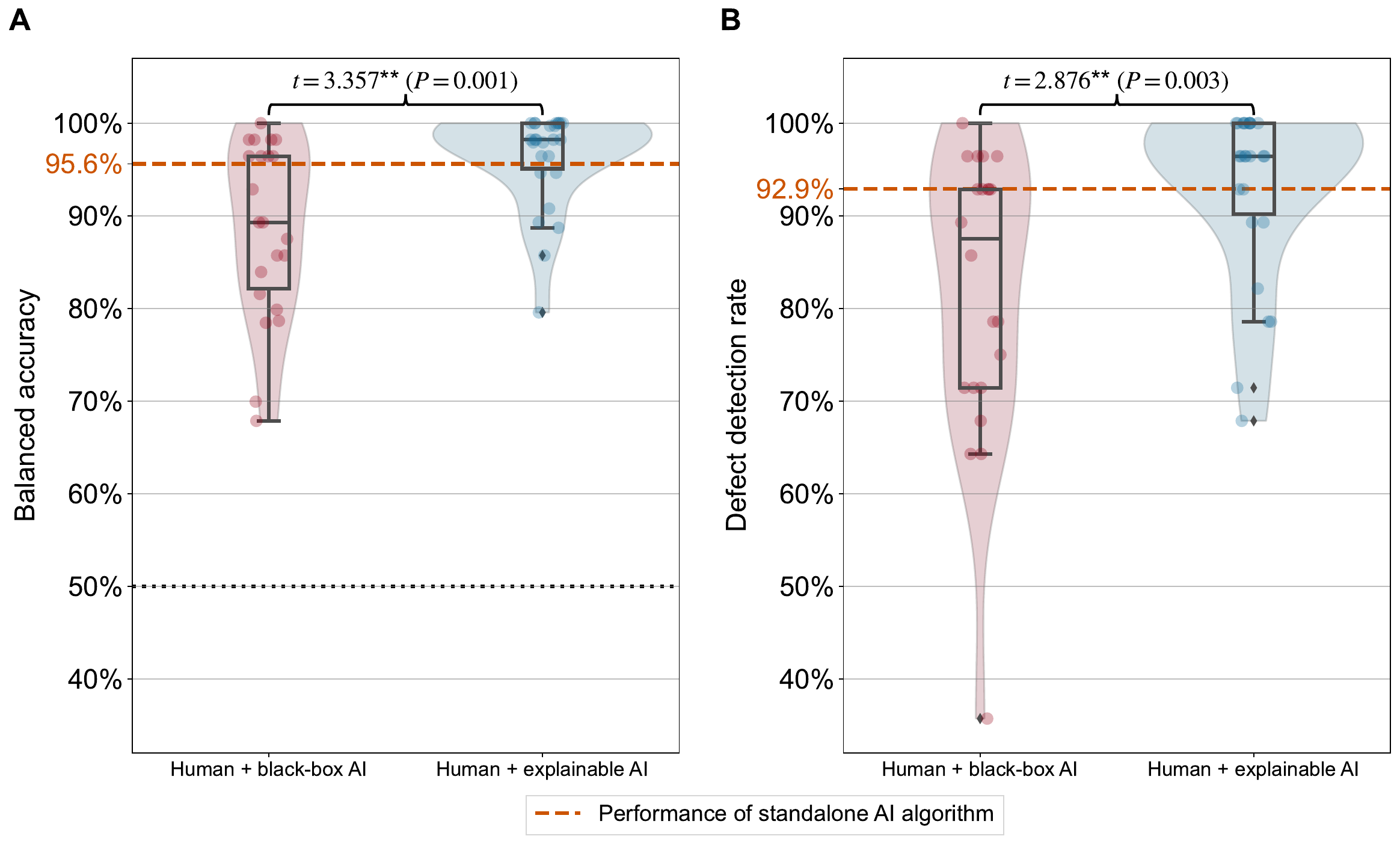}
\caption{\textbf{Results of manufacturing experiment.} The boxplots compare the task performance between the two treatments: black-box AI and explainable AI. The task performance is measured by the balanced accuracy (\textbf{\textsf{A}}) and the defect detection rate (\textbf{\textsf{B}}) based on the quality assessment of workers and the ground-truth labels of the product images. A balanced accuracy of 50\% provides a na{\"i}ve baseline corresponding to a random guess (black dotted line). The standalone AI algorithm attains a balanced accuracy of 95.6\% and a defect detection rate of 92.9\% (orange dashed lines). Statistical significance is based on a one-sided Welch's $t$-test (\textsuperscript{***}$P<0.001$, \textsuperscript{**}$P<0.01$, \textsuperscript{*}$P<0.05$). In the boxplots, the center line denotes the median; box limits are upper and lower quartiles; whiskers are defined as the 1.5x interquartile range.}
\label{fig:study_2_results}
\end{figure}

We found that workers supported by explainable AI achieved a better task performance than workers supported by black-box AI. Workers with black-box AI achieved a balanced accuracy with a mean of only 88.6\%, whereas workers with explainable AI treatment achieved a balanced accuracy with a mean of 96.3\% (Figure \hyperref[fig:study_2_results]{2\textbf{A}}). We then estimated the treatment effect of explainable AI by regressing the balanced accuracy on the treatment (black-box AI $= 0$, explainable AI $= 1$). The regression results show that the treatment effect of explainable AI is statistically significant and large ($\beta=7.653$, $\mathit{SE}=2.178$, $P=0.001$); that is, an improvement of 7.7 percentage points. Compared to the black-box AI, the explainable AI leads to a five-fold decrease in the median error rate. 

Workers with explainable AI outperformed workers with black-box AI also with respect to the defect detection rate with a mean of 93.0\% versus a mean of 82.0\% (Figure \hyperref[fig:study_2_results]{2\textbf{B}}). The regression results again confirm that the treatment effect of explainable AI is statistically significant and large ($\beta=11.014$, $\mathit{SE}=3.680$, $P=0.004$). All regression results remain statistically significant when including relevant control variables (demographics, tenure, self-reported IT skills, and decision speed) in the regression model (see Supplement~\ref{sec:regression_models_2}).

A detailed analysis of the workers' assessments revealed that workers with explainable AI followed accurate predictions more often than workers with black-box AI ($\textrm{mean} = 93.5\%$ for black-box AI, $\textrm{mean} = 98.6\%$ for explainable AI). In particular, workers supported by black-box AI were 3.6 times more likely to erroneously overrule an AI prediction, despite the prediction being accurate ($t=2.437$, $P=0.011$). Interestingly, 73.1\% of the workers with explainable AI performed even better than the standalone AI algorithm. This suggests that the explanations (i.e., the heatmaps) not only improve adherence to accurate AI predictions, but also help humans make correct assessments when the AI predictions are wrong. We found that workers with explainable AI were, on average, able to identify and overrule 96.9\% of the wrong AI predictions. For comparison, workers supported by black-box AI only overruled 86.4\% of the wrong AI predictions. These results are highly relevant since -- regardless of an AI's performance -- wrong AI predictions can always occur due to external factors such as dust or different light conditions. The difference between both treatments is again statistically significant ($t=2.631$, $P=0.007$). These findings underscore the effectiveness of augmenting humans with explainable AI.

Finally, we assessed whether workers with explainable AI spent more time on making their quality assessments. For this, we analyzed whether the workers' median decision speeds across the 200 product images differed. No statistically significant difference ($t=0.308$, $P=0.380$) was observed between both treatments ($\textrm{mean} = \SI{5.01}{s}$ for black-box AI, $\textrm{mean} = \SI{4.88}{s}$ for explainable AI). Therefore, explainable AI improved task performance without affecting the productivity of the workers.

\subsection*{Study 2: Medical experiment}


In the medical experiment, radiologists were asked to visually inspect 50 chest X-ray images and decide whether at least one lung lesion was visible (\Cref{fig:experiment_medical}). Visual inspection tasks like ours are common in medicine across various subdisciplines \cite{Pantanowitz.2011, Tschandl.2020}. Analogous to the manufacturing task, radiologists had 35 minutes to complete the task and were randomly assigned to be either supported by black-box AI or by explainable AI (\Cref{fig:experiment_medical}). Both types of AI provided a score between 0 (most certainly a lung lesion visible) and 100 (most unlikely a lung lesion visible), which was identical in both treatment arms. In addition to that, radiologists with explainable AI were provided with a heatmap that highlights regions in the chest X-ray image that the AI finds most relevant for predicting lung lesions.


The results are based on a sample of $N=5,650$ assessments of chest X-ray images performed by 113 radiologists from the United States. Again, task performance was analyzed using the balanced accuracy and the disease detection rate (i.e., the true negative rate, where chest X-ray images containing lung lesions were considered a negative sample) based on the assessments made in the visual inspection task.

\begin{figure}[H]
\centering
\includegraphics[width=0.95\linewidth]{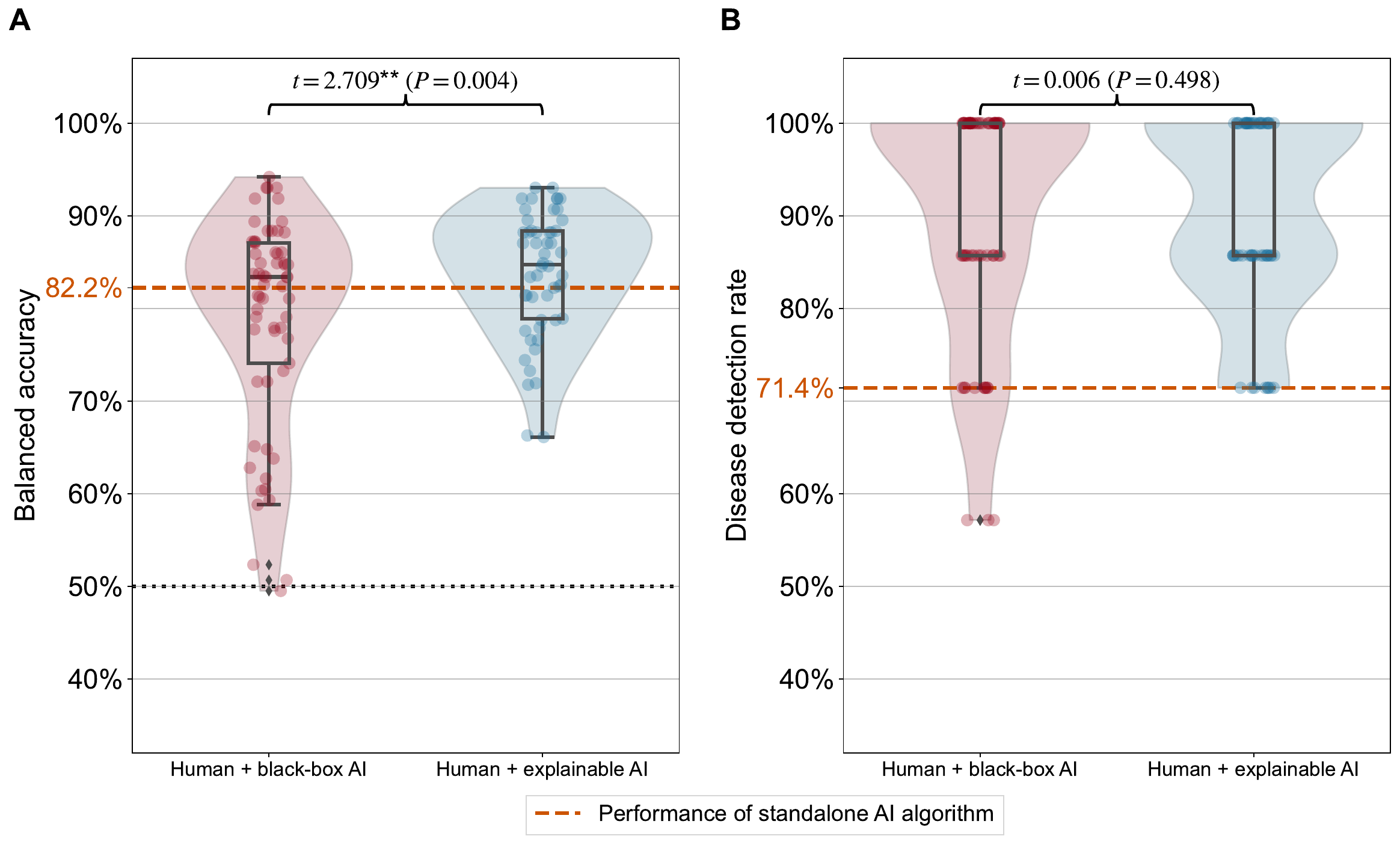}
\caption{\textbf{Results of medical experiment.} The boxplots compare the task performance between the two treatments: black-box AI and explainable AI. The task performance is measured by the balanced accuracy (\textbf{\textsf{A}}) and the disease detection rate (\textbf{\textsf{B}}) based on the quality assessment of radiologists and the ground-truth labels of the chest X-ray images. A balanced accuracy of 50\% provides a na{\"i}ve baseline corresponding to a random guess (black dotted line). The standalone AI algorithm attains a balanced accuracy of 82.2\% and a disease detection rate of 71.4\% (orange dashed lines). Statistical significance is based on a one-sided Welch's $t$-test (\textsuperscript{***}$P<0.001$, \textsuperscript{**}$P<0.01$, \textsuperscript{*}$P<0.05$). In the boxplots, the center line denotes the median; box limits are upper and lower quartiles; whiskers are defined as the 1.5x interquartile range.}
\label{fig:study_3_results}
\end{figure}

Radiologists augmented with explainable AI outperformed peers with black-box AI. Radiologists with black-box AI achieved a balanced accuracy with a mean of only 79.1\%, whereas radiologists with explainable AI achieved a balanced accuracy with a mean of 83.8\% (Figure \hyperref[fig:study_3_results]{3\textbf{A}}). We again estimated the treatment effect of explainable AI by regressing the balanced accuracy on the treatment (black-box AI $= 0$, explainable AI $= 1$). The regression results show that the treatment effect of explainable AI is statistically significant and large ($\beta=4.693$, $\mathit{SE}=1.800$, $P=0.01$); that is, an improvement of 4.7 percentage points. All results remain statistically significant when including relevant control variables (tenure, self-reported IT skills, and decision speed) in the regression model (see Supplement~\ref{sec:regression_models_3}). In contrast to the manufacturing experiment, no difference in task performance with respect to the disease detection rate was observed; radiologists in both treatment arms achieved a disease detection rate with a mean of 90.4\%  (Figure \hyperref[fig:study_3_results]{3\textbf{B}}). This was also observed when regressing the disease detection rate on the treatment ($\beta=-0.014$, $\mathit{SE}=2.244$, $P=0.995$). This can be expected since missing a lung lesion has more serious consequences than erroneously believing a lung lesion is visible; thus, leading to conservative decision-making of radiologists. Therefore, we additionally inspected precision as a task performance metric. We find that radiologists augmented with explainable AI were significantly more precise (improvement of 6.4 percentage points, $P=0.014$) in identifying lung lesions compared to radiologists with black-box AI (see Supplement~\ref{sec:precision}).

As in Study~1, we found that radiologists with explainable AI followed accurate AI predictions more often than radiologists with black-box AI treatment ($\textrm{mean} = 72.4\%$ for black-box AI, $\textrm{mean} = 82.1\%$ for explainable AI). In particular, radiologists supported by black-box AI were 54.2\% times more likely to erroneously overrule an AI prediction, although it was correct ($t=3.084$, $P=0.001$). We observed that radiologists with explainable AI only overruled 50.8\% of the wrong AI predictions compared to 57.7\% for radiologists with black-box AI treatment.

Again, we assessed the decision speed of radiologists in both treatment arms. We found no significant difference ($t=0.392$, $P=0.348$) between both treatments ($\textrm{mean} = \SI{10.71}{s}$ for black-box AI, $\textrm{mean} = \SI{10.29}{s}$ for explainable AI). Thus, task performance was improved by explainable AI without reducing the productivity of the radiologists.

\section*{Discussion}
\label{sec:discussion}


The \textquote{age of AI} redefines the way humans and machines collaborate, thus raising questions about how human-AI collaborations can be effectively designed. As we show, the effectiveness of human-AI collaboration largely depends on the extent to which humans incorporate correct AI predictions and overrule wrong ones. However, many state-of-the-art AI algorithms operate as black-box, thus making it difficult for humans to compare the reasoning of the AI to their own domain knowledge. In this paper, we contribute a unique perspective by studying the impact of AI explainability on task performance of domain experts in human-AI collaboration, presenting empirical evidence from different domains with robust and generalizable results.


We conducted two preregistered experiments to estimate the effect of explainability in human-AI collaboration in real-world visual inspection tasks. Our results demonstrate that domain experts make subpar decisions when they are supported by a black-box AI algorithm with opaque predictions. In contrast, we find that explanations from an explainable AI are a powerful decision aid. Explanations were provided in the form of heatmaps, which provide a clear and intuitive way of highlighting areas that are determinants of AI predictions. The explanations do not provide more information from an AI perspective (i.e., the prediction performance is identical), but rather make the information more accessible to domain experts. Specifically, compared to black-box AI, augmenting domain experts with explainable AI improved the task performance by 7.7 percentage points in a manufacturing experiment and by 4.7 percentage points in a medical experiment. In the manufacturing experiment, 73.1\% of the domain experts even outperformed the standalone AI algorithm when they were augmented with explainable AI. The prime reason was that domain experts supported by explainable AI were more likely to follow AI predictions when they were accurate and more likely to overrule them when they were wrong.


Improving the task performance of domain experts has practical implications in many fields such as manufacturing and medicine. For example, factory workers at \emph{Siemens} augmented with explainable AI were able to identify 13\% more defects than peers augmented with black-box AI. Thus, explainable AI could help to reduce downstream costs for manufacturing companies by filtering out defective products at the earliest possible stage. Similarly in medicine, task performance of physicians is crucial, especially for important tasks such as identifying possibly cancerogenous lung lesions. By showing that the results are consistent across different settings, we demonstrate that our insights are generalizable.


Our work contributes experimental evidence to the literature on human-AI collaboration \cite{Amershi.2019, DeArteaga.2020, Fuegener.2021, Cadario.2021, Dietvorst.2015, Dietvorst.2016, Dietvorst.2020, Sun.2020, Kawaguchi.2020, Castelo.2019, Yeomans.2019, Senoner.2022}. Algorithm aversion provides a barrier to the wider adoption of human-AI collaboration. Prior literature has presented several remedies, such as describing the functional logic of an algorithm \cite{Cadario.2021}, giving users permission to modify an algorithm \cite{Dietvorst.2016}, or letting users integrate their own forecasts into an algorithm \cite{Kawaguchi.2020}. This paper presents evidence of an effective alternative; that is, explaining individual predictions from an otherwise opaque AI algorithm. Such explanations allow domain experts to validate how an AI arrives at a certain prediction. Interestingly, while many studies advocate for complete automation, we also show the importance of human-AI collaboration: domain knowledge can help to identify errors in the AI and lead to a better performance than an AI-only system. 


A strength of our study is that we gather empirical evidence of improved task performance by explainable AI compared to black-box AI. In particular, by performing experiments in two different settings, we demonstrate that these results are generalizable. Unlike previous works on studying task performance in human-AI collaboration \cite{Bucinca.2020, Chu.2020, Alufaisan.2021, Schemmer.2023}, we (i) conducted experiments of two real-world job tasks in manufacturing and medicine and (ii) recruited domain experts for those tasks, i.e., factory workers and radiologists. When experiments use simplified decision tasks (e.g., object recognition) that are not representative of actual human work in the field, real-world validity is reduced. In contrast, our study has high external validity.


Our work is orthogonal to the literature on explainable AI in computer science, where the main goal is to develop and evaluate new methods for explaining black-box AI algorithms. Contrary, we are interested in a behavioral outcome, namely task performance in human-AI collaboration. A previous study on the effect of explainable AI on task performance found an improvement of 1.5 percentage points in accuracy relative to black-box AI \cite{Jabbour.2023}. However, their experiment was designed such that participants were not only shown the real explainable AI but also systemically biased explainable AI. This could have decreased the trust of the participants in the AI and, thus, explain the smaller treatment effect in comparison to our experiments. Prior work has also made use of expert annotations as a proxy for explainable AI \cite{Gaube.2023}. However, this prevents any conclusion on whether real explainable AI improves task performance.

 
One limitation of our research is that we, in both experiments, studied one specific human-AI work setting (a visual inspection task) with one specific form of explainability (a heatmap indicating the location of potential quality defects or lung lesions). However, this experimental task is representative of many real-world human-AI work settings and heatmaps are standard in explaining AI predictions of images. We also show that other heatmap algorithms lead to similar results (see Supplement~\ref{sec:heatmap_checks}). Still, we invite future research replicating our findings in other work settings using other methods for explainability. We also acknowledge reservations against using explainable AI in general. Explainable AI can be fooled by adversarial attacks \cite{Slack.2020} or may itself generate explanations that are unreliable and thus lead to misleading conclusions \cite{Rudin.2019}. Nevertheless, it is likely that performance improvements from explainable AI can be achieved in other settings, where explanations serve as a decision aid. Further, it is important to note that, in both experiments, 14\% of the images contained quality defects/lung lesions. Thus, quality defects/lung lesions were more prevalent than what domain experts would typically encounter in their respective job. This discrepancy might have influenced their prior expectations while performing the task. However, as participants in both treatment arms were shown exactly the same images, this factor likely had minimal impact on our findings.


Policy initiatives in many countries aim to promote transparency in AI algorithms (e.g., the United States \cite{AmericanAccountabilityAct.2019} and the European Union \cite{EuropeanCommission.2019}). These efforts are usually motivated from the perspective of ethics, regulation, and safety \cite{Guidotti.2018, Wachter.2017, Muller.2021, Jobin.2019}. Our research suggests that the benefits of algorithmic transparency are more profound: augmenting domain experts with explainable AI can enable better decisions with benefits for individuals, organizations, and society.

\newpage

\section*{Methods}
\label{sec:methods}


This work analyzes the effect of augmenting domain experts with explainable AI (as opposed to black-box AI) in human-AI collaboration. We preregistered our hypotheses (i.e., Study~1: \url{https://osf.io/7djxb} and Study~2: \url{https://osf.io/69yqt}; see also Supplement~\ref{sec:hypotheses}), which were tested in two randomized experiments. We comply with all ethical regulations and the research design was approved by the Ethics Commission of ETH Zurich (EK 2021-N-34). All participants provided informed consent.

\subsection*{Tasks}

In the following, details of both visual inspection tasks are provided.

\subsubsection*{Study 1: Manufacturing experiment}


For the manufacturing task, we designed a representative, real-world visual inspection task in collaboration with \emph{Siemens Smart Infrastructure} in Zug, Switzerland. The experimental task is representative of various domains in which workers have to make decisions under a limited time budget. During the experiment, workers were shown images of electronic products and were asked to label them as faultless or defective. Reassuringly, we emphasize that our experiment involved a real work scenario: we conducted it with real workers familiar with quality management practice, a real user interface for state-of-the-art quality management, realistic incentives, and real product images. All steps in the experiments were carried out via a computer interface that was designed analogously to the real-world quality inspection setup at \emph{Siemens} (see Supplement~\ref{sec:experimental_interface} for details). In the experiment, we made sure that all workers have the exact same conditions (exact same product images, same computer setup, same time limits, etc.). Thereby, we can rule out confounding variables that would arise naturally during the usual work routines and thus ensure that the experiment is scientifically sound.


We obtained 200 images of four different types of electronic products (printed circuit boards) from \emph{Siemens}. Example images are provided in Supplement~\ref{sec:research_setting}. All four different product types are of equal importance to \emph{Siemens}. Each product type comprised 43 images with faultless products and 7 images with defective products (e.g., missing components, wrong components, and faulty components). The different defects are all considered equally bad by the partner company, i.e., the products are considered to be either functional or non-functional. Hence, we considered defective products as scrap as best practice in quality management \cite{Juran.1979, Hoyle.2007}. 


We implemented an AI algorithm that computed an individual quality score for each image. The quality score gives a numerical value between 0 (most certainly defect) and 100 (most certainly faultless). Workers were instructed that a quality score below 90 suggests an increased likelihood of a quality defect and that the AI algorithm can make mistakes. As humans cannot understand how the AI algorithm arrives at the prediction, the quality score is regarded as opaque (\textquote{black-box AI}). When evaluating the quality score with a cutoff of 90 for mapping the numerical value onto a binary faultless/defect label, the standalone AI algorithm achieves a balanced accuracy of 95.6\% and a defect detection rate of 92.9\%. The prediction performance of the standalone AI algorithm was not communicated to the workers. The AI algorithm was trained on an additional set of product images that was not included in the experiment.


We used anomaly heatmaps \cite{Bergmann.2019a} to explain the opaque quality from the AI algorithm. The heatmaps were computed with standard computer vision methods and highlighted image regions with suspected quality defects (i.e., deviations from a faultless product). We chose heatmaps as the explanation technique for our AI algorithm as they provide a clear and intuitive way of highlighting areas with quality defects. This is especially important since the recruited domain experts are typically not familiar with explanation techniques for AI algorithms. Further, heatmaps are frequently used for images and are considered state-of-the-art with respect to their localization performance across various settings \cite{Bergmann.2019a, Bergmann.2019b, Binder.2021, Saporta.2022}. Details on the implementation of the AI algorithm and heatmaps are provided in Supplement~\ref{sec:ai_system}.


In the experiment, workers were randomly assigned to one of the two treatments: (a)~black-box AI or (b)~explainable AI. Workers in the black-box AI treatment arm were only supported by the opaque quality score. Workers in the explainable AI treatment arm had access to the same quality score but additionally received the heatmap that explained the otherwise opaque quality score. Of note, the explainable AI had the same accuracy as the black-box AI and did not carry more information from an AI perspective (i.e., the heatmaps were of the same predictive power).


The procedure of the experiment was as follows. Before starting the experiment, workers had to give written consent to participate and then pass a tutorial on how to use the interface. After that workers were randomly assigned to one of the two treatments, i.e., either black-box AI or explainable AI. During the experiment, the 200 product images were consecutively shown in random order. For each image, the workers had to assess the quality; that is, to \textquote{approve} or \textquote{reject} the shown product. We tracked the decision speed and the quality assessment (i.e., labeled as faultless or defective) made by the worker. To match real-world conditions, the workers were given a maximum of 35 minutes to finish the inspection task of 200 product images (around 10 seconds per image). In total, $N=9,600$ assessments of product images were performed by the workers. Finally, workers completed a post-experimental questionnaire (Supplement~\ref{sec:questionnaire}).

\subsubsection*{Study 2: Medical experiment}


For the medical task, radiologists had to identify lung lesions in real chest X-ray images. Lung lesions are common findings in chest X-ray images \cite{UTS.2024} and can be easily overlooked due to their frequently small size \cite{Oestmann.1988}. The radiologists were asked whether at least one lung lesion was visible in the X-ray image. The experiment was conducted via Qualtrics. To ensure a realistic experimental setup that resembles the same task in daily, medical practice, we implemented a zoom function, which allowed the radiologists to investigate an enlarged view of the image by moving their computer mouse over the image. Analogous to Study~1, we emphasize that our medical experiment involved a realistic work scenario: we conducted it with actual medical professionals, who were asked to investigate real chest X-ray images.


We used 50 chest X-ray images from the CheXpert dataset \cite{Irvin.2019}. The dataset comprised 7 images with at least one lung lesion and 43 images without lung lesions. Example images are provided in Supplement~\ref{sec:research_setting}. 


We implemented an AI algorithm that outputs the probability of whether a lung lesion is visible in the chest X-ray image. We transformed these probability outputs for lung lesions such that the AI score gives a numerical value between 0 (most certainly contains a lung lesion) and 100 (most certainly does not contain a lung lesion) to mirror the quality score from the manufacturing setting. Hence, the AI output can be interpreted as a risk score, which are widely used in medical practice. Radiologists were instructed that an AI score below 90 indicates that the AI algorithm suspects at least one lung lesion is visible and that the AI algorithm can make mistakes. When evaluating the AI score with a cutoff of 90 for mapping the numerical value onto a binary label (lung lesion visible yes/no), the standalone AI algorithm achieves a balanced accuracy of 82.2\% and a disease detection rate of 71.4\%. Analogously to the manufacturing task, the prediction performance of the standalone AI algorithm was not communicated to the participants.


As in the manufacturing task, the black-box AI algorithm was converted into an explainable AI by explaining the AI score via a heatmap, which is a state-of-the-art explanation technique for chest X-ray images in medicine \cite{Saporta.2022}. Further details about the implementation of the AI algorithm and the heatmap are provided in Supplement~\ref{sec:ai_system}.



The procedure was analogous to the manufacturing experiment. Before starting the experiment, radiologists had to confirm their area of specialization and give written consent to participate. Subsequently, the task was explained and the radiologists had to pass a tutorial on how to use the interface. After that radiologists were randomly assigned to one of the two treatments, i.e., either black-box AI or explainable AI. During the experiment, 50 chest X-ray images were randomly shown either in forward or reverse order. For each chest X-ray image, the radiologists had to answer whether at least one lung lesion is visible. The corresponding answers as well as the decision speed were tracked. Radiologists were given a maximum of 35 minutes to finish the inspection task of 50 chest X-ray images. Radiologists were given more time per image compared to factory workers in the manufacturing experiment to reflect the differences in manufacturing and clinical practice. In total, $N=5,650$ assessments of chest X-ray images were performed by the radiologists. Finally, all radiologists completed a post-experimental questionnaire (Supplement~\ref{sec:questionnaire}).

\subsection*{Study populations}

The inclusion were as follows. Participants had to be at least 18 years old. For the manufacturing task, participants additionally had to have no self-reported visual impairment. The exclusion criteria were preregistered and were as follows. In both studies, we excluded participants that failed the tutorial or did not finish the inspection task on time. Participants with obvious misbehavior were also excluded from our analyses. In the manufacturing task, this was the case for workers that approved all products (i.e., labeled no images as defective). In the medical task, this was the case for radiologists that assigned the same label for all 50 chest X-ray images. In both studies, participants whose performance with respect to balanced accuracy was more than three standard deviations worse than the mean of their respective treatment arm were excluded. We performed randomization checks to confirm that all treatment arms were demographically unbiased (Supplement~\ref{sec:randomization_checks}).

\subsubsection*{Study 1: Manufacturing experiment}

The manufacturing experiment was carried out from June 29 to July 8, 2021 on-site at a \emph{Siemens} factory in Zug, Switzerland. The objective of the field experiment was to get a real-world estimate of the treatment effect based on a representative sample of actual factory workers. Therefore, we only considered factory workers who were experienced in quality control practices. The factory workers were well familiar with the shown products and the visual inspection task. Overall, 56 factory workers (consisting of manufacturing employees, quality engineers, and team leaders) participated in our study. Out of them, all workers passed the tutorial; 6 did not finish on time; and 2 were excluded due to obvious misbehavior. The final sample consisted of 48 factory workers with an average working experience of 13.8 years. A larger sample size in our manufacturing experiment was not possible because the entire available workforce in one shift did not exceed 56 workers. Still, the experiment is well-powered as the treatment effect in the field experiment is considerably large. No additional financial incentive was given beyond the base salary to be representative of many real-world tasks from domain experts (e.g., as in manufacturing at \emph{Siemens}).

\subsubsection*{Study 2: Medical experiment}

The medical experiment was carried out via an online interface from February 27 to March 31, 2024. Actual radiologists based in the United States were recruited via MSI-ACI (https://site.msi-aci.com/). MSI-ACI paid a financial compensation to radiologists regardless of performance and adheres to the federal minimum wage in the United States. Overall, 122 radiologists started the study. Out of them, all passed the tutorial; 4 did not complete the study; 2 did not finish on time; and 3 were excluded due to obvious misbehavior. Hence, the final sample consisted of 113 radiologists with an average tenure as radiologist of 13.5 years.

\subsection*{Statistical analysis}


In manufacturing, it is common that all defects (e.g., missing components, wrong components, and faulty components) are considered equally bad \cite{Juran.1979, Hoyle.2007}, so that the product to inspect could be either functional or non-functional. Analogously, in the medical setting, either lung lesions were present or not. Hence, in both settings, the outcomes were binary. Therefore, task performance in the visual inspection tasks between the participants' assessments and the ground-truth labels was computed via (1)~balanced accuracy (i.e., average sensitivity across faultless and defective products) and (2)~defect/disease detection rate. For (1), the balanced accuracy is calculated via $0.5 \times [\mathit{TP}/P + \mathit{TN}/N]$ with true positives $\mathit{TP}$, positives $P$, true negatives $\mathit{TN}$, and negatives $N$. Here, we used balanced accuracy since it accounts for imbalanced distributions of labels by equally weighing the performance on each label, thus following best practice \cite{Hollon.2023}. In contrast, the standard accuracy score would not account for the imbalanced distribution of positive and negative labels encountered in both settings (i.e., 172 faultless products and 28 defective products in the manufacturing setting; 7 chest X-ray images with and 43 without lung lesions in the medical setting). For (2), the defect/disease detection rate is defined as $\mathit{TN}/N$, where defective products and chest X-ray images with lung lesions were defined as negatives. In our manufacturing setting, missing a defective product has more severe implications than labeling a faultless product as defective. In medicine, missing a lung lesion on a chest X-ray image has more severe implications than additionally performing a CT scan for a healthy patient. Hence, it is crucial to find the negative samples.


All statistical tests in the results are based on one-sided Welch's $t$-tests. We further used ordinary least square (OLS) regression models to estimate the treatment effect of explainable AI on task performance. The OLS models are estimated via
\begin{equation}
Y_{i} = \beta_{0} + \beta_{1}\,\mathit{Treatment}_{i} + \varepsilon_{i},
\label{eqn:regression_main}
\end{equation}
where $Y_{i}$ is the observed task performance (i.e., balanced accuracy or defect/disease detection rate), $Treatment_{i}$ is a binary variable which equals 0 if participant $i$ received the black-box AI treatment and 1 if participant $i$ received the explainable AI treatment. A significance level of $\alpha = 0.05$ was preregistered.

\subsection*{Robustness checks}

We conducted the following robustness checks. First, we repeated our analyses using precision as an additional task performance metric (Supplement~\ref{sec:precision}). Second, we repeated the OLS regression models with additional participant-specific controls to estimate the treatment effect of explainable AI (Supplement~\ref{sec:regression_models}). Third, we estimated the treatment effect with quasi-binomial regression (Supplement~\ref{sec:regression_models}). Fourth, we estimated the regression models including participants that were previously excluded due to obvious misbehavior or because they did not finish the inspection task in time (Supplement~\ref{sec:excluded_participants}). All robustness checks yielded conclusive findings.

To demonstrate that the heatmaps in our medical setting are robust with respect to the choice of algorithm, we used two additional, different algorithms to generate heatmaps. We find that different algorithms lead to similar heatmaps (Supplement~\ref{sec:heatmap_checks}). 

\subsection*{Comparison to non-experts}

Additionally, we repeated the manufacturing task with non-experts recruited from Amazon MTurk as a robustness check (Supplement~\ref{sec:online_experiment}). Typically, non-experts can not leverage explanations in the same way as domain experts due to missing domain knowledge. Hence, we were interested whether the large treatment effect of explainable AI on task performance we observed with domain experts transfers also to non-experts.

We found that non-experts supported by explainable AI also achieved a higher task performance than non-experts supported by black-box AI. Task performance of non-experts augmented with explainable AI was improved by 6.3 percentage points with respect to balanced accuracy. However, the treatment effect of explainable AI is slightly smaller as compared to the experiment with domain experts (where the balanced accuracy increased by 7.7 percentage points). Furthermore, non-experts with explainable AI achieved a higher defect detection rate with an improvement of 11.3 percentage points. This is of a similar effect size as in the real-world experiment (11.0 percentage points). The treatment effects in both metrics were again statistically significant (balanced accuracy: $\beta=6.252$, $\mathit{SE}=1.733$, $P<0.001$; defect detection rate: $\beta=11.271$, $\mathit{SE}=3.276$, $P=0.001$). For more details, see Supplement~\ref{sec:online_experiment}.

\bibliography{references}

\clearpage

\section*{Acknowledgements}
The authors thank all participants, as well as Davide Vecchione, Burak Seyid, and Alexander Dierolf for enabling the manufacturing experiment at \emph{Siemens}. SF acknowledges funding via Swiss National Science Foundation Grant 186932. TN acknowledges funding from \emph{Siemens}; however, without competing interest.

\section*{Author contributions}
All authors contributed to the research design, data analysis, interpretation of results, and writing of the paper. JS and BK performed the manufacturing experiment. SS performed the medical experiment. 

\section*{Competing interests}
The funding bodies had no control over design, conduct, data, analysis, review, reporting, or interpretation of the research conducted.

\section*{Data and code availability}
All analyses were conducted using Python (3.11) with \emph{numpy} (1.24.3), \emph{pandas} (1.5.3), \emph{scipy} (1.11.1), and \emph{statsmodels} (0.14.0). The data visualizations were created with \emph{seaborn} (0.12.2) and \emph{matplotlib} (3.7.1). The data and code to reproduce the results from all studies will be made publicly available at \url{https://osf.io/} upon publication.

\clearpage

\begin{center}
\LARGE Supplements
\end{center}
\vspace{1cm}
\begin{appendices}

\renewcommand\appendixname{Supplement}

\newcommand{\hbAppendixPreoverrule}{S}

\renewcommand{\thefigure}{\hbAppendixPreoverrule\arabic{figure}}
\setcounter{figure}{0}
\renewcommand{\thetable}{\hbAppendixPreoverrule\arabic{table}} 
\setcounter{table}{0}
\renewcommand{\theequation}{\hbAppendixPreoverrule\arabic{equation}} 
\setcounter{equation}{0}

\tableofcontents

\newpage
\section{Extended literature review}
\label{sec:literature_review}

In the following, we provide an extended literature review of explainable artificial intelligence (AI). In particular, we differentiate research on explainable AI in computer science (which is primarily focused on methodological outcomes) from our work (which is focused on behavioral science outcomes). An overview is provided in \Cref{tab:literature}.

\subsection{Explainable AI in computer science}
In the field of computer science, the primary objective concerning explainable AI is to develop and evaluate new methods to achieve better transparency of AI algorithms. For a general overview of explainable AI, see for example \cite{Guidotti.2018, Samek.2019, Linardatos.2020, Samek.2021}. AI algorithms can be broadly divided into two categories: algorithms that are considered to be inherently interpretable and algorithms that are not due to their complexity \cite{Rudin.2019}. The latter are often referred to as black-box algorithms.


An inherently interpretable model is linear regression, where the decision-making can be directly followed by inspecting the coefficients. Extensions of linear regression are generalized linear models (GLMs) and generalized additive models (GAMs) \cite{Nelder.1972, Hastie.1990}. GLMs were introduced as a unification of various methods that allow for different distributions of the dependent variable (e.g., a binary dependent variable as in logistic regression). GAMs were introduced to also allow for non-linear relationships between an independent and the dependent variable, which can be modeled for example with decision trees or shallow neural networks (see e.g., \cite{Lou.2012, Kraus.2023}). While GLMs and GAMs are still considered to be inherently interpretable, they are not as straightforward to interpret as linear regression.

Decision trees are also considered to be inherently interpretable by simply following the decision rules from the root to the leaf nodes. Other inherently interpretable models are na{\"i}ve Bayes classifier, k-nearest neighbor algorithm, rule-based learning, etc. (see \cite{Molnar.2019} for an introduction).


In contrast, post-hoc explanation techniques can be applied to better understand black-box algorithms such as neural networks. These explanation techniques are applied after the AI algorithm has been trained. Post-hoc explanation methods can be divided into global and local methods \cite{Linardatos.2020}, where the former aim at explaining the algorithm's overall decision-making process while the latter provide explanations for a single, specific input. An example of global methods are feature importance rankings, which rank the features based on their importance in predicting a model’s outcome, usually measured across the entire model rather than for individual predictions. A further example are partial dependence plots, which show the effect of a single feature on the predicted outcome of a model, averaged over a dataset. Prominent examples of local methods are local interpretable model-agnostic explanations (\emph{LIME}) and SHapley Additive exPlanations (\emph{SHAP}) \cite{Ribeiro.2016, Lundberg.2017}. \emph{LIME} approximates a black-box model locally around the prediction with an interpretable model (like a linear model) to explain individual predictions. \emph{SHAP} leverages a concept from cooperative game theory (Shapley values \cite{Shapley.1953}) to explain the output of a model by computing the contribution of each feature to the prediction while also considering possible interaction effects. Post-hoc explanation methods can be further categorized into model-specific and model-agnostic methods. Model-specific methods are designed for a specific class of AI algorithms or even for a single AI algorithm. Model-specific methods exist, for example, for convolutional neural networks \cite{Selvaraju.2017} or for kernel-based AI algorithms such as support vector machines \cite{Hansen.2011}. In contrast, model-agnostic methods can be applied to any AI algorithm; notable examples are \emph{LIME} and \emph{SHAP}. Another dimension to differentiate post-hoc explanation methods is for which data type (tabular, text, audio, images, etc.) the method was developed. In this study, we focus on images, and, in the following, we thus present some of the most relevant post-hoc explanation methods for AI algorithms in computer vision. For general overviews of explainable AI in computer vision, we refer to \cite{Buhrmester.2021, Ibrahim.2023}.


One of the earliest attempts to explain convolutional neural networks (CNN), which are nowadays widely used in computer vision, was made by Zeiler and Fergus \cite{Zeiler.2014}. Therein, the authors present the deconvolutional network (\emph{DeconvNet}) as a visualization method that maps feature activations back to the input image. Additionally, a simple technique called occlusion was discussed as a method for explaining the predictions of a CNN. For that, different portions of the input image are systematically occluded with a grey square, and the impact on the output of the network is observed. Significant changes in the output probabilities indicate the regions of the image most important for classification.

In \cite{Simonyan.2013}, a method was introduced for generating saliency maps by computing the gradient of the output category with respect to the input image. This technique highlights the regions of the image that contribute most to the model's classification decision, offering a straightforward visual explanation of where the network is \textquote{looking} to make its predictions.

Layer-wise relevance propagation \emph{(LRP}) backtracks the output decision of the network through the layers to assign relevance scores to individual pixels. This method helps in understanding which parts of the input image were most relevant for the model's decision, emphasizing a layer-by-layer decomposition of the prediction \cite{Bach.2015}.

Integrated gradients attribute the prediction of a neural network to its input features, calculating the gradients of the output prediction with respect to the input image. It integrates these gradients along the path from a baseline (zero input) to the actual input, offering a way to visualize the importance of each pixel \cite{Sundararajan.2017}.

Deep learning important features (\emph{DeepLIFT}) compares the activation of each neuron to its `reference activation' and assigns contribution scores according to the difference. This method can identify which features of the input contribute to differences in the output from some baseline, offering a more detailed view than simple gradient-based methods \cite{Shrikumar.2017}.

By using the global average pooling layers in CNNs, class activation mapping (\emph{CAM}) generates heatmaps that highlight the discriminative parts of the image used by the network to identify specific classes, facilitating visual explanations of model decisions \cite{Zhou.2016}. Several extensions to this approach exist \cite{Selvaraju.2017, Chattopadhay.2018, BanyMuhammad.2021}, with \emph{GradCAM} being one of the most often used approach \cite{Selvaraju.2017}. In contrast to \emph{CAM}, \emph{GradCAM} employs a gradient-based approach to generate heatmaps, and, as a result, no changes to the network architecture are required.

Counterfactual explanations provide insights by showing how a small change in the input image could change the classification result. This method helps in understanding model decisions by answering \textquote{what-if} scenarios, offering a direct way to comprehend how the model might react to different inputs \cite{Verma.2020}.

A special case of post-hoc explanation in computer vision are anomaly heatmaps. These were developed for unsupervised computer vision AI algorithms, i.e., algorithms that do not require a labeled dataset but rather aim at finding anomalies automatically \cite{Bergmann.2019a, Zipfel.2023, Liu.2024}.


A plethora of post-hoc explanation methods exists and often it is not obvious which method to choose. Thus, different evaluation measures for post-hoc explanation methods have been proposed \cite{DoshiVelez.2017}. These include fidelity, which measures how accurately the explanations reflect the decisions of the underlying model \cite{Ribeiro.2016, Plumb.2018, Hooker.2019, Saporta.2022}, and robustness, assessing stability under small changes in the input \cite{AlvarezMelis.2018}. Another measure is human-interpretability, which examines how understandable the explanations are to humans \cite{Narayanan.2018}. But also application-grounded measures have been proposed, where the evaluation metric is how humans perform in a certain task \cite{Jesus.2021, Amarasinghe.2024}.


In another stream of literature in computer science, tools that help AI engineers with designing and training AI algorithms are developed. Especially, deep neural networks are difficult to train and require a certain amount of experience. Therefore, visualization tools have emerged that facilitate this task (e.g., see \cite{Liu.2017, Ming.2017, Pezzotti.2018, Strobelt.2018}).

\subsection{Explainable AI in behavioral science}

In behavioral science, different outcomes of human-AI collaboration have been studied. Human delegation of tasks and decisions to AI algorithms has been intensively studied recently \cite{Fuegener.2021, Fuegener.2021b, Fuegener.2021c, Candrian.2022}. Specifically, it has been examined whether the use of explainable AI can increase the likelihood of humans delegating decisions to AI algorithms \cite{Bauer.2023}.


Algorithm aversion refers to the phenomenon where humans are reluctant to use algorithms \cite{Dietvorst.2015, Dietvorst.2016, Dietvorst.2020, Burton.2020, Ochmann.2020, Hou.2021, Bogert.2021, Cadario.2021, Sun.2020, Kawaguchi.2020, Castelo.2019, Yeomans.2019}. To overcome human aversion towards AI algorithms, it has been hypothesized that providing an explanation of an AI's decision may be beneficial. This has been tested with mixed findings \cite{David.2021, Flei.2024}. Missing trust in an AI's decision can lead to algorithm aversion \cite{Choung.2023}. Therefore, previous studies investigated whether explainable AI can increase the trust in AI algorithms \cite{Cai.2019b, Nourani.2019, BranleyBell.2020, Zhang.2020, LancasterFarrell.2022, Panigutti.2022, Panigutti.2023, Leichtmann.2023, Sivaraman.2023}.


A contrary phenomenon to algorithm aversion is overreliance \cite{Bucinca.2021, Vasconcelos.2023}, where humans place too much trust in AI algorithms, potentially overlooking or ignoring their limitations. Previous research has found mixed results on whether explainability of AI leads to decreased overreliance \cite{Bansal.2021, Vasconcelos.2023, Chen.2023}.


However, for a good task performance, it is crucial that humans only adhere to correct AI predictions and overrule wrong ones, which is also referred to as appropriate reliance \cite{Lee.2004}. The effect of explainable AI on task performance has been studied previously. In the majority of studies, however, non-experts (e.g., via Amazon Mechanical Turk or university students) were recruited and those oftentimes performed simplified, non-realistic, or even non-relevant tasks (see, e.g., \cite{Green.2019, Lage.2019, Lai.2019, Cai.2019, Bucinca.2020, Carton.2020, Lai.2020, Yang.2020, Chu.2020, Zhang.2020, Alqaraawi.2020, Wang.2021, PoursabziSangdeh.2021, vanderWaa.2021, Alufaisan.2021, Bansal.2021, Bucinca.2021, Kim.2022, Chen.2023, Leichtmann.2023b, Muller.2024}). It has been hypothesized that non-experts can not fully harness explanations due to a lack of domain knowledge \cite{Chen.2023}. Thus, an empirical evaluation of the effect of explainable AI on task performance in real job tasks, requires actual domain experts of those tasks \cite{DoshiVelez.2017}.

Previous studies that recruited domain experts to perform real-job tasks have other drawbacks. For example, prior work has compared the effect of explainable AI against humans alone \cite{Lundberg.2020, Das.2023}. Others have used expert annotations as a proxy for explainable AI \cite{Gaube.2023} or research designs that prevent isolating the treatment effect of explainable AI on task performance \cite{Jesus.2021, Jabbour.2023, Sivaraman.2023}. Finally, also other outcomes have been studied such as trust, confidence, and perceived usefulness \cite{Metta.2023, Nagendran.2023}, while, as our novelty, we add by focusing on task performance with domain experts.

\begin{table}[H]
\onehalfspacing
\footnotesize
\singlespacing
\begin{center}
\caption{\textbf{Overview of key literature on explainable AI.}} 
\label{tab:literature}
\begin{tabular}{p{1.2cm} p{2cm} p{3.5cm} p{1.5cm} p{4cm}}
\toprule
 Domain & Concept & Research summary & Dependent variable & References (examples) \\
\midrule
Computer science & New explanation methods & Derivations of new methods where the focus is on mathematical / algorithmic contributions & n/a & \cite{Nelder.1972, Hastie.1990, Lou.2012, Kraus.2023, Ribeiro.2016, Lundberg.2017, Hansen.2011, Zeiler.2014, Simonyan.2013, Bach.2015, Sundararajan.2017, Shrikumar.2017, Zhou.2016, Selvaraju.2017, Chattopadhay.2018, BanyMuhammad.2021, Verma.2020, Bergmann.2019a, Zipfel.2023} \\
 \cmidrule(lr){2-5}
 & Benchmarking methods/ datasets & Proposing new methods or datasets to benchmark the performance of explainable AI & n/a & \cite{Plumb.2018, Hooker.2019, Saporta.2022, AlvarezMelis.2018, Narayanan.2018, Jesus.2021, Amarasinghe.2024} \\
 \cmidrule(lr){2-5}
 & New visualization tools & Visualization tools that facilitate the development of complex machine learning models & n/a & \cite{Liu.2017, Ming.2017, Pezzotti.2018, Strobelt.2018} \\
\midrule
Behavioral science & Delegation between humans and AI & Humans avoid delegation to algorithms & Delegation frequency & \cite{Fuegener.2021, Fuegener.2021b, Fuegener.2021c, Candrian.2022, Bauer.2023} \\
\cmidrule(lr){2-5}
 & Algorithm aversion & Humans reject advice from algorithm & Adherence & \cite{Dietvorst.2015, Dietvorst.2016, Dietvorst.2020, Burton.2020, Ochmann.2020, Hou.2021, Bogert.2021, Cadario.2021, Sun.2020, Kawaguchi.2020, Castelo.2019, Yeomans.2019, David.2021, Flei.2024} \\
\cmidrule(lr){2-5}
 & Trust in AI & Humans do not trust AI algorithms & Trust & \cite{Cai.2019b, Nourani.2019, BranleyBell.2020, Zhang.2020, LancasterFarrell.2022, Panigutti.2022, Panigutti.2023, Leichtmann.2023, Sivaraman.2023} \\
\cmidrule(lr){2-5}
 & Overreliance on AI & Humans follow advice from algorithms blindly & Overreliance & \cite{Bansal.2021, Vasconcelos.2023, Chen.2023} \\
\cmidrule(lr){2-5}
 & Task performance in response to explainable AI & Comparison of black-box AI vs explainable AI for task performance using unrealistic tasks, non-experts, or non-causal research designs & Task performance & \cite{Green.2019, Lage.2019, Lai.2019, Cai.2019, Bucinca.2020, Carton.2020, Lai.2020, Yang.2020, Chu.2020, Zhang.2020, Alqaraawi.2020, Wang.2021, PoursabziSangdeh.2021, vanderWaa.2021, Alufaisan.2021, Bansal.2021, Bucinca.2021, Kim.2022, Chen.2023, Leichtmann.2023b, Muller.2024, Lundberg.2020, Das.2023, Gaube.2023, Jesus.2021, Jabbour.2023, Sivaraman.2023} \\
 \cmidrule(lr){3-5}
 & & Real-world job tasks with domain experts for estimating treatment effects of explainable AI vs black-box AI & Task performance & \textbf{ours} \\
\bottomrule
\end{tabular}
\end{center}
\end{table}

\clearpage
\section{Research setting}
\label{sec:research_setting}
\subsection{Manufacturing setting}

Poor quality generates 10\% to 15\% of the operating expenses in manufacturing.\footnote{\footnotesize American Society for Quality. \emph{Cost of Quality (COQ)}. URL: https://asq.org/quality-resources/cost-of-quality, last accessed on \today.} Identifying defective products before they move downstream in the value chain is essential to maintain a high operational performance. For this purpose, manufacturers conduct visual quality inspections to assess whether products have defects (e.g., assembly errors or surface damages) \cite{Bergmann.2019a}. In  manufacturing operations, many quality inspections are still conducted manually, which is often a tedious, tiring, and error-prone task. AI offers promising opportunities to overcome these drawbacks by supporting factory workers in automatically detecting quality defects before products are sold to customers. Specifically, AI can assist workers in detecting the location and type of error so rework can be conducted more effectively and efficiently. Therefore, AI algorithms enable factory workers to be more productive by focusing on their key value-creating work tasks.

\begin{figure}[H]
\centering
\begin{subfigure}{.5\textwidth}
  \centering
  \caption{\footnotesize{}}
  \includegraphics[width=0.95\linewidth]{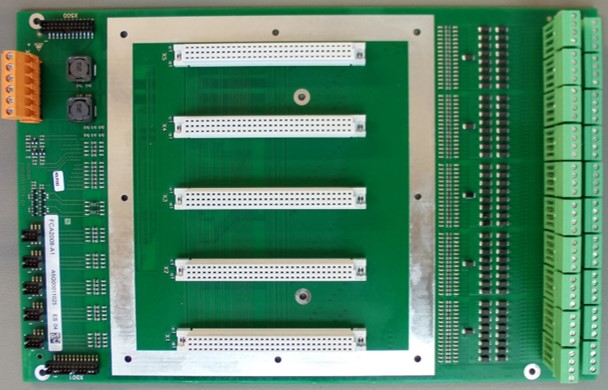}
\end{subfigure}%
\begin{subfigure}{.5\textwidth}
  \centering
  \caption{\footnotesize{}}
  \includegraphics[width=0.8\linewidth]{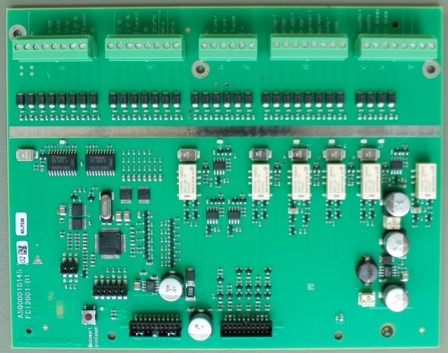}
\end{subfigure}
\\ \vspace{0.5cm}
\begin{subfigure}{.5\textwidth}
  \centering
  \caption{\footnotesize{}}
  \includegraphics[width=0.95\linewidth]{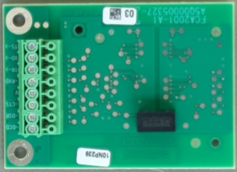}
\end{subfigure}%
\begin{subfigure}{.5\textwidth}
  \centering
  \caption{\footnotesize{}}
  \includegraphics[width=0.8\linewidth]{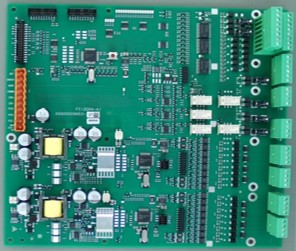}
\end{subfigure}
\caption{\textbf{Four types of electronic products (printed circuit boards)}. (\textbf{\textsf{A-D}})~Exemplary images of faultless products that were inspected during the experiment.}
\label{fig:products}
\end{figure}

\begin{figure}[H]
\centering
\begin{subfigure}{1\textwidth}
  \centering
  \caption{\footnotesize{}}
  \includegraphics[width=0.8\linewidth]{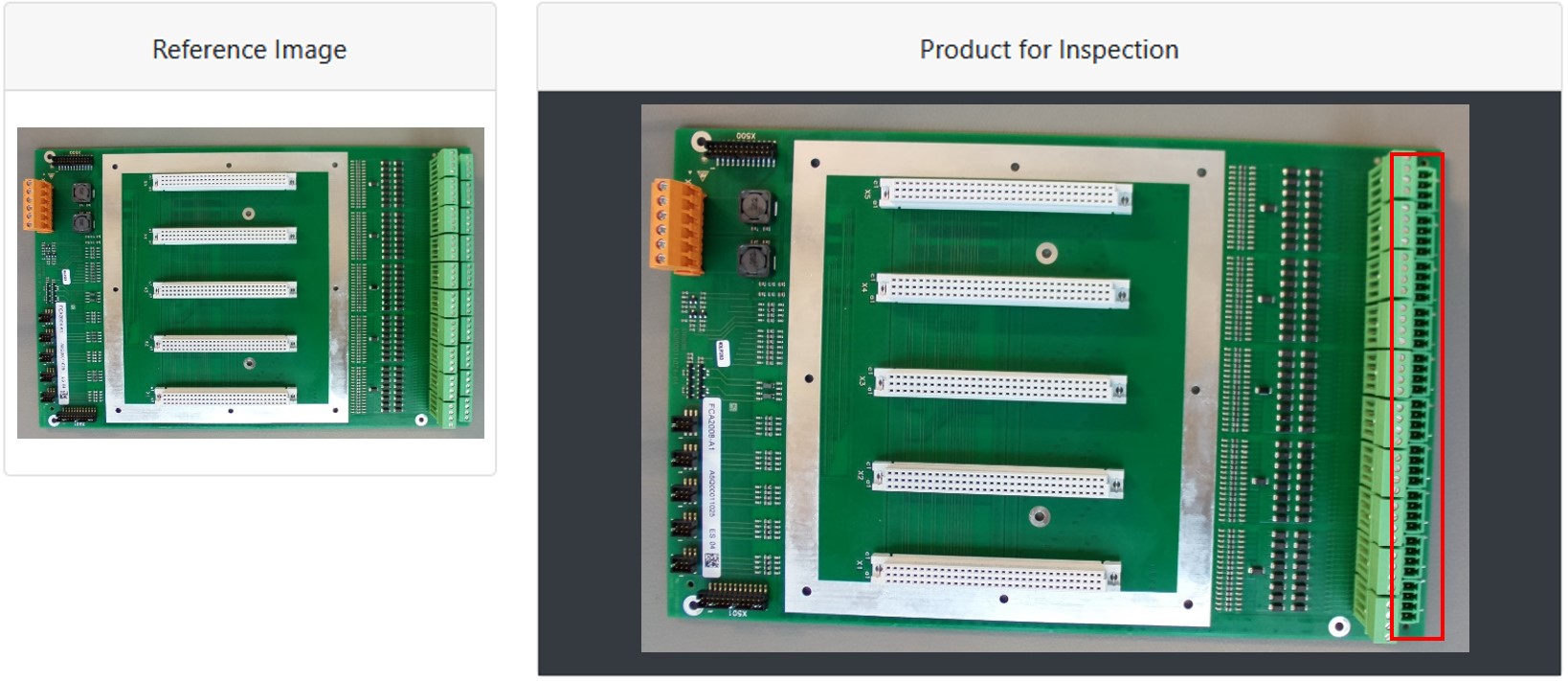}
\end{subfigure}%
\\ \vspace{0.5cm}
\begin{subfigure}{1\textwidth}
  \centering
  \caption{\footnotesize{}}
  \includegraphics[width=0.8\linewidth]{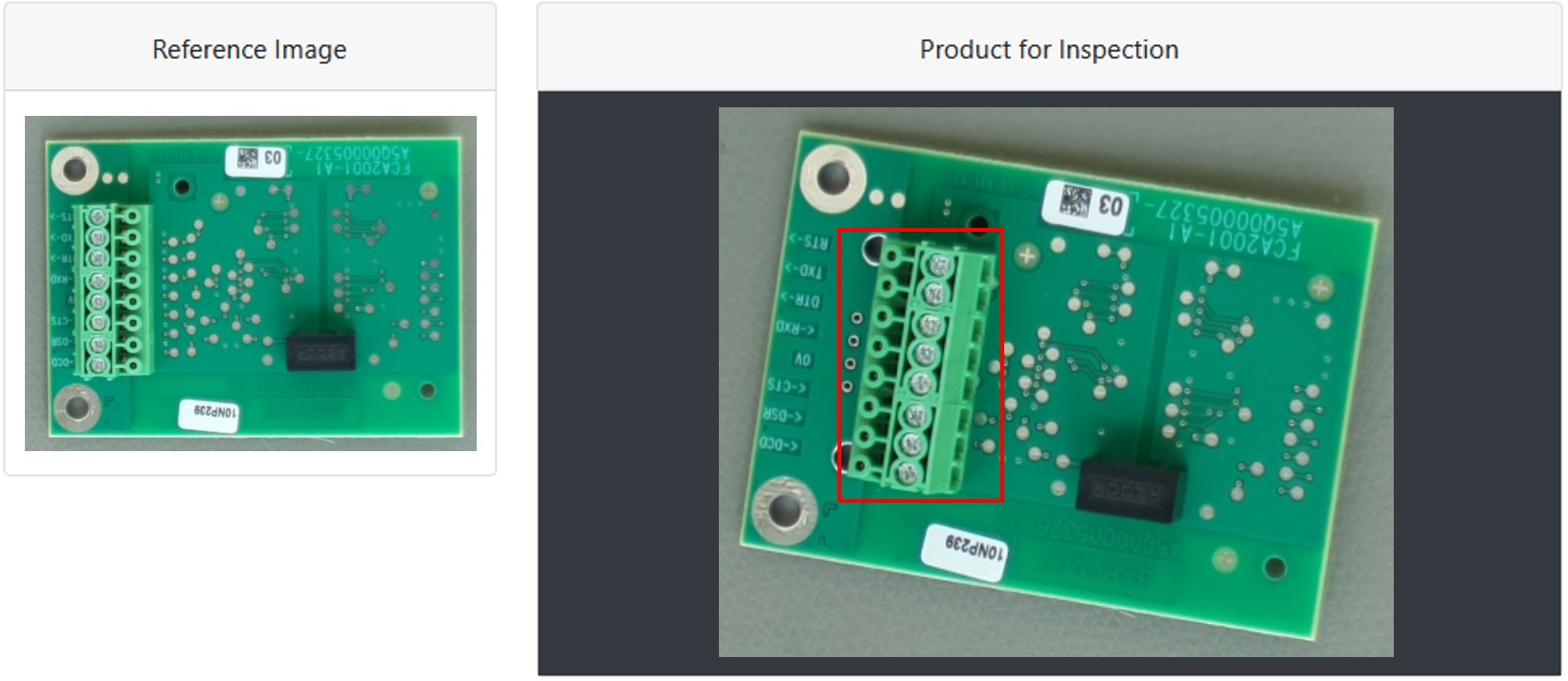}
\end{subfigure}
\\ \vspace{0.5cm}
\begin{subfigure}{1\textwidth}
  \centering
  \caption{\footnotesize{}}
  \includegraphics[width=0.8\linewidth]{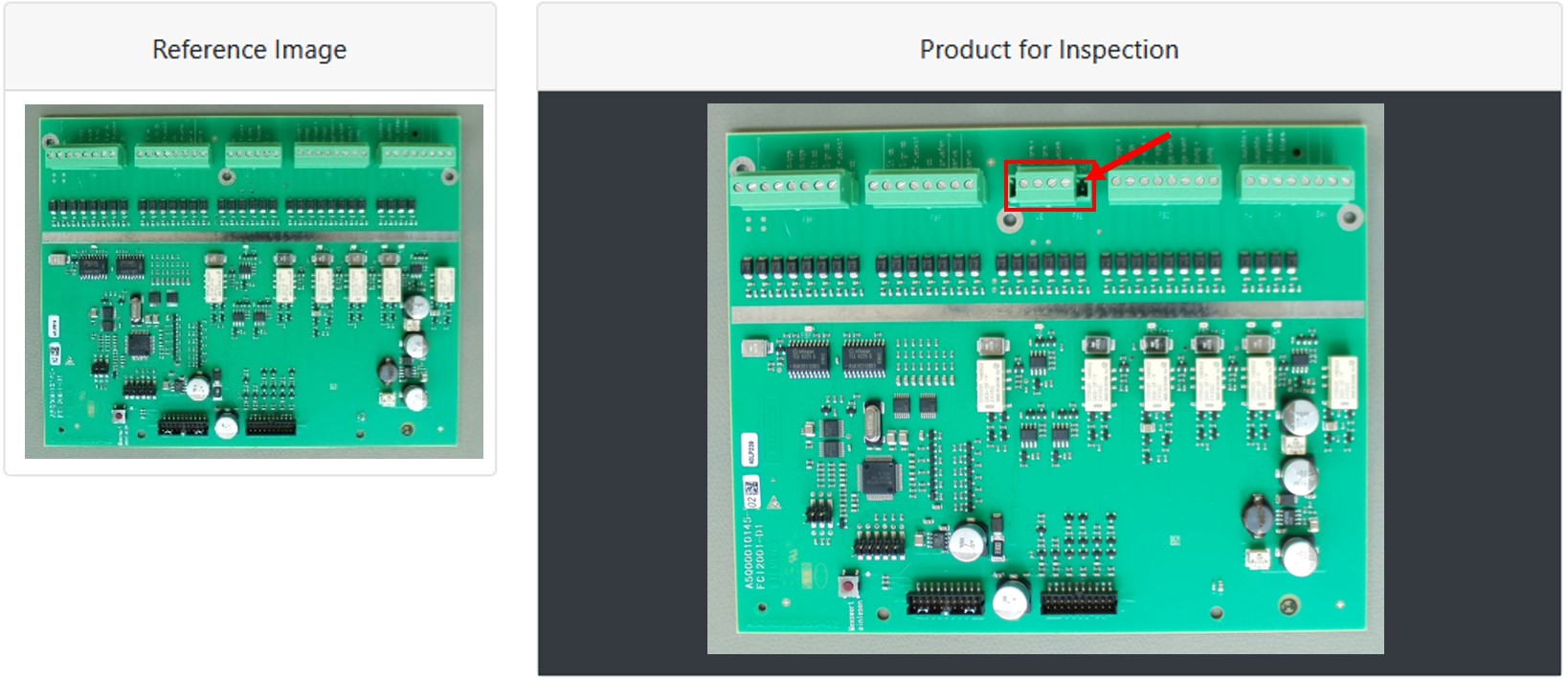}
\end{subfigure}%
\caption{\textbf{Examples of quality defects.} (\textbf{\textsf{A}})~Example of a defective product with wrong components. (\textbf{\textsf{B}})~Example of a defective product with a component assembled in the wrong orientation. (\textbf{\textsf{C}})~Example of a defective product with a faulty component.}
\label{fig:defects}
\end{figure}

Our research was carried out at \emph{Siemens} Smart Infrastructure in Zug, Switzerland. To test our hypotheses, the company provided us with real-world product images (each with $1920 \times 1080$ pixels) from their factory. The images comprise four different types of electronic products, all of which are printed circuit boards. \Cref{fig:products} shows example images of the four types of electronic products that were inspected during the experiment. \Cref{fig:defects} shows three examples of quality defects, which include products with wrong components, products with assembly errors, and products with faulty components.

Overall, we received two datasets. The first dataset comprised 200 images, including 43 correct products and 7 defective products for each of the four product types. All experiments (and thus the empirical results in the main analysis) are based on the first dataset. The second dataset comprised 200 additional faultless images (50 for each of the four product types). These images were used to train the AI algorithm that was used to compute the quality scores in the experiment (see Supplement~\ref{sec:ai_system}).  

\subsection{Medical setting}

Chest radiography (capturing X-ray images) is a widely performed diagnostic imaging test across the world and plays a crucial role in the screening, diagnosis, and management of numerous diseases that pose a threat to life \cite{Irvin.2019}. One of such diseases are lung lesions, which include lung nodules and masses in our experiment. Lung nodules are common and are encountered on roughly one out of 500 chest X-ray images \cite{UTS.2024}. 

Overlooking a lung lesion on a chest X-ray can have serious, potentially life-threatening consequences for patients. The failure to detect a lesion at an early stage can lead to a delay in diagnosis and treatment, allowing diseases to progress to more advanced stages. This can significantly worsen the prognosis for conditions such as lung cancer, tuberculosis, and pneumonia, where early intervention can often lead to better outcomes. Beyond the immediate health risks, there are also implications for patient care, including increased medical costs due to more complex and prolonged treatment that may become necessary as a disease progresses. However, subtle lung lesions can be easily overlooked even without any constraints on how long radiologists are allowed to inspect the chest X-ray image \cite{Oestmann.1988}. Given these reasons, identifying lung lesions on chest X-ray images is an important, non-trivial task in daily, medical care. To that end, giving physicians a decision aid for this task is crucial.

\begin{figure}[H]
\centering
\begin{subfigure}{.5\textwidth}
  \centering
  \caption{\footnotesize{}}
  \includegraphics[width=0.8\linewidth]{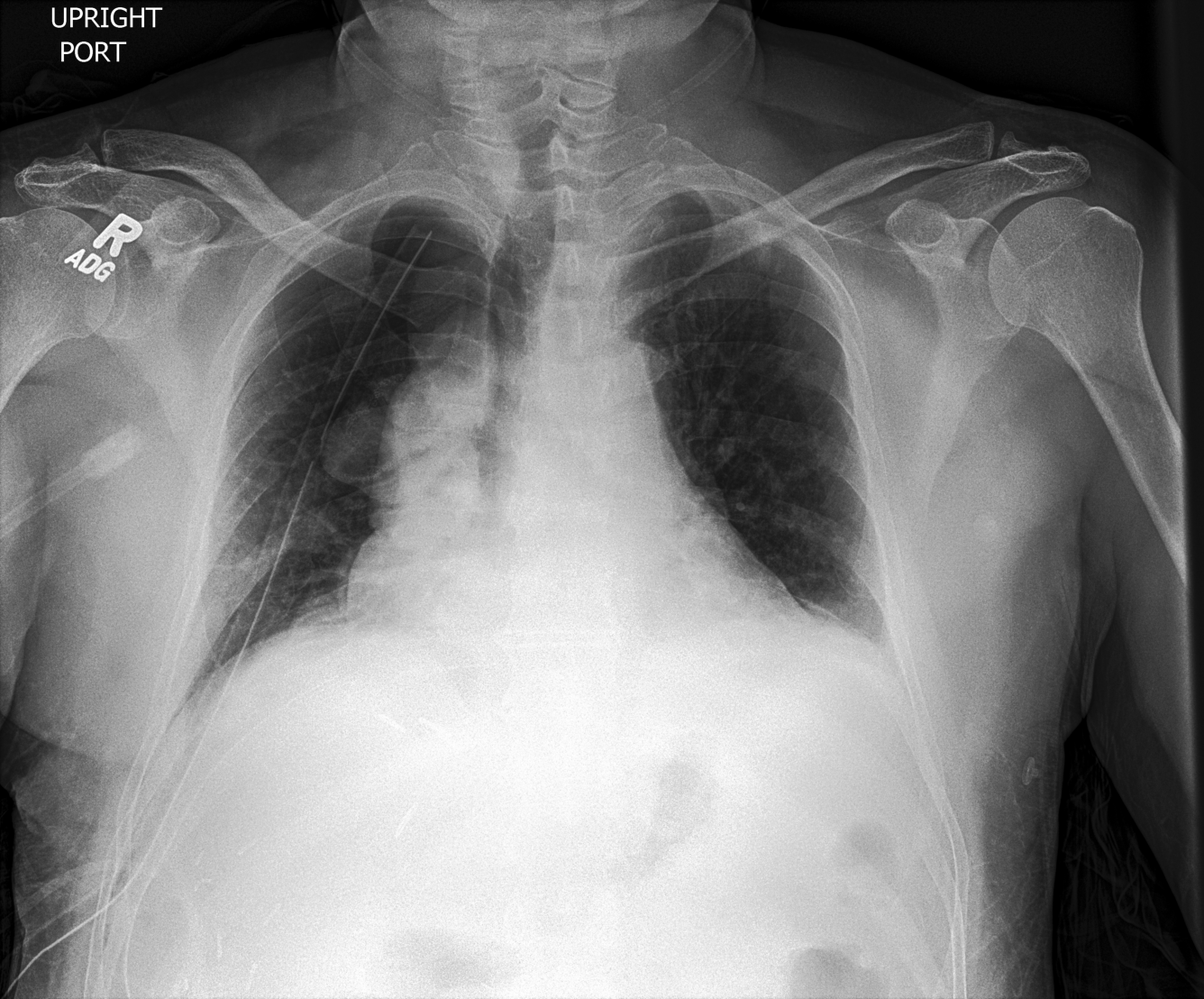}
\end{subfigure}%
\begin{subfigure}{.5\textwidth}
  \centering
  \caption{\footnotesize{}}
  \includegraphics[width=0.8\linewidth]{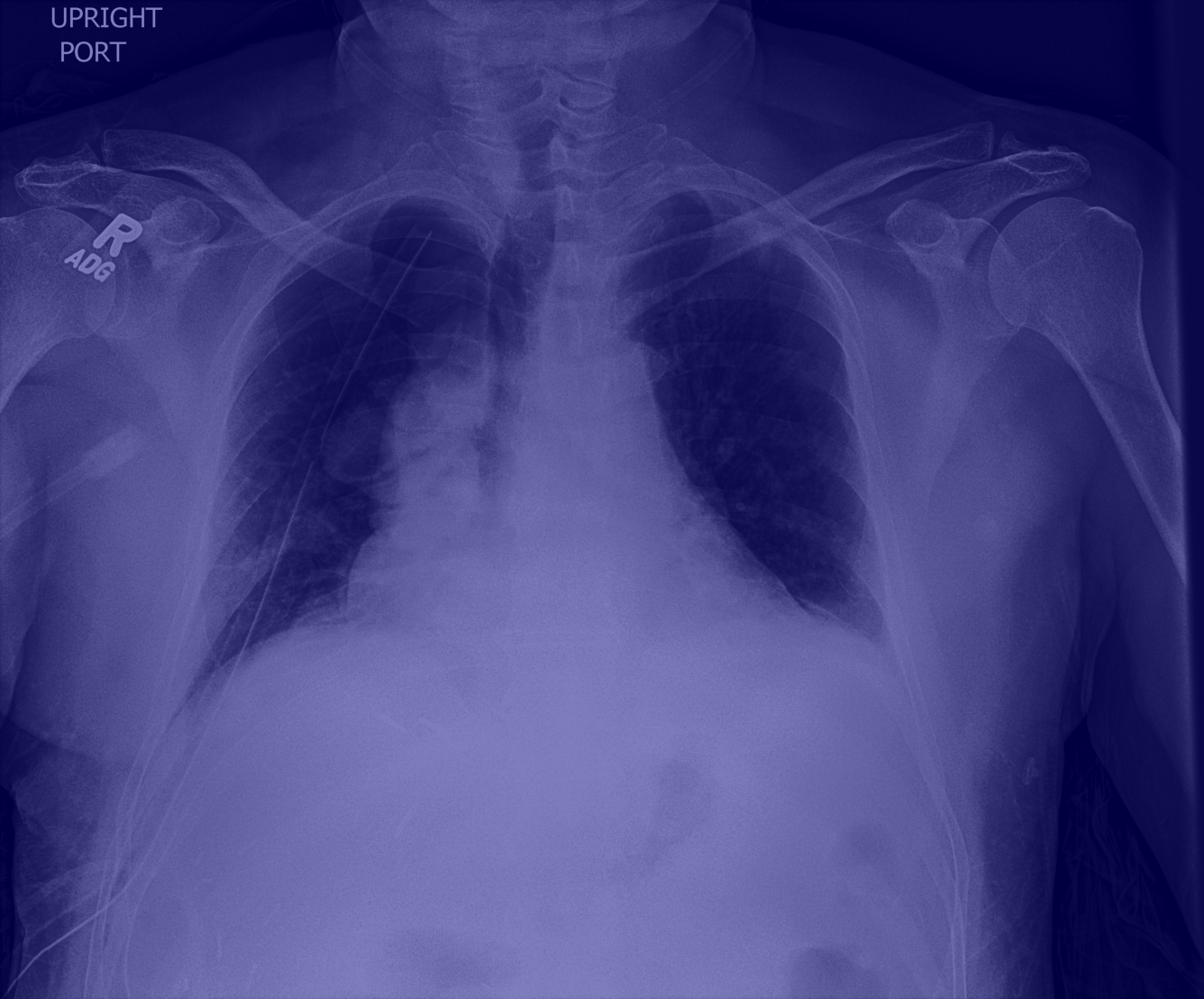}
\end{subfigure}
\begin{subfigure}{.5\textwidth}
  \centering
  \caption{\footnotesize{}}
  \includegraphics[width=0.8\linewidth]{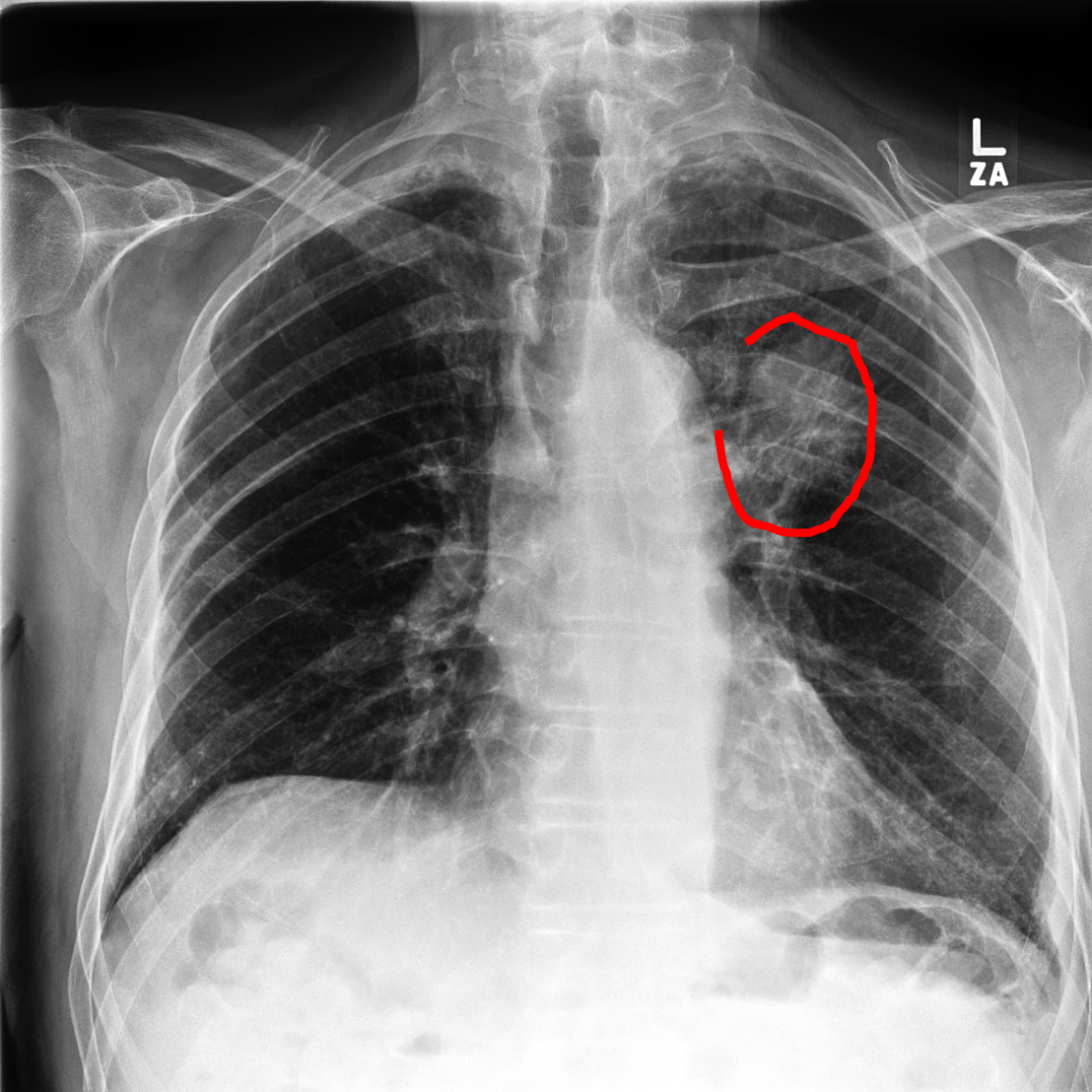}
\end{subfigure}%
\begin{subfigure}{.5\textwidth}
  \centering
  \caption{\footnotesize{}}
  \includegraphics[width=0.8\linewidth]{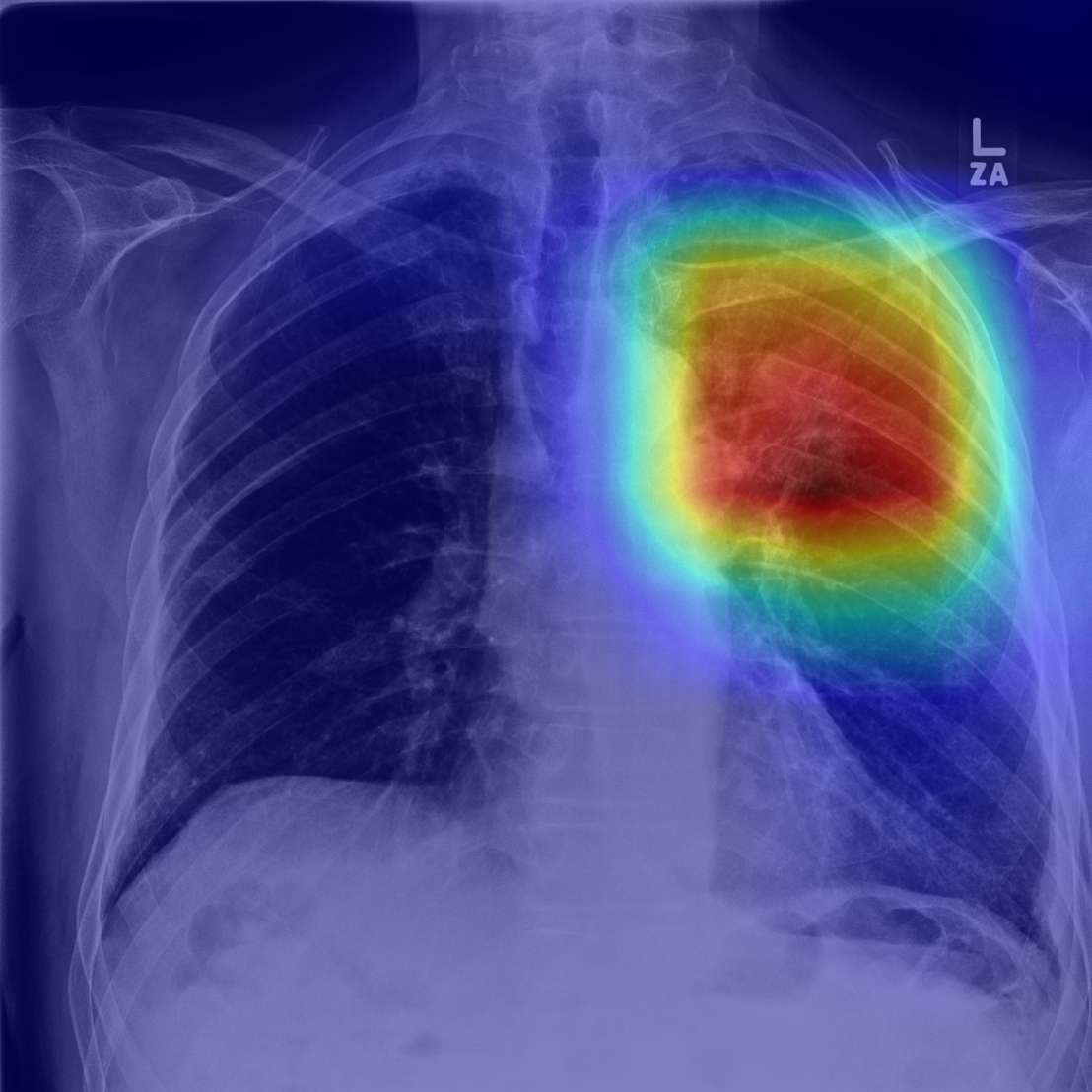}
\end{subfigure}
\caption{\textbf{Example chest X-ray images with corresponding heatmaps}. (\textbf{\textsf{A}})~Chest X-ray image without lung lesions. (\textbf{\textsf{B}})~Heatmap overlaid over the chest X-ray image from \textbf{\textsf{A}}. (\textbf{\textsf{C}})~Chest X-ray image with a lung lesion annotated in red by an experienced radiologist. (\textbf{\textsf{D}})~Heatmap overlaid over the chest X-ray image from \textbf{\textsf{C}}.}
\label{fig:heatmaps_medical}
\end{figure}

Example chest X-ray images including the corresponding heatmaps are provided in \Cref{fig:heatmaps_medical}.

\clearpage

\section{Implementation of AI algorithm}
\label{sec:ai_system}

\subsection{Manufacturing setting}
As part of this research, we implemented an AI algorithm that provided the predictions (i.e., quality scores) that were shown to the participants during the manufacturing experiment. Our AI algorithm builds upon unsupervised anomaly detection \cite{Pimentel.2014} and, as such, follows common standards in industry for visually analyzing the quality of product images \cite{Bergmann.2019a, Bergmann.2019b}. Algorithms based on unsupervised anomaly detection are particularly suitable for industrial settings because they only require a set of faultless product images to be trained. Therefore, there is no need to specify defect types beforehand, which also allows identifying quality defects that have never been observed before. This is reflected in anomaly detection where product images with sufficiently large deviations from \textquote{normal} products are labeled as defective. 

\textbf{AI algorithm.} In our case, the AI algorithm performs unsupervised anomaly detection as follows \cite{Pimentel.2014}. First, we are given an existing training set $T$ with images of faultless products. Upon deployment, an out-of-sample product image $x$ is subject to assessment; that is, whether it is similar to any of the images $t \in T$ and thus likely faultless or whether it is highly dissimilar and thus likely defective. Here, anomaly detection compares the similarity (with regard to some similarity function $d$) between the new image $x$ and the existing images $t \in T$. For each, a similarity $d(x, t)$, for all $t \in T$ is computed. If the similarity $d$ falls below a certain threshold $\theta^*$, an image is labeled as defective.  

\textbf{AI predictions.} The similarity of product images is computed by following best practice in computer vision. As such, we refrain from simply computing the L2-norm (or some other norm) between $x$ and $t$. The reason is that such distance would give equal weight to all pixels and cannot properly account for the semantic similarity in images. Rather, we follow established practice and compute the similarity via the so-called structural similarity index \cite{Wang.2004}. The structural similarity index is a standard computer vision method for quantifying the similarity of images between 0 (i.e., no similarity at all) and 1 (i.e., perfect similarity). Product images with a low structural similarity indicate an increased probability of a quality defect because they are less similar to the training data (i.e., images of faultless products). For details on the computation of the structural similarity index, we refer to \cite{Wang.2004}. Eventually, we scaled the structural similarity index of all images between 0 and 100 and rounded the values to the nearest integer to enhance readability. The resulting similarity measure corresponds to the quality scores that were shown to the participants in the experiment.


\textbf{Prediction performance.} We evaluated the out-of-sample prediction accuracy of the AI algorithm as follows. In a first step, we mapped the quality scores onto a binary faultless/defective label. For this, we introduced a quality score of $\theta^* = 90$ as a cutoff (i.e., predicting that a product is defective if the quality score is below 90 and faultless otherwise). We then compared the predictions of the algorithm against the ground-truth quality labels provided by \emph{Siemens}. \Cref{tab:confusion_matrix} gives the confusion matrix for the 200 out-of-sample images used in the experiment. We measure the prediction performance via the balanced accuracy (i.e., average sensitivity across faultless and defective products). We choose the balanced accuracy as our main performance metric because it accounts for the unbalanced distribution between faultless and defective products (i.e., 172 products are faultless and 28 products are defective). The standalone AI algorithm achieves a balanced accuracy of 95.6\% (i.e., $0.5 \times [169/172 + 26/28]$). Additionally, we evaluated the defect detection rate (true negative rate) of the AI algorithm, i.e., how many of the defective products were identified as such. The defect detection rate of the standalone AI algorithm was 92.9\% ($26/28$).

\begin{table}[H]
\onehalfspacing
\footnotesize
\renewcommand*{\arraystretch}{1.5}
\centering
\caption{\textbf{Confusion matrix comparing AI predictions with ground-truth labels in the manufacturing setting}}
\begin{tabular}{p{0.5cm} | l c c}
& & \multicolumn{2}{c}{\emph{\textbf{Predicted label}}} \\
\midrule
 \multirow{3}{*}{\rotatebox[origin=c]{90}{\parbox[c]{1.7cm}{\centering 
\emph{\textbf{Actual label}}}}}  & & Faultless & Defective \\
& Faultless & 169 & 3  \\
& Defective & 2 & 26 \\
\end{tabular}
\label{tab:confusion_matrix}
\end{table}

\begin{figure}[H]
\centering
\begin{subfigure}{.5\textwidth}
  \centering
  \caption{\footnotesize{}}
  \includegraphics[width=0.8\linewidth]{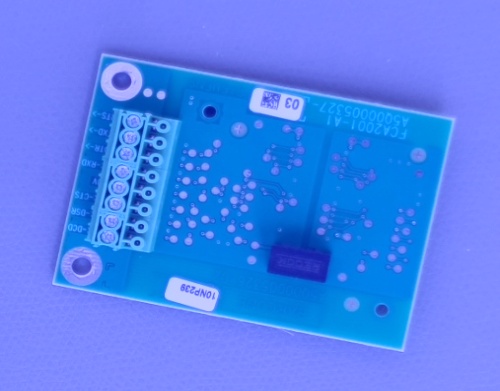}
\end{subfigure}
\begin{subfigure}{.5\textwidth}
  \centering
  \caption{\footnotesize{}}
  \includegraphics[width=0.8\linewidth]{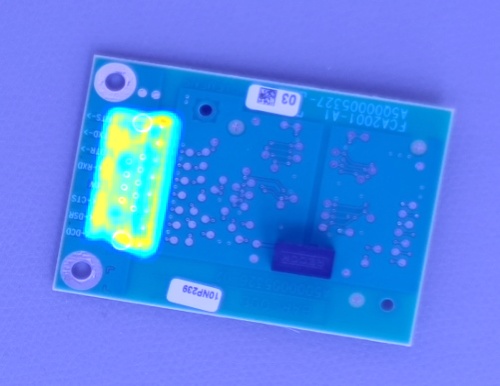}
\end{subfigure}
\caption{\textbf{Anomaly heatmaps for AI predictions}. (\textbf{\textsf{A}})~Example anomaly heatmap for a faultless product. (\textbf{\textsf{B}})~Example anomaly heatmap for a defective product.}
\label{fig:heatmaps}
\end{figure}

\textbf{Explainable AI.} We extended the above AI algorithm to produce explanations for each prediction as follows. We followed other research in computer vision that generates so-called \textquote{anomaly heatmaps} \cite{Bergmann.2019a}. Anomaly heatmaps visualize in what area of an image a quality defect is predicted to be. Formally, in an anomaly heatmap, each pixel $x_i$ is associated with a score measuring the likelihood of a defect at that location. Pixels that receive a bright color (yellow, orange, red, etc.) correspond to \textquote{anomalous} regions because they have a large distance to the training data (i.e., the pixel is dissimilar to the one in a faultless product). In contrast, pixels that are colored in blue have a small distance to the training data and should thus be considered as \textquote{normal.} For better usability, we overlay the anomaly heatmap over the actual product image (with partially transparent colors). Examples of two anomaly heatmaps are shown in \Cref{fig:heatmaps}. In the experiment, the heatmaps were shown to the participants in the explainable AI treatment arm as an additional decision aid.

\subsection{Medical setting}

\textbf{AI algorithm.} We used an already trained DenseNet121 from \cite{Saporta.2022}. DenseNet121 is a convolutional neural network that is part of the DenseNet family, known for its dense connectivity pattern where each layer is connected to every other layer in a feed-forward fashion \cite{Huang.2017}. The \textquote{121} in DenseNet121 stands for the total number of layers in the network, including convolutional layers, pooling layers, and fully connected layers, summing up to 121. The DenseNet121 was set up as a multi-label classifier, which takes a chest X-ray image as input and outputs probabilities for the following 10 labels: airspace opacity, atelectasis, cardiomegaly, consolidation, edema, enlarged cardiomediastinum, lung lesion, pleural effusion, pneumothorax, and support devices. It was trained on 224,316 chest X-ray images from 65,240 patients.

\textbf{AI predictions.} The probabilities returned by the DenseNet121 were mapped onto binary yes/no labels by finding the probability threshold that maximized the balanced accuracy on a validation set of chest X-ray images, which were not used during training. In order to have a score identical to the quality score from the manufacturing setting, where a smaller score indicates a greater likelihood of showing a defect and with a cutoff of 90 that divides the quality scores into defective and faultless, the following transformations were performed: (i)~the probabilities were inverted, (ii)~the threshold that divides the two classes was set to 90, (iii)~the inverted probabilities larger than that threshold were rescaled using min-max scaling on a scale from 90 to 100, and (iv)~the inverted probabilities smaller than that threshold were rescaled using min-max scaling on a scale from 0 to 90.

\textbf{Prediction performance.} As in the manufacturing setting, we evaluate performance of the AI algorithm on the 50 chest X-ray images by calculating the balanced accuracy and the disease detection rate. Those 50 images were neither used for training nor for finding the class dividing threshold. The standalone AI algorithm achieved a balanced accuracy of 82.2\% (i.e., $0.5 \times [40/43 + 5/7]$) and a disease detection rate of 71.4\% (i.e., $5/7$). Additionally, the confusion matrix for the 50 images is shown in \Cref{tab:confusion_matrix_medical}.

\begin{table}[H]
\onehalfspacing
\footnotesize
\renewcommand*{\arraystretch}{1.5}
\centering
\caption{\textbf{Confusion matrix comparing AI predictions with ground-truth labels in the medical setting}}
\begin{tabular}{p{0.5cm} | l c c}
& & \multicolumn{2}{c}{\emph{\textbf{Predicted label}}} \\
\midrule
 \multirow{3}{*}{\rotatebox[origin=c]{90}{\parbox[c]{1.7cm}{\centering 
\emph{\textbf{Actual label}}}}}  & & No lung lesion & At least one lung lesion \\
& No lung lesion & 40 & 3  \\
& At least one lung lesion & 2 & 5 \\
\end{tabular}
\label{tab:confusion_matrix_medical}
\end{table}

\textbf{Explainable AI.} As explanation technique for the above AI algorithm, we used \emph{GradCAM} \cite{Selvaraju.2017}. \emph{GradCAM} outputs heatmaps similar to the anomaly heatmaps from the manufacturing setting and showed state-of-the-art localization performance on chest X-ray images across a variety of diagnoses \cite{Saporta.2022}. Analogous to the manufacturing setting, pixels with bright colors (yellow, orange, red, etc.) correspond to regions that were most relevant for predicting lung lesions, whereas blue pixels were least relevant. To increase usability, heatmaps were overlaid over the raw chest X-ray images with partially transparent colors. Examples of two heatmaps next to the original chest X-ray images are shown in \Cref{fig:heatmaps_medical}. Heatmaps were only provided to radiologists in the explainable AI treatment arm.

\newpage
\section{Experimental interface}
\label{sec:experimental_interface}

\subsection{Manufacturing setting}

The experiment was carried out via a computer interface that was analogously designed to the real-world quality inspection setup at \emph{Siemens}. The experiment comprises the following steps: (1) the study description and study consent, (2) a tutorial on how to use the application, (3) the visual inspection task involving 200 images, (4) a post-experimental questionnaire.  

\begin{figure}[H]
\centering
\begin{subfigure}{.5\textwidth}
  \centering
  \caption{\footnotesize{}}
  \includegraphics[width=0.95\linewidth]{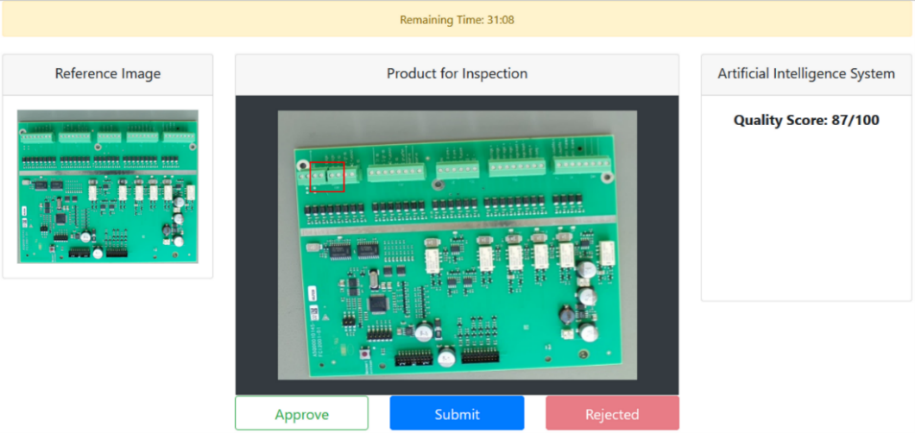}
\end{subfigure}%
\begin{subfigure}{.5\textwidth}
  \centering
  \caption{\footnotesize{}}
  \includegraphics[width=0.95\linewidth]{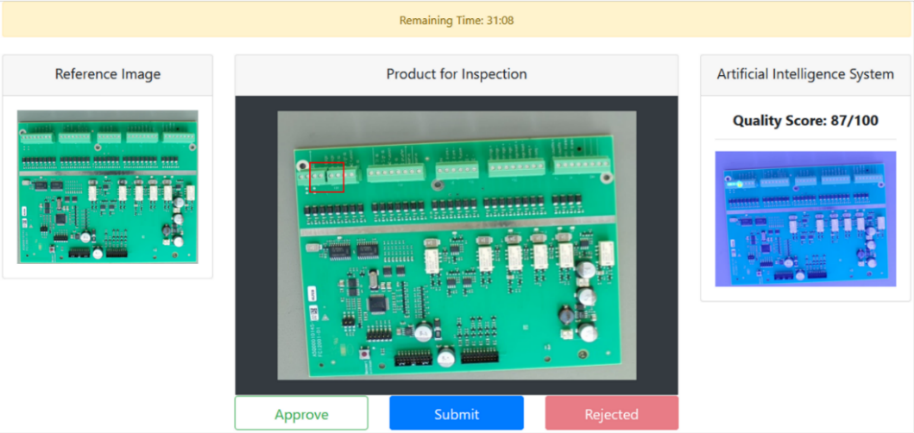}
\end{subfigure}
\\ \vspace{0.5cm}
\begin{subfigure}{.5\textwidth}
  \centering
  \caption{\footnotesize{}}
  \includegraphics[width=0.95\linewidth]{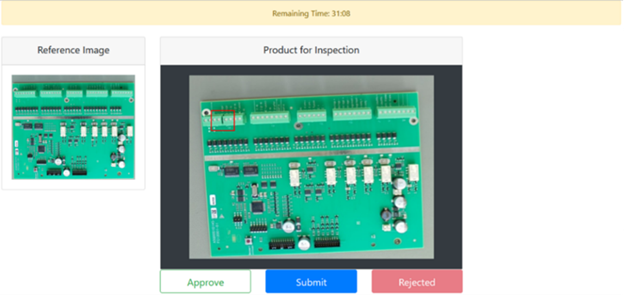}
\end{subfigure}%
\begin{subfigure}{.5\textwidth}
  \centering
  \caption{\footnotesize{}}
  \includegraphics[width=0.8\linewidth]{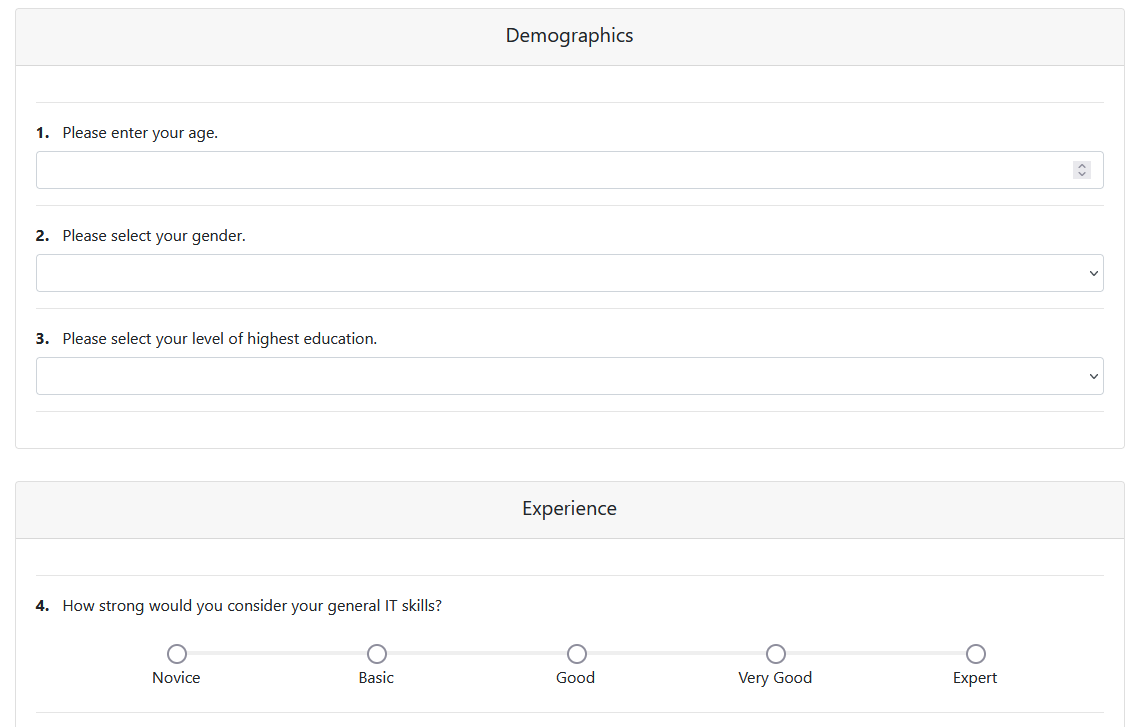}
\end{subfigure}
\caption{\textbf{Different interfaces depending on treatment arm.} (\textbf{\textsf{A}})~The interface for the black-box AI. (\textbf{\textsf{B}})~The interface for the explainble AI. (\textbf{\textsf{C}})~The interface for the human without AI treatment. (\textbf{\textsf{D}})~The interface for the post-experimental questionnaire.}
\label{fig:web_application}
\end{figure}

Depending on the randomly assigned treatment, different versions of the quality inspection interface were shown to participants (\Cref{fig:web_application}). Similar to the real-world setting at \emph{Siemens}, all participants had access to a reference image, which showed a faultless product. The participants were asked to evaluate each of the 200 images individually and to make an \textquote{approve} (faultless product) or \textquote{reject} (defective product) decision by clicking the respective buttons. This represents the quality assessments that we use for all analyses. The participants were allowed to change their quality assessment before submitting their decision and proceeding to the next image. Once a decision was submitted, participants could no longer return to the previous image. Overall, the participants were given 35 minutes to solve the task, which corresponds to realistic field conditions. The remaining time was always shown on the top of the interface.

We tracked several metrics during the experiment. In the tutorial, we tracked whether participants were following the steps correctly and screened out those that did not complete the tutorial successfully. During the visual inspection task, we tracked the final quality assessment (i.e., faultless or defective) and the decision speed of the users. In the post-experimental questionnaire, we saved the answers to individual questions. The aggregated user data were stored in a database and later converted into a comma-separated values (CSV) file.

\subsection{Medical setting}
The experiment was conducted via Qualtrics. The experiment was divided in the following steps: (1)~physician confirmation and study consent, (2)~a tutorial on how to perform the experiment, (3)~the visual inspection task consisting of 50 chest X-ray images, and (4)~a post-experimental questionnaire. 

\begin{figure}[H]
\centering
\begin{subfigure}{.5\textwidth}
  \centering
  \caption{\footnotesize{}}
  \includegraphics[width=0.95\linewidth]{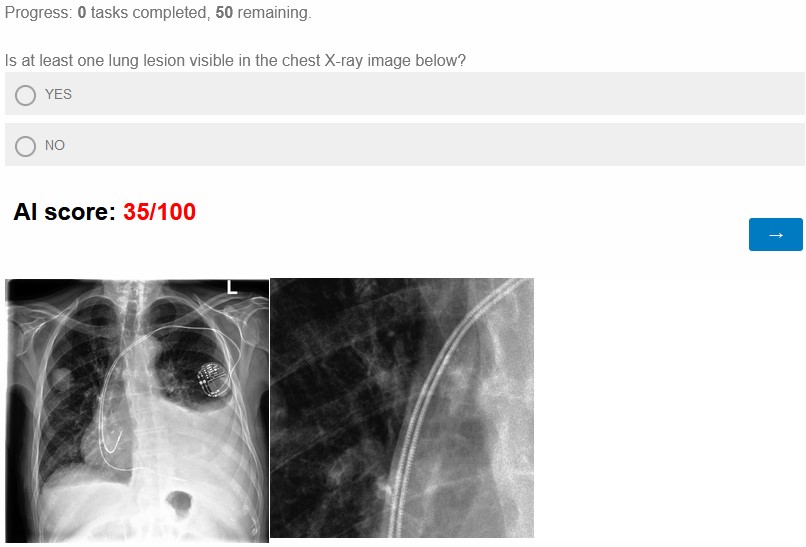}
\end{subfigure}
\begin{subfigure}{.5\textwidth}
  \centering
  \caption{\footnotesize{}}
  \includegraphics[width=0.95\linewidth]{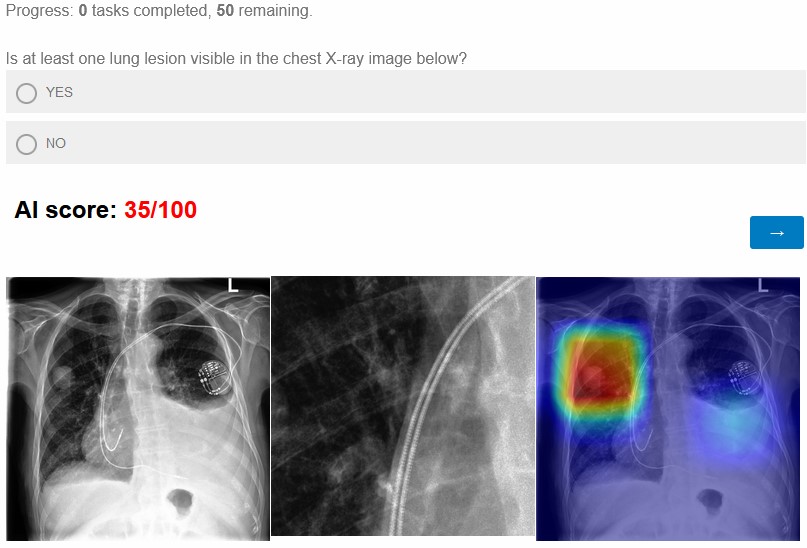}
\end{subfigure}
\caption{\textbf{Different interfaces depending on treatment arm.} (\textbf{\textsf{A}})~The interface for the black-box AI. (\textbf{\textsf{B}})~The interface for the explainble AI.}
\label{fig:qualtrics_setup}
\end{figure}

A different version of the chest X-ray inspection interface was shown to radiologists depending on the randomly assigned treatment (\Cref{fig:qualtrics_setup}). For both treatment arms, the chest X-ray image to inspect was shown on the left. To the right of it, an enlarged view was shown, which could be altered by moving the mouse over the chest X-ray image. Radiologists in the treatment arm with explainable AI additionally received a heatmap, which was displayed right of the enlarged view. The radiologists were asked to inspect 50 chest X-ray images and to answer the question \textquote{Is at least one lung lesion visible in the chest X-ray image below?} with either \textquote{YES} or \textquote{NO} for each image. The radiologists were allowed to change their assessment before submitting their decision (clicking the blue arrow button to proceed to the next page). Each chest X-ray image was shown on a separate page and radiologists were not allowed to go back to a previous image once a decision was submitted. The radiologists had 35 minutes to complete the task and the remaining time was always shown on the top-left of the page.

Several metrics were recorded during the experiment: In the tutorial, we tracked whether radiologists understood how to perform the task. In the visual inspection task, the final assessment for each image as well as the corresponding decision speed were recorded. In the post-hoc questionnaire, we saved the answers to the individual questions. The data was stored on Qualtrics and exported as a CSV file.

\clearpage
\section{Randomization checks}
\label{sec:randomization_checks}

\subsection{Study 1: Manufacturing experiment}
\label{sec:randomization_checks_2}

We performed randomization checks to confirm that the distribution of workers in the two treatment arms of the manufacturing experiment was unbiased. The following demographic variables were collected: age bracket [$<$20, 20--30, 30--40, 40--50, 50--60, 60--70, $>$70], gender [male, female, not listed], and highest level of education [ISCED1, ISCED2, ISCED3, ISCED4, ISCED5, ISCED6, ISCED7].\footnote{UNESCO. \emph{International Standard Classification of Education (ISCED)}. URL: http://uis.unesco.org/en/topic/international-standard-classification-education-isced, last accessed on \today.} We further collected the participant-specific tenure at \emph{Siemens} measured in years since start of employment. \Cref{tab:randomization_check_2} reports the observed frequencies and the mean tenure (with standard deviation in parentheses) for both treatment arms. The randomization checks for age, gender, and education are based on $\mathcal{X}^2$-tests of independence. The randomization check for tenure is based on a two-sided Welch's $t$-test. The results suggest no statistically significant differences between the participants in the two treatment arms. 

\begin{table}[H]
\onehalfspacing
\footnotesize
\begin{center}
\caption{\textbf{Randomization checks for manufacturing experiment}} 
\label{tab:randomization_check_2}
\begin{tabular}{p{1.5cm} x{2.75cm} x{0cm} x{2.75cm} x{0cm} x{1.5cm}}
\toprule
& \textbf{Human with black-box AI} & & \textbf{Human with explainable AI} & & \textbf{$\bm{P}$-value}\\

\cmidrule(lr){2-2}\cmidrule(lr){4-4}\cmidrule(lr){6-6}
Age & 0\,\textbar\,1\,\textbar\,7\,\textbar\,8\,\textbar\,4\,\textbar\,2\,\textbar\,0 & & 0\,\textbar\,1\,\textbar\,3\,\textbar\,9\,\textbar\,12\,\textbar\,1\,\textbar\,0 & & 0.223\\
Gender & 16\,\textbar\,6\,\textbar\,0 & & 17\,\textbar\,9\,\textbar\,0 & & 0.815\\
Education & 0\,\textbar\,9\,\textbar\,4\,\textbar\,2\,\textbar\,3\,\textbar\,3\,\textbar\,1 & & 0\,\textbar\,14\,\textbar\,2\,\textbar\,3\,\textbar\,3\,\textbar\,3\,\textbar\,1 & & 0.897\\
Tenure & 11.91 (8.83) & & 15.38 (10.42) & & 0.217\\
\midrule
Observations & 22 & & 26 & & -- \\
\bottomrule
\end{tabular}
\end{center}
\vspace{-0.25cm}
Notes: The table reports the frequency of participants that fall in the specific subgroups of age, gender, and education (separated by vertical bars) and the average tenure per treatment arm (standard deviation in parentheses). The $P$-values for the randomization checks are computed based on  $\mathcal{X}^2$-tests of independence (age, gender, education) and a two-sided Welch's $t$-test (tenure).\\
\end{table}

\subsection{Study 2: Medical experiment}
\label{sec:randomization_checks_3}

We performed a randomization check to confirm that the distribution of radiologists with respect to tenure in the two treatment arms of the medical experiment was unbiased. The randomization check is based on a two-sided Welch's $t$-test. The result suggests no statistically significant differences between the radiologists in the two treatment arms. 

\begin{table}[H]
\onehalfspacing
\footnotesize
\begin{center}
\caption{\textbf{Randomization checks for medical experiment}} 
\label{tab:randomization_check_3}
\begin{tabular}{p{1.5cm} x{2.75cm} x{0cm} x{2.75cm} x{0cm} x{1.5cm}}
\toprule
& \textbf{Human with black-box AI} & & \textbf{Human with explainable AI} & & \textbf{$\bm{P}$-value}\\

\cmidrule(lr){2-2}\cmidrule(lr){4-4}\cmidrule(lr){6-6}
Tenure & 11.89 (8.96) & & 15.4 (11.71) & & 0.08\\
\midrule
Observations & 61 & & 52 & & -- \\
\bottomrule
\end{tabular}
\end{center}
\vspace{-0.25cm}
Notes: The table reports the average tenure per treatment arm (standard deviation in parentheses). The $P$-value is computed based on a two-sided Welch's $t$-test.\\
\end{table}

\clearpage
\section{Robustness of the heatmap}
\label{sec:heatmap_checks}

\begin{figure}[H]
\centering
\renewcommand\thesubfigure{}
\begin{subfigure}{.113\textwidth}
  \centering
  \caption{\footnotesize{}}
  \includegraphics[width=1\linewidth]{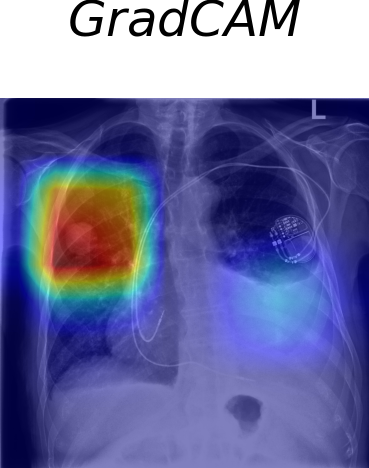}
\end{subfigure}
\vspace{-5.5mm}
\begin{subfigure}{.113\textwidth}
  \centering
  \caption{\footnotesize{}}
  \includegraphics[width=1\linewidth]{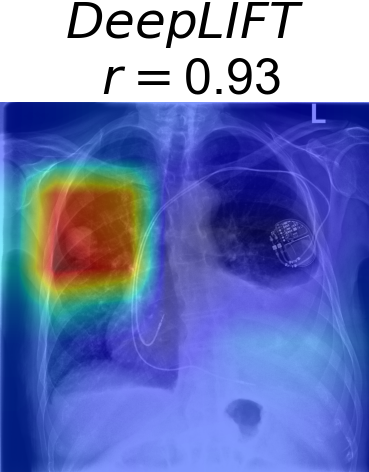}
\end{subfigure}
\begin{subfigure}{.113\textwidth}
  \centering
  \caption{\footnotesize{}}
  \includegraphics[width=1\linewidth]{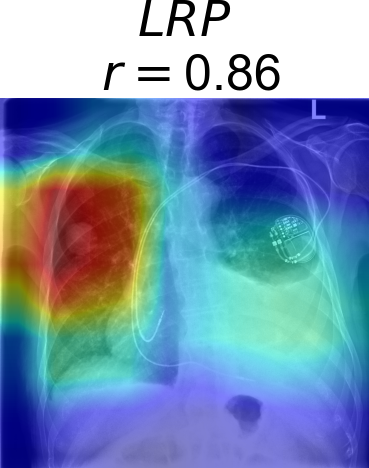}
\end{subfigure} \\
\begin{subfigure}{.113\textwidth}
  \centering
  \caption{\footnotesize{}}
  \includegraphics[width=1\linewidth]{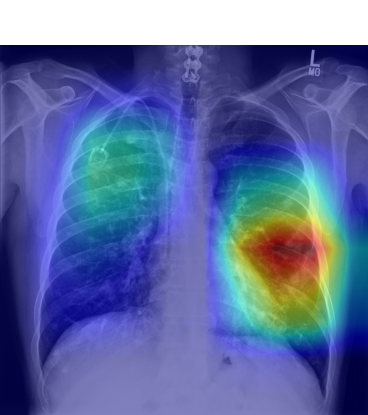}
\end{subfigure}
\vspace{-5.5mm}
\begin{subfigure}{.113\textwidth}
  \centering
  \caption{\footnotesize{}}
  \includegraphics[width=1\linewidth]{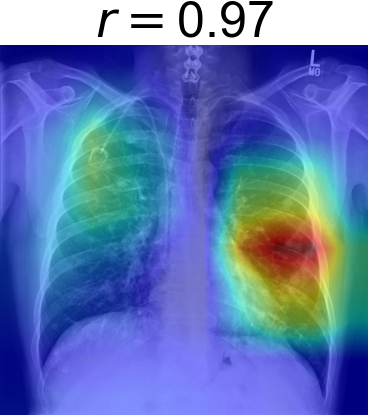}
\end{subfigure}
\begin{subfigure}{.113\textwidth}
  \centering
  \caption{\footnotesize{}}
  \includegraphics[width=1\linewidth]{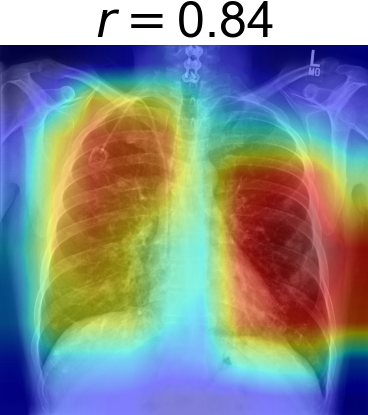}
\end{subfigure} \\
\begin{subfigure}{.113\textwidth}
  \centering
  \caption{\footnotesize{}}
  \includegraphics[width=1\linewidth]{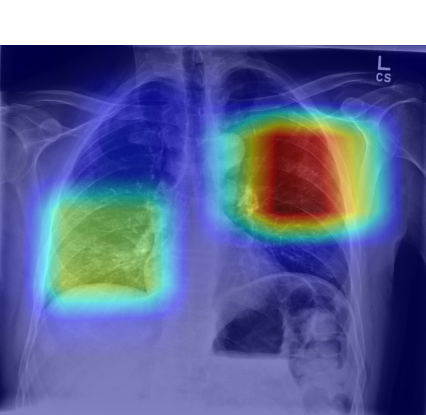}
\end{subfigure}
\vspace{-5.5mm}
\begin{subfigure}{.113\textwidth}
  \centering
  \caption{\footnotesize{}}
  \includegraphics[width=1\linewidth]{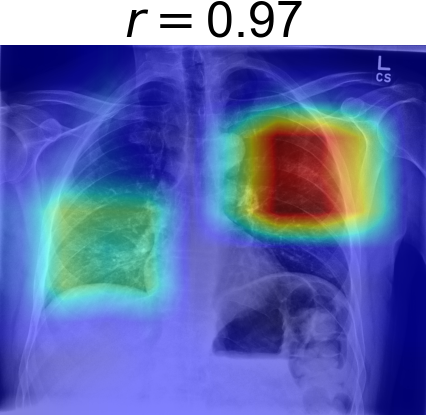}
\end{subfigure}
\begin{subfigure}{.113\textwidth}
  \centering
  \caption{\footnotesize{}}
  \includegraphics[width=1\linewidth]{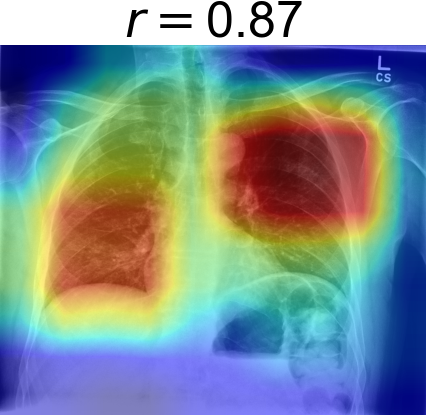}
\end{subfigure} \\
\begin{subfigure}{.113\textwidth}
  \centering
  \caption{\footnotesize{}}
  \includegraphics[width=1\linewidth]{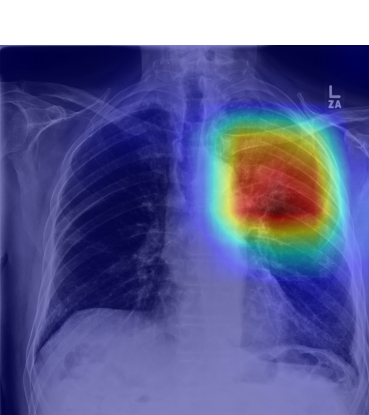}
\end{subfigure}
\vspace{-5.5mm}
\begin{subfigure}{.113\textwidth}
  \centering
  \caption{\footnotesize{}}
  \includegraphics[width=1\linewidth]{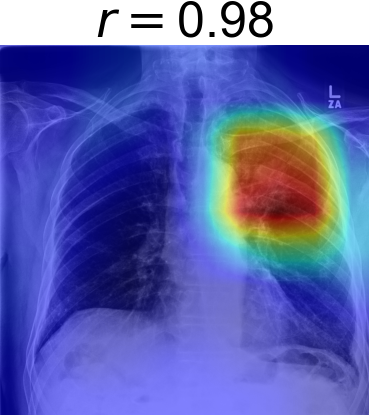}
\end{subfigure}
\begin{subfigure}{.113\textwidth}
  \centering
  \caption{\footnotesize{}}
  \includegraphics[width=1\linewidth]{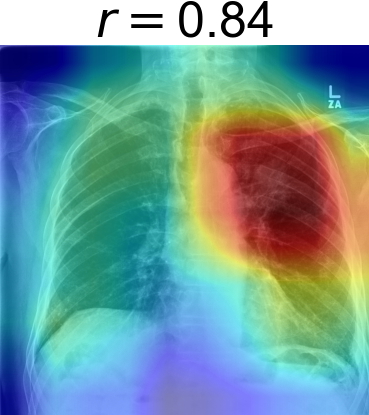}
\end{subfigure} \\
\begin{subfigure}{.113\textwidth}
  \centering
  \caption{\footnotesize{}}
  \includegraphics[width=1\linewidth]{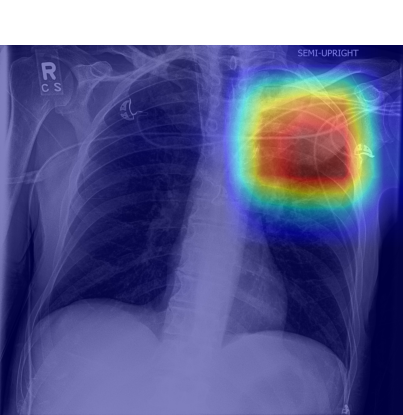}
\end{subfigure}
\vspace{-5.5mm}
\begin{subfigure}{.113\textwidth}
  \centering
  \caption{\footnotesize{}}
  \includegraphics[width=1\linewidth]{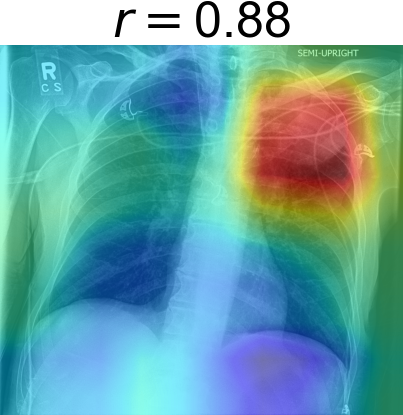}
\end{subfigure}
\begin{subfigure}{.113\textwidth}
  \centering
  \caption{\footnotesize{}}
  \includegraphics[width=1\linewidth]{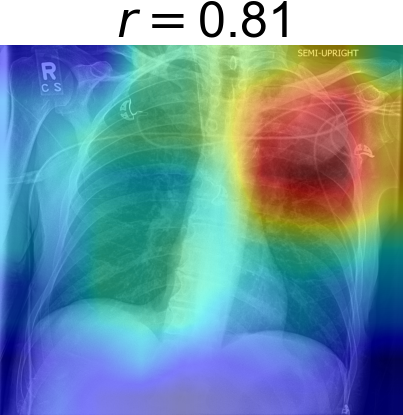}
\end{subfigure} \\
\begin{subfigure}{.113\textwidth}
  \centering
  \caption{\footnotesize{}}
  \includegraphics[width=1\linewidth]{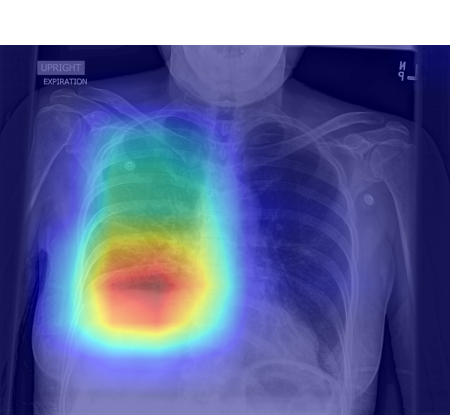}
\end{subfigure}
\vspace{-5.5mm}
\begin{subfigure}{.113\textwidth}
  \centering
  \caption{\footnotesize{}}
  \includegraphics[width=1\linewidth]{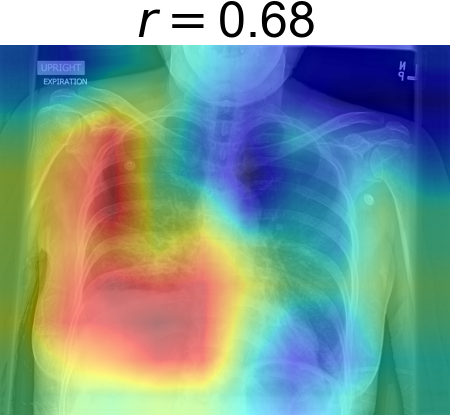}
\end{subfigure}
\begin{subfigure}{.113\textwidth}
  \centering
  \caption{\footnotesize{}}
  \includegraphics[width=1\linewidth]{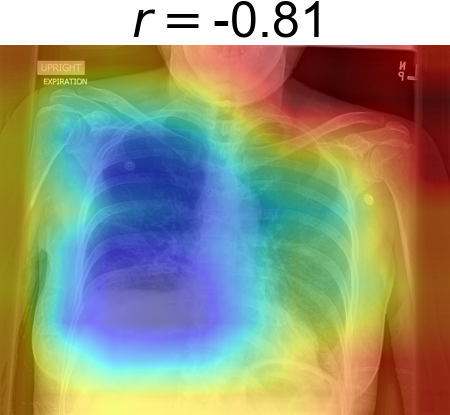}
\end{subfigure} \\
\begin{subfigure}{.113\textwidth}
  \centering
  \caption{\footnotesize{}}
  \includegraphics[width=1\linewidth]{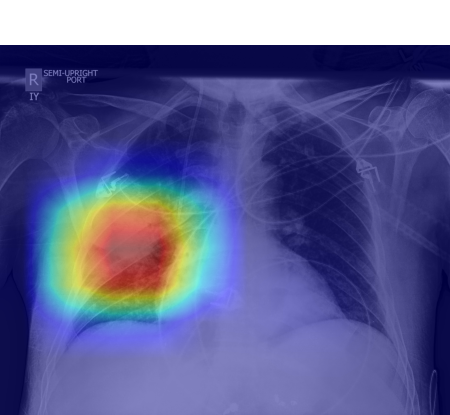}
\end{subfigure}
\vspace{-5.5mm}
\begin{subfigure}{.113\textwidth}
  \centering
  \caption{\footnotesize{}}
  \includegraphics[width=1\linewidth]{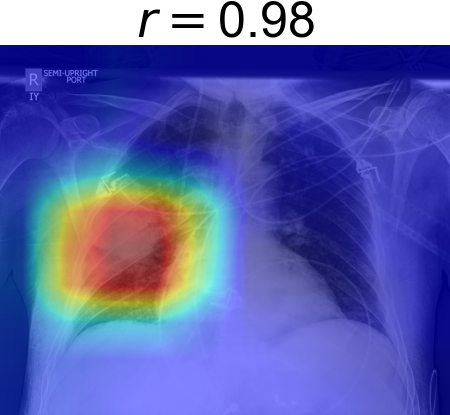}
\end{subfigure}
\begin{subfigure}{.113\textwidth}
  \centering
  \caption{\footnotesize{}}
  \includegraphics[width=1\linewidth]{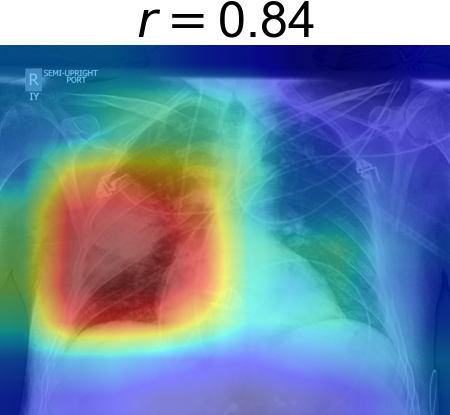}
\end{subfigure} \\
\begin{subfigure}{.113\textwidth}
  \centering
  \caption{\footnotesize{}}
  \includegraphics[width=1\linewidth]{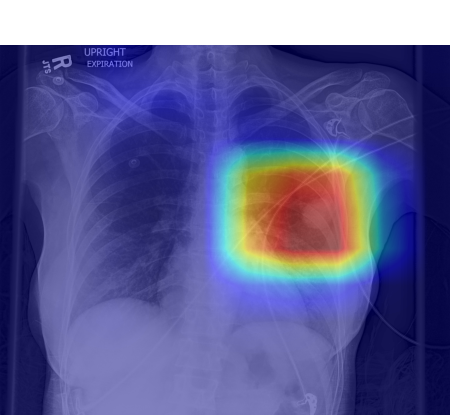}
\end{subfigure}
\vspace{-2mm}
\begin{subfigure}{.113\textwidth}
  \centering
  \caption{\footnotesize{}}
  \includegraphics[width=1\linewidth]{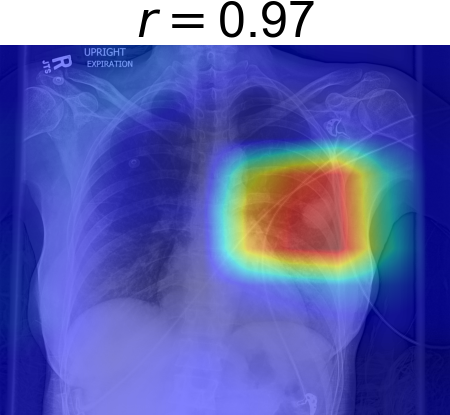}
\end{subfigure}
\begin{subfigure}{.113\textwidth}
  \centering
  \caption{\footnotesize{}}
  \includegraphics[width=1\linewidth]{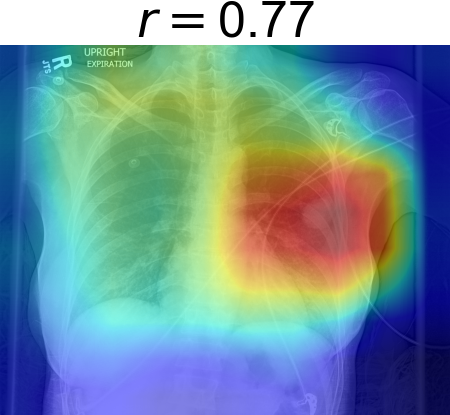}
\end{subfigure}
\caption{\textbf{Heatmaps generated by three different algorithms.} The left column shows the heatmaps generated by \emph{GradCAM} (the algorithm we used for our medical experiment). The middle columns shows the heatmaps generated by \emph{DeepLIFT}. The right column shows the heatmaps generated by \emph{LRP}. $r$ denotes the Pearson correlation coefficient between the heatmaps generated by \emph{GradCAM} (the algorithm we used) and \emph{DeepLIFT}/\emph{LRP}, respectively.}
\label{fig:heatmap_robustness}
\end{figure}

We applied two additional algorithms to generate the heatmaps in the medical setting in order to show that different algorithms lead to similar heatmaps. In particular, we compared our heatmaps generated by \emph{GradCAM} to heatmaps generated by \emph{DeepLIFT} and \emph{LRP} (for an introduction to these two methods see Supplement~\ref{sec:literature_review}) \cite{Shrikumar.2017, Bach.2015}. \Cref{fig:heatmap_robustness} shows the heatmaps generated by the three distinct algorithms for the eight chest X-ray images, where the AI algorithm predicted that lung lesions are visible.

Additionally, we calculated Pearson correlation coefficients between the heatmaps generated by different algorithms to quantify whether they highlight similar regions. The Pearson correlation coefficient is calculated via $r=\frac{\mathrm{cov}(x_i, x_j)}{\sqrt{\mathrm{Var}(x_i)\,\mathrm{Var}(x_j)}}$, where $x_i$ and $x_j$ denote the flattened array of heatmaps generated by algorithm $i$ and $j$. The average Pearson correlation coefficient between \emph{GradCAM} and \emph{DeepLIFT} is $r=0.92$, and the average Pearson correlation coefficient between \emph{GradCAM} and \emph{LRP} is $r=0.63$. The exact Pearson correlation coefficient for each heatmap pair is reported in \Cref{fig:heatmap_robustness}. The Pearson correlation coefficient was statistically significant for each pair of heatmaps ($P<0.001$). In general, we observe that all three algorithms produce similar heatmaps. The heatmaps we used in our medical setting (generated by \emph{GradCAM}) are more similar to the heatmaps generated by \emph{DeepLIFT} than to those generated by \emph{LRP}. 

\clearpage
\section{Results with precision as task performance metric}
\label{sec:precision}

\begin{figure}[H]
\centering
\includegraphics[width=0.95\linewidth]{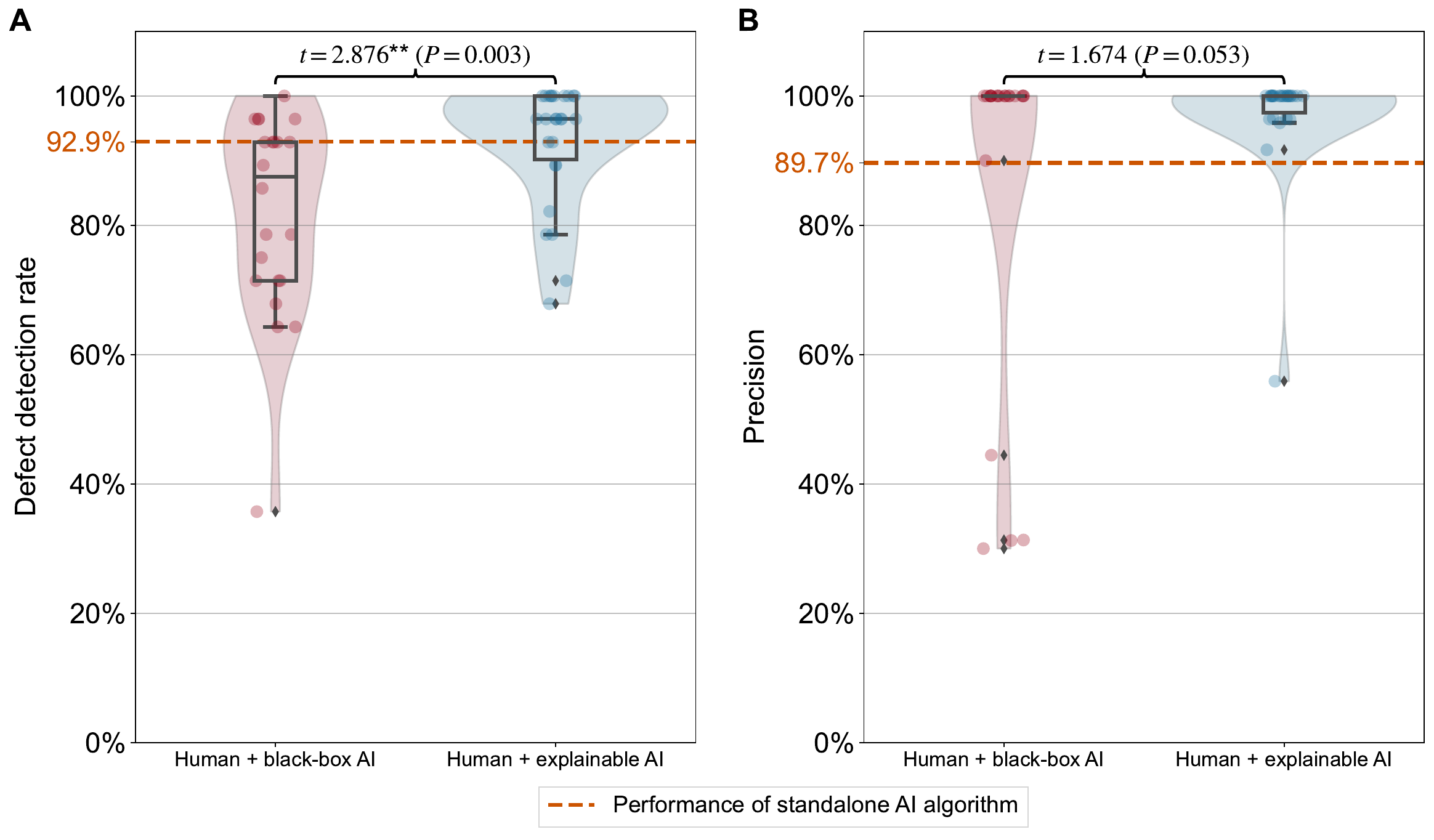}
\caption{\textbf{Results of manufacturing experiment.} The boxplots compare the task performance between the two treatments: black-box AI and explainable AI. The task performance is measured by the defect detection rate (\textbf{\textsf{A}}) and the precision (\textbf{\textsf{B}}) based on the quality assessment of workers and the ground-truth labels of the product images. The standalone AI algorithm attains a defect detection rate of 92.9\% and a precision of 89.7\% (orange dashed lines). Statistical significance is based on a one-sided Welch's $t$-test (\textsuperscript{***}$P<0.001$, \textsuperscript{**}$P<0.01$, \textsuperscript{*}$P<0.05$). In the boxplots, the center line denotes the median; box limits are upper and lower quartiles; whiskers are defined as the 1.5x interquartile range.}
\label{fig:study_2_precision_results}
\end{figure}

\begin{figure}[H]
\centering
\includegraphics[width=0.95\linewidth]{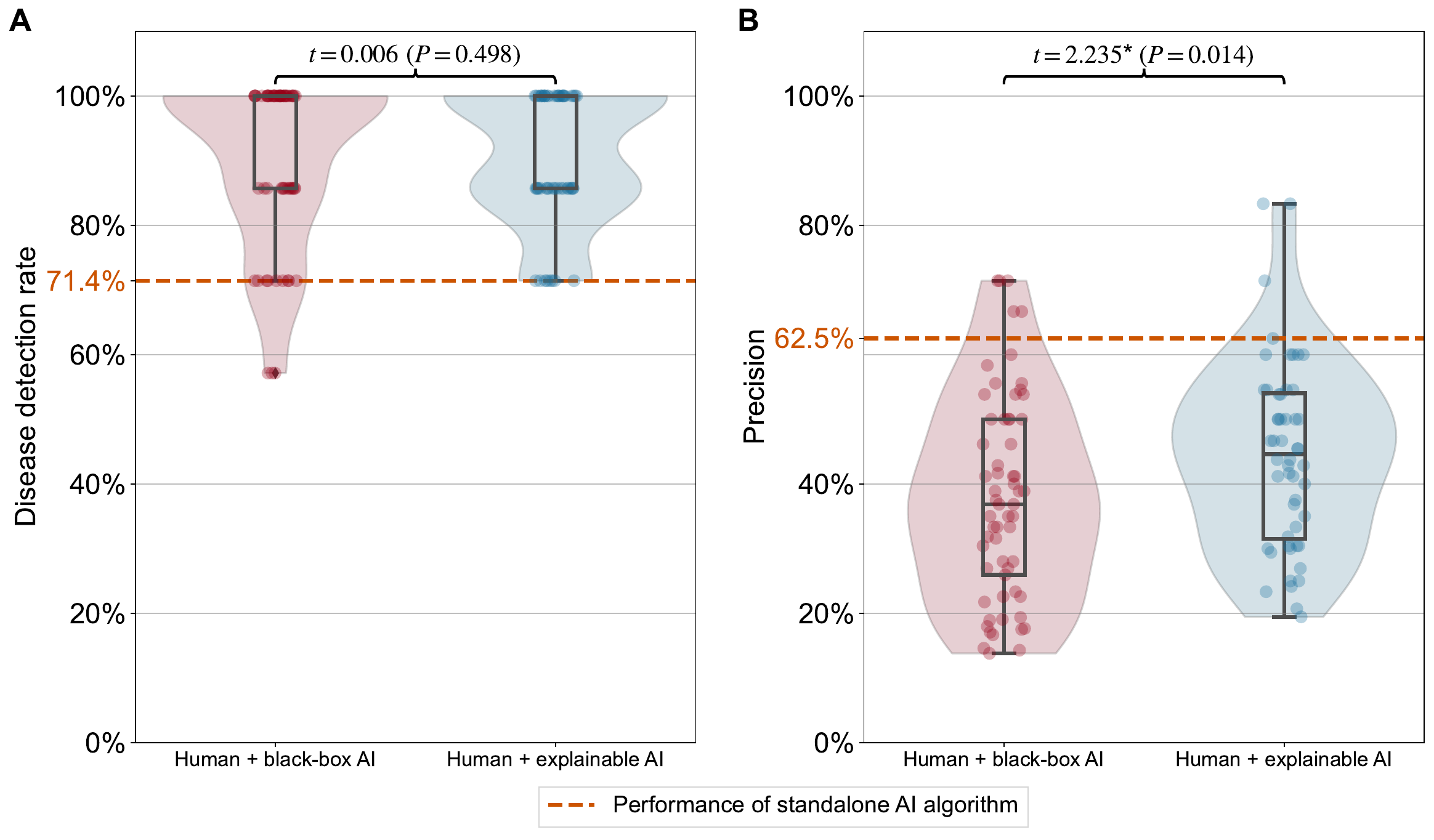}
\caption{\textbf{Results of medical experiment.} The boxplots compare the task performance between the two treatments: black-box AI and explainable AI. The task performance is measured by the disease detection rate (\textbf{\textsf{A}}) and the precision (\textbf{\textsf{B}}) based on the quality assessment of radiologists and the ground-truth labels of the chest X-ray images. The standalone AI algorithm attains a disease detection rate of 71.4\% and a precision of 62.5\% (orange dashed lines). Statistical significance is based on a one-sided Welch's $t$-test (\textsuperscript{***}$P<0.001$, \textsuperscript{**}$P<0.01$, \textsuperscript{*}$P<0.05$). In the boxplots, the center line denotes the median; box limits are upper and lower quartiles; whiskers are defined as the 1.5x interquartile range.}
\label{fig:study_3_precision_results}
\end{figure}

In addition to the balanced accuracy and defect detection rate, we also report precision as a metric for task performance of the participants in combination with the defect/disease detection rate. Formally, precision is computed via $\mathit{TN}/\mathit{PN}$ with true negatives $\mathit{TN}$ and predicted negatives $\mathit{PN}$. We again compare the effect of augmenting humans with explainable AI versus black-box AI. \Cref{fig:study_2_precision_results} reports the results for the manufacturing experiment (Study~1) and \Cref{fig:study_3_precision_results} for the medical experiment (Study~2)

\clearpage
\section{Regression models}
\label{sec:regression_models}

This section reports various regression models estimating the treatment effect of augmenting humans with explainable AI. The models are estimated via
\begin{equation}
Y_{i} = \beta_{0} + \beta_{1}\,\mathit{Treatment}_{i} + \beta_{2}\,X_{i} + \varepsilon_{i},
\label{eqn:regression}
\end{equation}
where $Y_{i}$ is the observed task performance (i.e., balanced accuracy or defect/disease detection rate), $Treatment_{i}$ is a binary variable which equals 0 if participant $i$ received the black-box AI treatment and 1 if participant $i$ received the explainable AI treatment, and $X_{i}$ is the vector of participant-specific control variables. The above regression is estimated via ordinary least squares (OLS). 

We acknowledge that the balanced accuracy is only defined between 0 and 100. Because OLS regression models could return values below 0 and above 100, we additionally estimate quasi-binomial regression models with a logit link function. For this, we set the scale parameter of the regression models to the Pearson $\mathcal{X}^2$-statistic divided by the residual degrees of freedom.

\subsection{Study 1: Manufacturing experiment}
\label{sec:regression_models_2}

\Cref{tab:regression_results_2_ols} reports three OLS regression models estimating the treatment effect with different control variables. Model~(1) estimates the treatment effect for explainable AI with demographic controls (age, gender, and highest level of education) and the tenure at \emph{Siemens} measured in years from start of employment. Model~(2) estimates the treatment effect for explainable AI with demographic controls, tenure, and self-reported IT skills (ranging from 1: \textquote{novice} to 5: \textquote{expert}). Model~(3) estimates the treatment effect for explainable AI with demographic controls, tenure, self-reported IT skills, and the decision speed (median across the 200 images). All three models return a significant treatment effect for both metrics (balanced accuracy and defect detection rate) as dependent variables.

\begin{table}[H]
\onehalfspacing
\footnotesize
\begin{center}
\caption{\textbf{OLS regression results for treatment effect (manufacturing experiment)}} 
\label{tab:regression_results_2_ols}
\begin{tabular}{p{2.5cm} x{1.5cm} x{1.5cm} x{1.5cm} x{1.5cm} x{1.5cm} x{1.5cm}}
\toprule
& \multicolumn{3}{c}{\textbf{Balanced accuracy}} & \multicolumn{3}{c}{\textbf{Defect detection rate}} \\
\cmidrule(lr){2-4} \cmidrule(lr){5-7}
& \textbf{Model (1)} & \textbf{Model (2)} & \textbf{Model (3)} & \textbf{Model (1)} & \textbf{Model (2)} & \textbf{Model (3)}\\

\cmidrule(lr){2-2}\cmidrule(lr){3-3}\cmidrule(lr){4-4}\cmidrule(lr){5-5}\cmidrule(lr){6-6}\cmidrule(lr){7-7}
Treatment & 8.131*** & 7.513*** & 7.508** & 11.783** & 10.914** & 10.888** \\
(explainable AI) & (2.087) & (2.098) & (2.117) & (3.732) & (3.790) & (3.717) \\
\midrule
Demographics & Yes & Yes & Yes & Yes & Yes & Yes \\
Tenure & Yes & Yes & Yes & Yes & Yes & Yes \\
IT skills & No & Yes & Yes & No & Yes & Yes \\
Decision speed & No & No & Yes & No & No & Yes \\
\midrule
Observations & 48 & 48 & 48  & 48 & 48 & 48 \\
\bottomrule
\end{tabular}
\end{center}
\vspace{-0.25cm}
Notes: The table reports three OLS regression models with different sets of control variables and two different metrics as dependent variables. The standard errors of the treatment effect are reported in parentheses. \\
Statistical significance: \textsuperscript{***}$P<0.001$, \textsuperscript{**}$P<0.01$, \textsuperscript{*}$P<0.05$.
\end{table}

\Cref{tab:regression_results_2_qb} reports three quasi-binomial regression models estimating the treatment effect with the same control variables as before. Again, all three models return a significant treatment effect for both metrics as dependent variables.

\begin{table}[H]
\onehalfspacing
\footnotesize
\begin{center}
\caption{\textbf{Quasi-binomial regression results for treatment effect (manufacturing experiment)}} 
\label{tab:regression_results_2_qb}
\begin{tabular}{p{2.5cm} x{1.5cm} x{1.5cm} x{1.5cm} x{1.5cm} x{1.5cm} x{1.5cm}}
\toprule
& \multicolumn{3}{c}{\textbf{Balanced accuracy}} & \multicolumn{3}{c}{\textbf{Defect detection rate}} \\
\cmidrule(lr){2-4} \cmidrule(lr){5-7}
& \textbf{Model (1)} & \textbf{Model (2)} & \textbf{Model (3)} & \textbf{Model (1)} & \textbf{Model (2)} & \textbf{Model (3)}\\

\cmidrule(lr){2-2}\cmidrule(lr){3-3}\cmidrule(lr){4-4}\cmidrule(lr){5-5}\cmidrule(lr){6-6}\cmidrule(lr){7-7}
Treatment & 1.295*** & 1.175** & 1.184*** & 1.157** & 1.060**  & 1.089** \\
(explainable AI) & (0.340) & (0.358) & (0.357) & (0.369) & (0.386) & (0.375) \\
\midrule
Demographics & Yes & Yes & Yes & Yes & Yes & Yes\\
Tenure & Yes & Yes & Yes & Yes & Yes & Yes \\
IT skills & No & Yes & Yes & No & Yes & Yes \\
Decision speed & No & No & Yes & No & No & Yes \\
\midrule
Observations & 48 & 48 & 48 & 48 & 48 & 48 \\
\bottomrule
\end{tabular}
\end{center}
\vspace{-0.25cm}
Notes: The table reports three quasi-binomial regression models with different sets of control variables and two different metrics as dependent variables. The standard errors of the treatment effect are reported in parentheses. \\
Statistical significance: \textsuperscript{***}$P<0.001$, \textsuperscript{**}$P<0.01$, \textsuperscript{*}$P<0.05$.
\end{table}

\subsection{Study 2: Medical experiment}
\label{sec:regression_models_3}

\Cref{tab:regression_results_3_ols} reports three OLS regression models estimating the treatment effect with different control variables. Model~(1) estimates the treatment effect for explainable AI with tenure measured in years as a control variable. Model~(2) estimates the treatment effect for explainable AI with tenure and self-reported IT skills (ranging from 1: \textquote{novice} to 5: \textquote{expert}). Model~(3) estimates the treatment effect for explainable AI with tenure, self-reported IT skills, and the decision speed (median across the 50 images). All three models return a significant treatment effect for balanced accuracy as a dependent variable.

\begin{table}[H]
\onehalfspacing
\footnotesize
\begin{center}
\caption{\textbf{OLS regression results for treatment effect (medical experiment)}} 
\label{tab:regression_results_3_ols}
\begin{tabular}{p{2.5cm} x{1.5cm} x{1.5cm} x{1.5cm} x{1.5cm} x{1.5cm} x{1.5cm}}
\toprule
& \multicolumn{3}{c}{\textbf{Balanced accuracy}} & \multicolumn{3}{c}{\textbf{Defect detection rate}} \\
\cmidrule(lr){2-4} \cmidrule(lr){5-7}
& \textbf{Model (1)} & \textbf{Model (2)} & \textbf{Model (3)} & \textbf{Model (1)} & \textbf{Model (2)} & \textbf{Model (3)}\\

\cmidrule(lr){2-2}\cmidrule(lr){3-3}\cmidrule(lr){4-4}\cmidrule(lr){5-5}\cmidrule(lr){6-6}\cmidrule(lr){7-7}
Treatment & 4.637* & 4.452* & 4.473* & 0.129 & 0.304 & 0.645 \\
(explainable AI) & (1.834) & (1.853) & (1.863) & (2.286) & (2.312) & (2.215) \\
\midrule
Tenure & Yes & Yes & Yes & Yes & Yes & Yes \\
IT skills & No & Yes & Yes & No & Yes & Yes \\
Decision speed & No & No & Yes & No & No & Yes \\
\midrule
Observations & 113 & 113 & 113 & 113 & 113 & 113 \\
\bottomrule
\end{tabular}
\end{center}
\vspace{-0.25cm}
Notes: The table reports three OLS regression models with different sets of control variables and two different metrics as dependent variables. The standard errors of the treatment effect are reported in parentheses. \\
Statistical significance: \textsuperscript{***}$P<0.001$, \textsuperscript{**}$P<0.01$, \textsuperscript{*}$P<0.05$.
\end{table}

\Cref{tab:regression_results_3_qb} reports three quasi-binomial regression models estimating the treatment effect with the same control variables as before. Again, all three models return a significant treatment effect for balanced accuracy as dependent variable.

\begin{table}[H]
\onehalfspacing
\footnotesize
\begin{center}
\caption{\textbf{Quasi-binomial regression results for treatment effect (medical experiment)}} 
\label{tab:regression_results_3_qb}
\begin{tabular}{p{2.5cm} x{1.5cm} x{1.5cm} x{1.5cm} x{1.5cm} x{1.5cm} x{1.5cm}}
\toprule
& \multicolumn{3}{c}{\textbf{Balanced accuracy}} & \multicolumn{3}{c}{\textbf{Defect detection rate}} \\
\cmidrule(lr){2-4} \cmidrule(lr){5-7}
& \textbf{Model (1)} & \textbf{Model (2)} & \textbf{Model (3)} & \textbf{Model (1)} & \textbf{Model (2)} & \textbf{Model (3)}\\

\cmidrule(lr){2-2}\cmidrule(lr){3-3}\cmidrule(lr){4-4}\cmidrule(lr){5-5}\cmidrule(lr){6-6}\cmidrule(lr){7-7}
Treatment & 0.308* & 0.296* & 0.297* & 0.015 & 0.035 & 0.073 \\
(explainable AI) & (0.120) & (0.121) & (0.122) & (0.264) & (0.268) & (0.259) \\
\midrule
Tenure & Yes & Yes & Yes & Yes & Yes & Yes \\
IT skills & No & Yes & Yes & No & Yes & Yes \\
Decision speed & No & No & Yes & No & No & Yes \\
\midrule
Observations & 113 & 113 & 113 & 113 & 113 & 113 \\
\bottomrule
\end{tabular}
\end{center}
\vspace{-0.25cm}
Notes: The table reports three quasi-binomial regression models with different sets of control variables and two different metrics as dependent variables. The standard errors of the treatment effect are reported in parentheses. \\
Statistical significance: \textsuperscript{***}$P<0.001$, \textsuperscript{**}$P<0.01$, \textsuperscript{*}$P<0.05$.
\end{table}

\clearpage
\section{Analysis with excluded participants}
\label{sec:excluded_participants}

In our data analyses, we followed our preregistration and excluded participants who did not finish the task in time or participants with obvious misbehavior. Specifically, six and two participants were excluded from Study~1 and Study~2, respectively, because they did not finish the task in time. Further, in Study~1, we excluded participants who did not label a single product as defective (which corresponds to one participant) and in Study~2, radiologists were excluded if they assigned only one label to all chest X-ray images (which corresponds to 1 radiologist). Further, participants whose performance was more than three standard deviations worse than the mean of their respective treatment arm were excluded (which corresponds to one worker in Study~1 and two radiologists in Study~2). This section repeats the OLS regression from the main paper (without control variables) with participants who were excluded due to obvious misbehavior. Overall, we arrive at consistent findings. 

\begin{table}[H]
\onehalfspacing
\footnotesize
\begin{center}
\caption{\textbf{Excluded participants across treatment arms}} 
\label{tab:excluded_participants}
\begin{tabular}{p{2cm} x{2cm} x{2cm} x{2cm} x{2cm}}
\toprule
& \multicolumn{2}{c}{\textbf{Study 1: Manufacturing}} & \multicolumn{2}{c}{\textbf{Study 2: Medical}}\\
\cmidrule(lr){2-3}\cmidrule(lr){4-5}
& Black-box AI & Explainable AI & Black-box AI & Explainable AI\\

\midrule
Time-out & 5 & 1 & 0 & 2 \\
No defective & 1 & 0 & -- & -- \\
Single label & -- & -- & 1 & 0 \\
Worse than $3\sigma$ & 0 & 1 & 1 & 1 \\
\bottomrule
\end{tabular}
\end{center}
\end{table}

\subsection{Study 1: Manufacturing experiment}
\label{sec:excluded_participants_2}

\Cref{tab:regression_results_excluded_1} reports the OLS regression model estimating the treatment effect for all different combinations of exclusion criteria. As in our main analysis, the effect of explainable AI is statistically significant for both metrics, balanced accuracy and defect detection rate. The only exception is for the defect detection rate when workers that timed-out and did not label a single product as defective were excluded while including those that were worse than three standard deviations than the mean.

\begin{table}[h]
\onehalfspacing
\footnotesize
\begin{center}
\caption{\textbf{OLS regression results with excluded participants (manufacturing experiment)}}
\label{tab:regression_results_excluded_1}
\begin{tabular}{x{1.6cm} x{1.6cm} x{1.6cm} x{2.2cm} x{2.5cm} x{2.5cm}}
\toprule
 \multicolumn{3}{c}{\textbf{Excluded:}} & \textbf{Observations} & \textbf{Balanced accuracy} & \textbf{Defect detection rate} \\
\cmidrule(lr){1-3} \cmidrule(lr){4-4} \cmidrule(lr){5-5} \cmidrule(lr){6-6}
\tiny{Time-out} & \tiny{No defective} & \tiny{Worse than $3\sigma$} & & & \\
\cmidrule(lr){1-1}\cmidrule(lr){2-2}\cmidrule(lr){3-3}\cmidrule(lr){4-4}\cmidrule(lr){5-5}\cmidrule(lr){6-6}
\xmark &       \xmark &     \xmark &          56 &   8.143** (2.754) &       11.590* (4.965) \\
\cmark &       \xmark &     \xmark &          50 &    7.944* (3.016) &       12.192* (5.472) \\
\xmark &       \cmark &     \xmark &          55 &   6.769** (2.432) &        8.651* (4.077) \\
\xmark &       \xmark &     \cmark &          54 &  8.120*** (2.042) &      10.961** (3.391) \\
\cmark &       \cmark &     \xmark &          49 &    6.253* (2.642) &         8.628 (4.477) \\
\cmark &       \xmark &     \cmark &          48 &   7.653** (2.178) &      11.014** (3.680) \\
\xmark &       \cmark &     \cmark &          54 &  8.120*** (2.042) &      10.961** (3.391) \\
\bottomrule
\end{tabular}
\end{center}
\vspace{-0.25cm}
Notes: The table reports the OLS regression model with two different metrics as dependent variables for all combinations of exclusion criteria. The standard errors of the treatment effect are reported in parentheses. \\
Statistical significance: \textsuperscript{***}$P<0.001$, \textsuperscript{**}$P<0.01$, \textsuperscript{*}$P<0.05$.
\end{table}

\subsection{Study 2: Medical experiment}
\label{sec:excluded_participants_3}

\Cref{tab:regression_results_excluded_2} reports the OLS regression model estimating the treatment effect for all different combinations of exclusion criteria. As in our main analysis, the effect of explainable AI is statistically significant for balanced accuracy, irrespective of the exclusion criteria. Only when radiologists that assigned one label to all chest X-ray images are excluded the treatment effect for balanced accuracy was not statistically significant. For the disease detection rate, the treatment effect was not statistically significant in the main analysis. This did not change when different exclusion criteria are considered.

\begin{table}[h]
\onehalfspacing
\footnotesize
\begin{center}
\caption{\textbf{OLS regression results with excluded participants (medical experiment)}}
\label{tab:regression_results_excluded_2}
\begin{tabular}{x{1.6cm} x{1.6cm} x{1.6cm} x{2.2cm} x{2.5cm} x{2.5cm}}
\toprule
 \multicolumn{3}{c}{\textbf{Excluded:}} & \textbf{Observations} & \textbf{Balanced accuracy} & \textbf{Disease detection rate} \\
\cmidrule(lr){1-3} \cmidrule(lr){4-4} \cmidrule(lr){5-5} \cmidrule(lr){6-6}
\tiny{Time-out} & \tiny{Single label} & \tiny{Worse than $3\sigma$} & & & \\
\cmidrule(lr){1-1}\cmidrule(lr){2-2}\cmidrule(lr){3-3}\cmidrule(lr){4-4}\cmidrule(lr){5-5}\cmidrule(lr){6-6}
  \xmark &       \xmark &     \xmark &           118 &    4.418* (2.067) &        -0.282 (2.337) \\
  \cmark &       \xmark &     \xmark &           116 &   5.266** (1.996) &         0.543 (2.274) \\
  \xmark &       \cmark &     \xmark &           117 &     3.967 (2.026) &        -0.121 (2.349) \\
  \xmark &       \xmark &     \cmark &           115 &   4.988** (1.846) &        -0.256 (2.212) \\
  \cmark &       \cmark &     \xmark &           115 &    4.815* (1.950) &         0.704 (2.286) \\
  \cmark &       \xmark &     \cmark &           114 &   5.162** (1.857) &        -0.168 (2.232) \\
  \xmark &       \cmark &     \cmark &           114 &    4.520* (1.790) &        -0.102 (2.224) \\
\bottomrule
\end{tabular}
\end{center}
\vspace{-0.25cm}
Notes: The table reports the OLS regression model with two different metrics as dependent variables for all combinations of exclusion criteria. The standard errors of the treatment effect are reported in parentheses. \\
Statistical significance: \textsuperscript{***}$P<0.001$, \textsuperscript{**}$P<0.01$, \textsuperscript{*}$P<0.05$.
\end{table}

\clearpage
\section{Experiment with non-experts}
\label{sec:online_experiment}

To extend our findings to non-experts, we conducted a third experiment with participants recruited via Amazon MTurk to perform the visual inspection task in the manufacturing setting. We chose the manufacturing task for this because, in principle, non-experts could compare electronic products against a reference image (a faultless product looks always identical). For chest X-ray images, this is hardly possible since these can look very different across different healthy patients. We followed common practice by only admitting MTurk workers with an approval rating above 95\% \cite{Peer.2013}. We prevented double participation by tracking the IP of participants. The participants received a base compensation (\$5) and had the opportunity to earn a performance-dependent bonus proportional to the correctly labeled quality defects (\$3). Participants were randomly assigned to one of the following three treatments: (a)~human with black-box AI, (b)~human with explainable AI, and (c)~human without AI. Following the preregistration, we aimed to include approximately 600 participants excluding dropouts. We thus recruited 861 participants (U.S. residents) who started the study between July 19 and July 21, 2021. Out of them, 117 participants did not complete the study; 152 failed the tutorial; 92 did not finish on time; and 70 participants were excluded due to obvious misbehavior. The final sample consisted of $430$ participants, out of which 288 were assigned to treatment arms (a) or (b), performing $N=57,600$ assessments of electronic products.

\begin{figure}[H]
  \centering
  \includegraphics[width=0.95\linewidth]{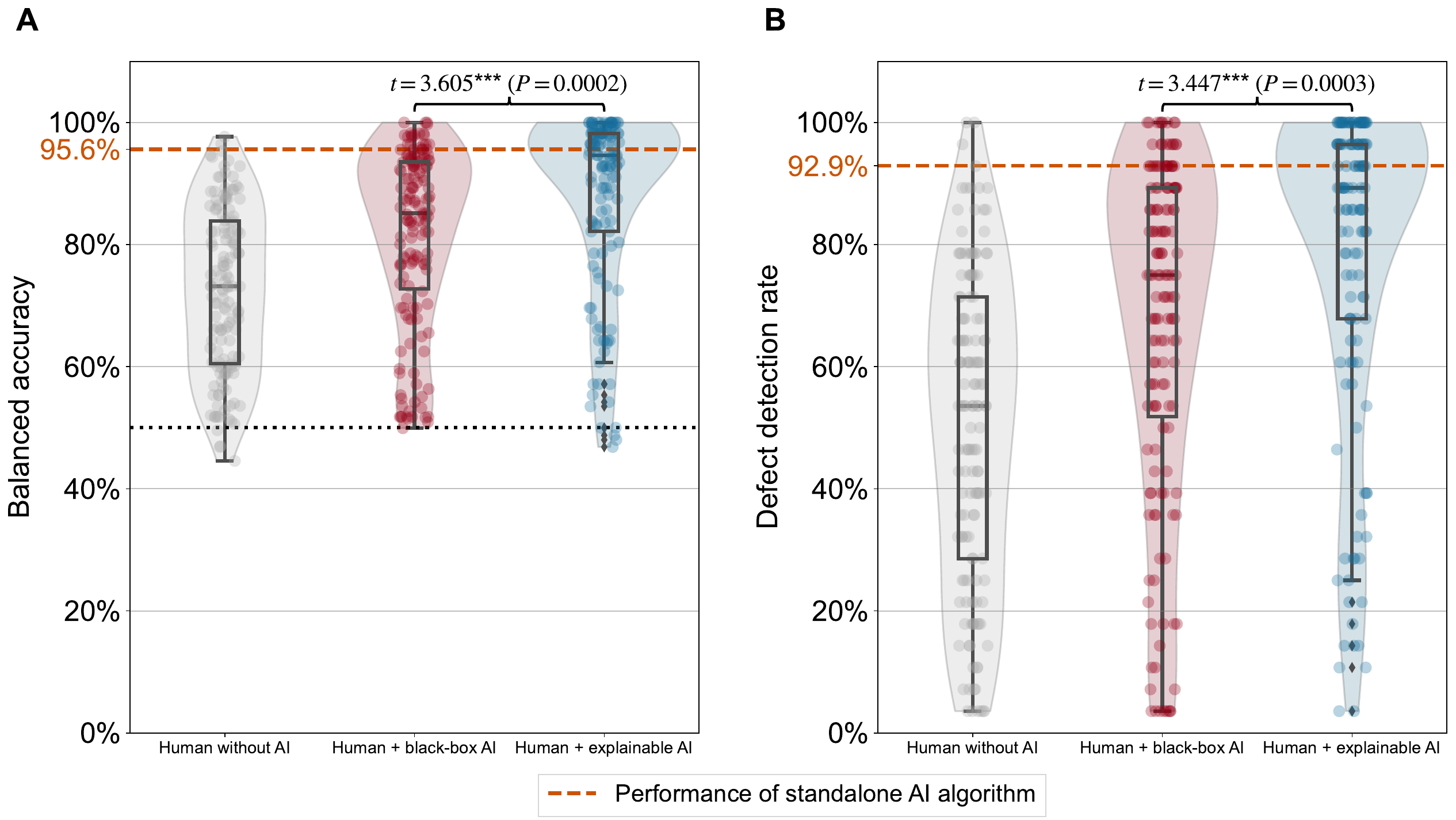}
\caption{\textbf{Results of non-experts experiment.} The boxplots compare the task performance between humans without AI, with black-box AI, and with explainable AI. The task performance is measured by the balanced accuracy (\textbf{\textsf{A}}) and the defect detection rate (\textbf{\textsf{B}}) based on the quality assessment of participants and the ground-truth labels of the product images. A balanced accuracy of 50\% provides a na{\"i}ve baseline corresponding to a random guess (black dotted line). The standalone AI algorithm attains a balanced accuracy of 95.6\% and a defect detection rate of 92.9\% (orange dashed lines). Statistical significance is based on a one-sided Welch's $t$-test (\textsuperscript{***}$P<0.001$, \textsuperscript{**}$P<0.01$, \textsuperscript{*}$P<0.05$). In the boxplots, the center line denotes the median; box limits are upper and lower quartiles; whiskers are defined as the 1.5x interquartile range.}
\label{fig:study_1_results}
\end{figure}

We found that participants supported by explainable AI reached a higher task performance than the participants supported by black-box AI across both metrics (\Cref{fig:study_1_results}). Participants with black-box AI treatment only achieved a balanced accuracy with a mean of 81.4\%, whereas participants with explainable AI treatment achieved a balanced accuracy with a mean of 87.6\%. We then estimated the overall treatment effect on the task performance by regressing the balanced accuracy on the treatment (black-box AI $= 0$, explainable AI $= 1$). The regression results suggest that the treatment effect of explainable AI is statistically significant and large ($\beta=6.252$, $\mathit{SE}=1.733$, $P<0.001$); that is, an improvement of 6.3 percentage points. Accordingly, participants equipped with explainable AI achieved a higher defect detection rate with mean of 77.7\% compared to participants with black-box AI with a mean of 66.4\%. Again, the regression results showed a large and statistically significant treatment effect of explainable AI ($\beta=11.271$, $\mathit{SE}=3.276$, $P=0.001$). The regression results remain statistically significant for both metrics when including relevant control variables (demographics, self-reported IT skills, and decision speed) in the regression model (Supplement~\ref{sec:regression_models_1}).

We additionally compared how humans without AI support performed relative to humans with black-box AI or explainable AI. For this, we further recruited 142 participants and assigned them to a third treatment: human without AI. Here, participants only got images of the to-be-inspected products and the corresponding reference images of faultless products, but not the AI-based quality scores or the heatmaps. We found that participants without AI support only achieved a balanced accuracy with a mean of 72.4\% (\Cref{fig:study_1_results}) and were significantly outperformed by participants with both black-box AI ($t = 5.507$, $P < 0.001$) and explainable AI ($t = 9.017$, $P < 0.001$). Similar results were found for the defect detection rate, where participants without AI achived a mean of 53.6\% and were outperformed by participants with both black-box AI ($t = 5.202$, $P < 0.001$) and explainable AI ($t = 8.733$, $P < 0.001$).

We further explored whether the performance difference between the treatments (black-box AI versus explainable AI) was associated with adherence to AI predictions. For this, we compared how likely participants were to follow quality scores that were accurate (i.e., the AI prediction for the inspected product was correct). The results suggest that participants with explainable AI were more likely to adhere to accurate quality scores than participants with black-box AI ($\textrm{mean} = 92.9\%$ for black-box AI, $\textrm{mean} = 95.2\%$ for explainable AI). Overall, participants supported by black-box AI were 47.9\% more likely to erroneously overrule an AI prediction, despite the prediction being accurate ($t=2.377$, $P=0.009$). We also analyzed whether participants were able to identify and overrule AI predictions that were wrong. Here, we found that participants supported by black-box AI only overruled 65.8\% of the wrong AI predictions, whereas participants supported by explainable AI overruled 79.1\% of the wrong AI predictions. The difference between both treatments is statistically significant ($t=4.563$, $P<0.001$). Evidently, explainable AI gives a powerful decision aid: it made participants not only less averse to following accurate AI predictions but also helped them overrule wrong AI predictions. 

We also assessed whether participants with explainable AI invested more time for the visual inspection task. For this, we compared participants' median decision speeds across the 200 product images. No  significant differences ($t=0.584$, $P=0.280$) between both treatments ($\textrm{mean} =\SI{4.61}{s}$ for black-box AI, $\textrm{mean} =\SI{4.50}{s}$ for explainable AI) were observed. Hence, explainable AI improved task performance, but not at the cost of decision speed.

\subsection{Results with precision as task performance metric}

\begin{figure}[H]
\centering
\includegraphics[width=0.95\linewidth]{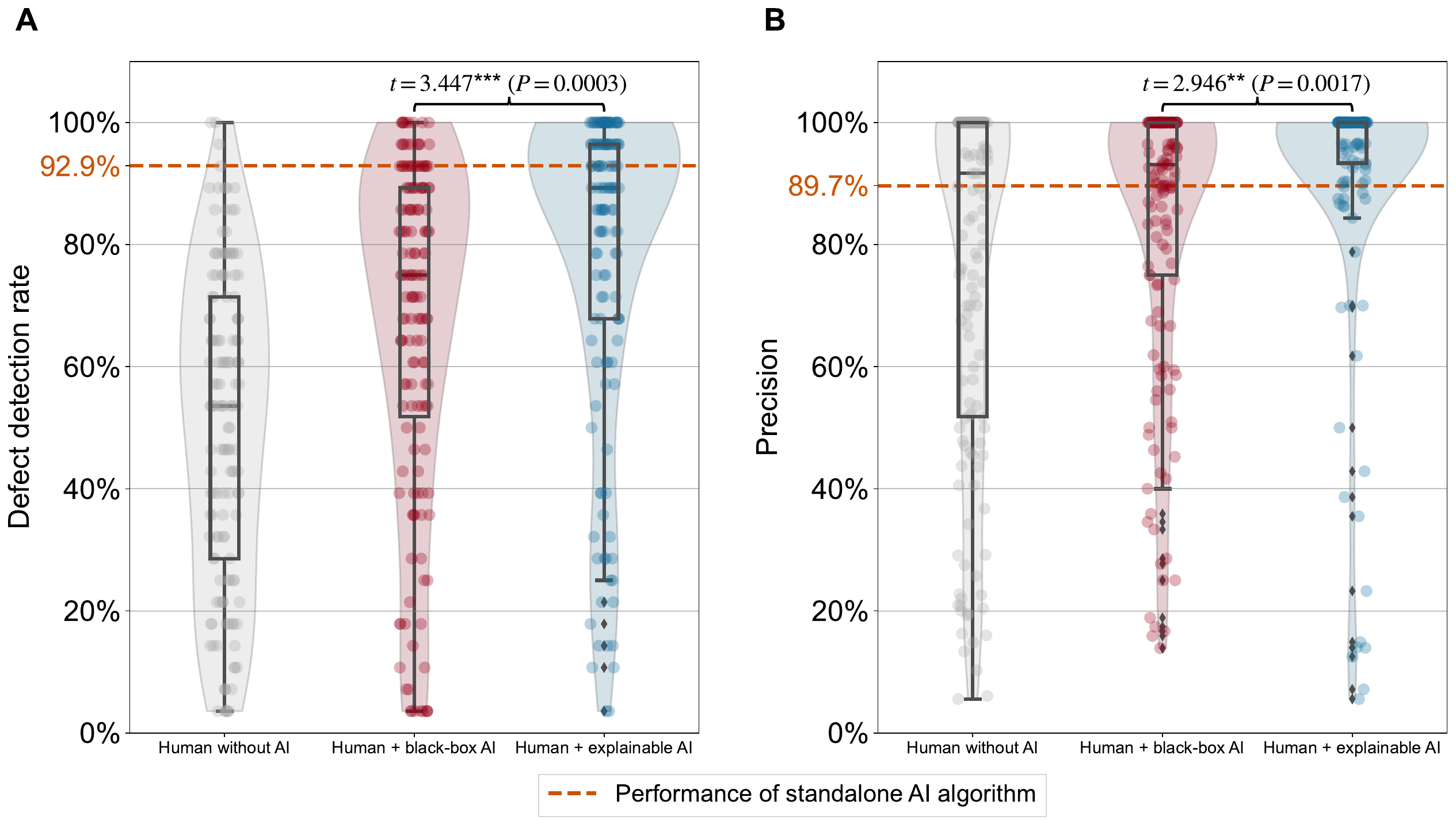}
\caption{\textbf{Results of non-expert experiment.} The boxplots compare the task performance between the two treatments: black-box AI and explainable AI. The task performance is measured by the defect detection rate (\textbf{\textsf{A}}) and the precision (\textbf{\textsf{B}}) based on the quality assessment of radiologists and the ground-truth labels of the chest X-ray images. The standalone AI algorithm attains a defect detection rate of 92.9\% and a precision of 89.7\% (orange dashed lines). Statistical significance is based on a one-sided Welch's $t$-test (\textsuperscript{***}$P<0.001$, \textsuperscript{**}$P<0.01$, \textsuperscript{*}$P<0.05$). In the boxplots, the center line denotes the median; box limits are upper and lower quartiles; whiskers are defined as the 1.5x interquartile range.}
\label{fig:study_1_precision_results}
\end{figure}

In \Cref{fig:study_1_precision_results}, the results with precision as task performance metric are shown. We find that non-experts augmented by explainable AI are more precise in identifying defective electronic products in comparison to peers supported by black-box AI.

\subsection{Randomization checks}
\label{sec:randomization_checks_1}

We performed randomization checks to confirm that the distribution of participants in the three treatment arms of the non-experts experiment was unbiased. The following demographic variables were collected: age bracket [$<$20, 20--30, 30--40, 40--50, 50--60, 60--70, $>$70], gender [male, female, not listed], and highest level of education [no schooling, primary school, some high-school; no degree, high school degree, Bachelor's degree, Master's degree, doctorate]. \Cref{tab:randomization_check_1} reports the observed frequencies in the three treatment arms. The randomization checks are based on $\mathcal{X}^2$-tests of independence. The results suggest no statistically significant differences between the participants in the three treatment arms. 

\begin{table}[H]
\onehalfspacing
\footnotesize
\begin{center}
\caption{\textbf{Randomization checks for non-experts experiment}} 
\label{tab:randomization_check_1}
\begin{tabular}{p{1.5cm} x{2.75cm} x{0cm} x{2.75cm} x{0cm} x{2.75cm} x{0cm} x{1.5cm}}
\toprule
& \textbf{Human with black-box AI} & & \textbf{Human with explainable AI} & & \textbf{Human without AI} & & \textbf{$\bm{P}$-value}\\

\cmidrule(lr){2-2}\cmidrule(lr){4-4}\cmidrule(lr){6-6}\cmidrule(lr){8-8}
Age & 0\,\textbar\,39\,\textbar\,58\,\textbar\,36\,\textbar\,16\,\textbar\,1\,\textbar\,1 & & 0\,\textbar\,26\,\textbar\,56\,\textbar\,34\,\textbar\,13\,\textbar\,8\,\textbar\,0 & & 1\,\textbar\,46\,\textbar\,49\,\textbar\,28\,\textbar\,13\,\textbar\,5\,\textbar\,0 & & 0.168\\
Gender & 104\,\textbar\,47\,\textbar\,0 & & 86\,\textbar\,51\,\textbar\,0 & & 89\,\textbar\,53\,\textbar\,0 & & 0.443\\
Education & 0\,\textbar\,0\,\textbar\,1\,\textbar\,30\,\textbar\,88\,\textbar\,32\,\textbar\,0 & & 0\,\textbar\,0\,\textbar\,1\,\textbar\,24\,\textbar\,91\,\textbar\,21\,\textbar\,0 & & 0\,\textbar\,0\,\textbar\,2\,\textbar\,16\,\textbar\,90\,\textbar\,34\,\textbar\,0 & & 0.278\\
\midrule
Observations & 151 & & 137 & & 142 & & -- \\
\bottomrule
\end{tabular}
\end{center}
Notes: The table reports the frequency of participants that fall in the specific subgroups of age, gender, and education (separated by vertical bars). The $P$-values for the randomization checks are computed based on  $\mathcal{X}^2$-tests of independence.\\
\end{table}

\subsection{Regression models}
\label{sec:regression_models_1}

\Cref{tab:regression_results_1_ols} reports three OLS regression models estimating the treatment effect with different control variables.  Model~(1) estimates the treatment effect for explainable AI with demographic controls (age, gender, and highest level of education). Model~(2) estimates the treatment effect for explainable AI with demographic controls and self-reported IT skills (ranging from 1: \textquote{novice} to 5: \textquote{expert}). Model~(3) estimates the treatment effect for explainable AI with demographic controls, self-reported IT skills, and the decision speed (median across the 200 images). All three models return a significant treatment effect for both metrics (balanced accuracy and defect detection rate) as dependent variables.

\begin{table}[H]
\onehalfspacing
\footnotesize
\begin{center}
\caption{\textbf{OLS regression results for treatment effect (non-experts experiment)}} 
\label{tab:regression_results_1_ols}
\begin{tabular}{p{2.5cm} x{1.5cm} x{1.5cm} x{1.5cm} x{1.5cm} x{1.5cm} x{1.5cm}}
\toprule
& \multicolumn{3}{c}{\textbf{Balanced accuracy}} & \multicolumn{3}{c}{\textbf{Defect detection rate}} \\
\cmidrule(lr){2-4} \cmidrule(lr){5-7}
& \textbf{Model (1)} & \textbf{Model (2)} & \textbf{Model (3)} & \textbf{Model (1)} & \textbf{Model (2)} & \textbf{Model (3)}\\

\cmidrule(lr){2-2}\cmidrule(lr){3-3}\cmidrule(lr){4-4}\cmidrule(lr){5-5}\cmidrule(lr){6-6}\cmidrule(lr){7-7}
Treatment & 5.792*** & 5.832*** & 5.570** & 10.435** & 10.509** & 10.299** \\
(explainable AI) & (1.729) & (1.720) & (1.707) & (3.274) & (3.258) & (3.263) \\
\midrule
Demographics & Yes & Yes & Yes & Yes & Yes & Yes \\
IT skills & No & Yes & Yes & No & Yes & Yes \\
Decision speed & No & No & Yes & No & No & Yes \\
\midrule
Observations & 288 & 288 & 288 & 288 & 288 & 288 \\
\bottomrule
\end{tabular}
\end{center}
\vspace{-0.25cm}
Notes: The table reports three OLS regression models with different sets of control variables and two different metrics as dependent variables. The standard errors of the treatment effect are reported in parentheses. \\
Statistical significance: \textsuperscript{***}$P<0.001$, \textsuperscript{**}$P<0.01$, \textsuperscript{*}$P<0.05$.
\end{table}

\Cref{tab:regression_results_1_qb} reports three quasi-binomial regression models estimating the treatment effect with the same control variables as before. Again, all three models return a significant treatment effect for both metrics as dependent variables.

\begin{table}[H]
\onehalfspacing
\footnotesize
\begin{center}
\caption{\textbf{Quasi-binomial regression results for treatment effect (non-experts experiment)}} 
\label{tab:regression_results_1_qb}
\begin{tabular}{p{2.5cm} x{1.5cm} x{1.5cm} x{1.5cm} x{1.5cm} x{1.5cm} x{1.5cm}}
\toprule
& \multicolumn{3}{c}{\textbf{Balanced accuracy}} & \multicolumn{3}{c}{\textbf{Defect detection rate}} \\
\cmidrule(lr){2-4} \cmidrule(lr){5-7}
& \textbf{Model (1)} & \textbf{Model (2)} & \textbf{Model (3)} & \textbf{Model (1)} & \textbf{Model (2)} & \textbf{Model (3)}\\

\cmidrule(lr){2-2}\cmidrule(lr){3-3}\cmidrule(lr){4-4}\cmidrule(lr){5-5}\cmidrule(lr){6-6}\cmidrule(lr){7-7}
Treatment & 0.449** & 0.455*** &  0.442** & 0.528** & 0.536**  & 0.528** \\
(explainable AI) & (0.137) & (0.136) & (0.136) & (0.168) & (0.167) & (0.168) \\
\midrule
Demographics & Yes & Yes & Yes & Yes & Yes & Yes \\
IT skills & No & Yes & Yes & No & Yes & Yes \\
Decision speed & No & No & Yes & No & No & Yes \\
\midrule
Observations & 288 & 288 & 288 & 288 & 288 & 288 \\
\bottomrule
\end{tabular}
\end{center}
\vspace{-0.25cm}
Notes: The table reports three quasi-binomial regression models with different sets of control variables and two different metrics as dependent variables. The standard errors of the treatment effect are reported in parentheses. \\
Statistical significance: \textsuperscript{***}$P<0.001$, \textsuperscript{**}$P<0.01$, \textsuperscript{*}$P<0.05$.
\end{table}

\subsection{Analysis with excluded participants}
\label{sec:excluded_participants_1}

\Cref{tab:excluded_participants_non_experts} reports the number of patients that were excluded according to the three different criteria that we have preregistered.

\begin{table}[H]
\onehalfspacing
\footnotesize
\begin{center}
\caption{\textbf{Excluded participants across treatment arms}} 
\label{tab:excluded_participants_non_experts}
\begin{tabular}{p{2.5cm} x{1.4cm} x{1.4cm} x{1.4cm}}
\toprule
 & Black-box AI & Explainable AI & Without AI \\

\midrule
Time-out & 26 & 31 & 35 \\
No defective & 21 & 21 & 27 \\
Worse than $3\sigma$ & 0 & 1 & 0 \\
\bottomrule
\end{tabular}
\end{center}
\end{table}

\Cref{tab:regression_results_excluded_3} reports the OLS regression model estimating the treatment effect for all different combinations of exclusion criteria. As in our main analysis, the effect of explainable AI is statistically significant for both metrics, balanced accuracy and defect detection rate, irrespective of the exclusion criteria.

\begin{table}[h]
\onehalfspacing
\footnotesize
\begin{center}
\caption{\textbf{OLS regression results with excluded participants (non-experts experiment)}}
\label{tab:regression_results_excluded_3}
\begin{tabular}{x{1.6cm} x{1.6cm} x{1.6cm} x{2.2cm} x{2.5cm} x{2.5cm}}
\toprule
 \multicolumn{3}{c}{\textbf{Excluded:}} & \textbf{Observations} & \textbf{Balanced accuracy} & \textbf{Defect detection rate} \\
\cmidrule(lr){1-3} \cmidrule(lr){4-4} \cmidrule(lr){5-5} \cmidrule(lr){6-6}
\tiny{Time-out} & \tiny{No defective} & \tiny{Worse than $3\sigma$} & & & \\
\cmidrule(lr){1-1}\cmidrule(lr){2-2}\cmidrule(lr){3-3}\cmidrule(lr){4-4}\cmidrule(lr){5-5}\cmidrule(lr){6-6}
  \xmark &       \xmark &     \xmark &           388 &  6.344*** (1.825) &      11.539** (3.548) \\
  \cmark &       \xmark &     \xmark &           331 &    4.843* (1.990) &        8.765* (3.902) \\
  \xmark &       \cmark &     \xmark &           342 &  6.966*** (1.648) &     12.652*** (3.023) \\
  \xmark &       \xmark &     \cmark &           387 &  6.561*** (1.817) &     11.788*** (3.549) \\
  \cmark &       \cmark &     \xmark &           289 &  5.918*** (1.756) &      10.863** (3.288) \\
  \cmark &       \xmark &     \cmark &           330 &    5.102* (1.980) &        9.054* (3.904) \\
  \xmark &       \cmark &     \cmark &           341 &  7.238*** (1.631) &     12.988*** (3.014) \\
\bottomrule
\end{tabular}
\end{center}
\vspace{-0.25cm}
Notes: The table reports the OLS regression model with two different metrics as dependent variables for all combinations of exclusion criteria. The standard errors of the treatment effect are reported in parentheses. \\
Statistical significance: \textsuperscript{***}$P<0.001$, \textsuperscript{**}$P<0.01$, \textsuperscript{*}$P<0.05$.
\end{table}

\clearpage
\section{Preregistered hypotheses}
\label{sec:hypotheses}

The following hypotheses were preregistered at \url{https://osf.io/7djxb} (Study 1) and \url{https://osf.io/69yqt} (Study 2):

\begin{quote}
\onehalfspacing
\small
\begin{hyp}[H1] \label{hyp:first}
Explainable AI improves the overall decision performance (measured by the balanced accuracy and defect detection rate) compared to humans without AI (i.e., manual inspection) ($\alpha=0.05$).
\end{hyp}

\begin{hyp}[H2] \label{hyp:second}
Explainable AI improves the overall decision performance (measured by the balanced accuracy and defect detection rate) compared to black-box AI ($\alpha=0.05$).
\end{hyp}

\begin{hyp}[H3] \label{hyp:third}
Explainable AI reduces variation in decision performance (measured by the variance in the balanced accuracy and defect detection rate) compared to black-box AI ($\alpha=0.05$).
\end{hyp}

\begin{hyp}[H4] \label{hyp:fourth}
Explainable AI increases the trust in model decisions (measured by the rate of correct model decisions that are not overruled by the user) compared to black-box AI ($\alpha=0.05$).
\end{hyp}
\end{quote}

\noindent
\Cref{tab:hypotheses} summarizes the results from all three studies: (1)~the manufacturing experiment at \emph{Siemens}, (2)~the medical experiment, and (3)~the manufacturing experiments with non-experts from Amazon MTurk. We report the $P$-values for both the balanced accuracy and defect detection rate for hypotheses H1, H2, and H3. The statistical testing for hypotheses H1, H2, and H4 are based on one-sided Welch's $t$-tests. The statistical testing for Hypothesis~H3 is based on Levene's test for equality of variances. As specified in our preregistration, we refrained from testing Hypothesis~H1 in our manufacturing field experiment and our medical experiment. The reason is that we wanted sufficient power in our main treatment arms of interest (i.e., black-box AI versus explainable AI). All hypotheses except for Hypothesis~H3 in the non-experts experiment and Hypotheses~H2 and H3 for the disease detection rate in the medical experiment were confirmed at a significance level of $\alpha = 0.05$. The latter can be expected since missing a lung lesion has more serious consequences than erroneously believing a lung lesion is visible; thus, leading to conservative decision-making of radiologists. Therefore, we additionally inspected precision as a task performance metric. We find that radiologists augmented with explainable AI were significantly more precise in identifying lung lesions compared to radiologists with black-box AI (see Supplement~\ref{sec:precision}).

\begin{table}[H]
\onehalfspacing
\footnotesize
\begin{center}
\caption{\textbf{Comparison of results against preregistered hypotheses.}} 
\label{tab:hypotheses}
\begin{tabular}{l ccc}
\toprule
& \textbf{Study 1: Manufacturing} & \textbf{Study 2: Medical} & \textbf{Study 3: Non-experts}  \\

\midrule
\textbf{H1} (BACC) & \emph{not part of preregistration} & \emph{not part of preregistration} & \cmark \, ($P<0.001$)\\
\textbf{H1} (DDR) & \emph{not part of preregistration} & \emph{not part of preregistration} & \cmark \, ($P<0.001$)\\
\midrule
\textbf{H2} (BACC) & \cmark \, ($P=0.001$) & \cmark \, ($P=0.004$) & \cmark \, ($P<0.001$)\\
\textbf{H2} (DDR) & \cmark \, ($P=0.004$) & \xmark \, ($P=0.498$) & \cmark \, ($P<0.001$)\\
\midrule
\textbf{H3} (BACC) & \cmark \, ($P=0.002$) & \cmark \, ($P=0.033$) & \xmark \, ($P=0.356$) \\
\textbf{H3} (DDR) & \cmark \, ($P=0.023$) & \xmark \, ($P=0.790$) & \xmark \, ($P=0.217$) \\
\midrule
\textbf{H4} & \cmark \, ($P=0.011$) & \cmark \, ($P=0.001$) & \cmark \, ($P=0.009$) \\
\bottomrule
\end{tabular}
\end{center}
Notes: The significance level was preregistered at $\alpha=0.05$ and the marks denote whether the corresponding $P$-value was significant at this level. BACC refers to balanced accuracy and DDR to defect/disease detection rate.\\
\end{table}

\clearpage
\section{Post-experimental questionnaire}
\label{sec:questionnaire}

For post-hoc exploratory analyses, we asked participants to complete a questionnaire. The questions involved established constructs, such as self-reported task load \cite{NASA.1986}, perceived usefulness \cite{Davis.1989}, perceived ease of use \cite{Davis.1989}, and self-reported trust \cite{Jian.2000}. We further asked participants about their previous experience and the perceived performance of the AI algorithm. In the manufacturing experiment, the questions were translated to German. In the medical experiment, the questions were adapted for the medical setting (for the exact wording, see our preregistration \url{https://osf.io/69yqt}). Participants in the non-experts experiment were asked to answer the questions from the viewpoint of a factory worker (\textquote{Imagine you work in a factory with a similar job task as you just did.}). The results for all three studies are provided in \Crefrange{tab:task_load}{tab:trust}.

\begin{table}[H]
\onehalfspacing
\tiny
\begin{center}
\caption{\textbf{Self-reported task load}} 
\label{tab:task_load}
\begin{tabular}{p{2.4cm} x{1.3cm} x{1.5cm} x{0cm} x{1.3cm} x{1.5cm} x{0cm} x{1.3cm} x{1.5cm}}
\toprule
& \multicolumn{2}{c}{\textbf{Study 1: Manufacturing}} && \multicolumn{2}{c}{\textbf{Study 2: Medical}} && \multicolumn{2}{c}{\textbf{Study 3: Non-experts}}\\
\cmidrule(lr){2-3}\cmidrule(lr){5-6}\cmidrule(lr){8-9}
\textbf{Question} & \textbf{Human with black-box AI} & \textbf{Human with explainable AI} && \textbf{Human with black-box AI} & \textbf{Human with explainable AI} && \textbf{Human with black-box AI} & \textbf{Human with explainable AI} \\
\midrule
How mentally demanding was the task? (1 = very low, 7 = very high) & 3.41 (1.37) & 3.54 (1.17) & & 3.80 (1.34) & 3.73 (1.60) && 4.72 (1.58) & 4.87 (1.62) \\ 
\midrule
How physically demanding was the task? (1 = very low, 7 = very high) & 3.09 (1.54) & 2.85 (1.38) && 2.18 (1.32) & 2.52 (1.42) && 4.09 (2.10) & 3.95 (2.06) \\ 
\midrule
How hurried or rushed was the pace of the task? (1 = very low, 7 = very high) & 3.59 (1.10) & 3.92 (1.23) && 2.76 (1.35) & 3.16 (1.66) && 4.56 (1.55) & 4.44 (1.68) \\ 
\midrule
How successful were you in accomplishing what you were asked to do? (1 = very poor, 7 = very good) & 5.27 (0.94) & 5.81 (0.85) && 5.64 (1.09) & 5.57 (0.93) && 5.68 (1.08) & 5.87 (1.02) \\ 
\midrule
How hard did you have to work to accomplish your level of performance? (1 = very low, 7 = very high) & 4.00 (1.35) & 4.00 (0.94) && 3.33 (1.28) & 3.36 (1.30) && 5.18 (1.38) & 5.29 (1.46) \\ 
\midrule
How insecure, discouraged, irritated, stressed, and annoyed were you? (1 = very low, 7 = very high) & 2.77 (1.31) & 2.77 (1.58) && 2.76 (1.28) & 2.91 (1.43) && 3.52 (1.98) & 3.31 (1.95) \\
\bottomrule
\end{tabular}
\end{center}
Notes: The table reports the average scores for the self-reported task load. Standard deviations are reported in parentheses. \\
\end{table}

\begin{table}[H]
\onehalfspacing
\tiny
\begin{center}
\caption{\textbf{Perceived usefulness}} 
\label{tab:usefulness}
\begin{tabular}{p{2.4cm} x{1.3cm} x{1.5cm} x{0cm} x{1.3cm} x{1.5cm} x{0cm} x{1.3cm} x{1.5cm}}
\toprule
& \multicolumn{2}{c}{\textbf{Study 1: Manufacturing}} && \multicolumn{2}{c}{\textbf{Study 2: Medical}} && \multicolumn{2}{c}{\textbf{Study 3: Non-experts}}\\
\cmidrule(lr){2-3}\cmidrule(lr){5-6}\cmidrule(lr){8-9}
\textbf{Question} & \textbf{Human with black-box AI} & \textbf{Human with explainable AI} && \textbf{Human with black-box AI} & \textbf{Human with explainable AI} && \textbf{Human with black-box AI} & \textbf{Human with explainable AI} \\
\midrule
Using the Artificial Intelligence System in my job would enable me to accomplish tasks more quickly. (1 = very unlikely, 7 = extremely likely) & 4.95 (1.70) & 5.12 (1.37) && 5.00 (1.40) & 5.25 (1.06) && 5.68 (1.02) & 5.86 (1.08) \\ 
\midrule
Using the Artificial Intelligence System would improve my job performance. (1 = very unlikely, 7 = extremely likely) & 4.91 (1.54) & 5.23 (0.99) && 5.04 (1.24) & 5.09 (1.14) && 5.68 (1.04) & 5.98 (1.06) \\ 
\midrule
Using the Artificial Intelligence System in my job would increase my productivity. (1 = very unlikely, 7 = extremely likely) & 4.95 (1.43) & 4.96 (1.22) && 5.04 (1.38) & 5.39 (1.15) && 5.71 (1.11) & 5.98 (1.01) \\ 
\midrule
Using the Artificial Intelligence System in my job would enhance my effectiveness on the job. (1 = very poor, 7 = very good) & 5.05 (1.53) & 4.88 (1.14) && 4.98 (1.34) & 5.20 (1.25) && 5.66 (1.12) & 5.91 (0.97) \\ 
\midrule
Using the Artificial Intelligence System would make it easier to do my job. (1 = very unlikely, 7 = extremely likely) & 5.18 (1.26) & 5.08 (1.32) && 4.93 (1.45) & 5.23 (1.08) && 5.65 (1.23) & 6.01 (1.00) \\ 
\midrule
I would find the Artificial Intelligence System useful in my job. (1 = very unlikely, 7 = extremely likely) & 5.27 (1.42) & 5.23 (1.24) && 5.00 (1.41) & 5.09 (1.22) && 5.79 (1.09) & 6.07 (1.04) \\
\bottomrule
\end{tabular}
\end{center}
Notes: The table reports the average scores for the perceived usefulness of the AI algorithm. Standard deviations are reported in parentheses. \\
\end{table}

\begin{table}[H]
\onehalfspacing
\tiny
\begin{center}
\caption{\textbf{Perceived ease of use}} 
\label{tab:ease_of_use}
\begin{tabular}{p{2.4cm} x{1.3cm} x{1.5cm} x{0cm} x{1.3cm} x{1.5cm} x{0cm} x{1.3cm} x{1.5cm}}
\toprule
& \multicolumn{2}{c}{\textbf{Study 1: Manufacturing}} && \multicolumn{2}{c}{\textbf{Study 2: Medical}} && \multicolumn{2}{c}{\textbf{Study 3: Non-experts}}\\
\cmidrule(lr){2-3}\cmidrule(lr){5-6}\cmidrule(lr){8-9}
\textbf{Question} & \textbf{Human with black-box AI} & \textbf{Human with explainable AI} && \textbf{Human with black-box AI} & \textbf{Human with explainable AI} && \textbf{Human with black-box AI} & \textbf{Human with explainable AI} \\
\midrule
Learning to operate the Artificial Intelligence System would be easy for me. (1 = very unlikely, 7 = extremely likely) & 5.00 (1.45) & 5.35 (1.20) && 5.78 (0.88) & 6.00 (0.86) && 5.69 (1.11) & 5.90 (0.95) \\ 
\midrule
I would find it easy to get the Artificial Intelligence System to do what I want it to do. (1 = very unlikely, 7 = extremely likely) & 4.05 (1.36) & 4.46 (0.90) && 4.87 (1.22) & 5.09 (1.03) && 5.61 (1.02) & 5.78 (0.97) \\ 
\midrule
My interaction with the Artificial Intelligence System would be clear and understandable. (1 = very unlikely, 7 = extremely likely) & 5.09 (0.97) & 5.27 (0.96) && 5.16 (1.07) & 5.20 (1.21) && 5.65 (1.08) & 5.99 (0.93) \\ 
\midrule
I would find the Artificial Intelligence System to be flexible to interact with. (1 = very poor, 7 = very good) & 4.82 (1.22) & 5.00 (1.06) && 4.82 (1.28) & 4.84 (1.27) && 5.36 (1.24) & 5.55 (1.10) \\ 
\midrule
It would be easy for me to become skillful at using the Artificial Intelligence System. (1 = very unlikely, 7 = extremely likely) & 5.14 (1.21) & 5.38 (0.85) && 5.38 (1.21) & 5.61 (0.84) && 5.73 (0.97) & 5.93 (0.99) \\ 
\midrule
I would find the Artificial Intelligence System easy to use. (1 = very unlikely, 7 = extremely likely) & 5.14 (0.94) & 5.42 (1.03) && 5.16 (1.17) & 5.66 (0.83) && 5.71 (1.03) & 6.05 (0.96) \\
\bottomrule
\end{tabular}
\end{center}
Notes: The table reports the average scores for the perceived ease of use of the AI algorithm. Standard deviations are reported in parentheses. \\
\end{table}

\begin{table}[H]
\onehalfspacing
\tiny
\begin{center}
\caption{\textbf{Previous experience and perceived performance}} 
\label{tab:perceived_performance}
\begin{tabular}{p{2.4cm} x{1.3cm} x{1.5cm} x{0cm} x{1.3cm} x{1.5cm} x{0cm} x{1.3cm} x{1.5cm}}
\toprule
& \multicolumn{2}{c}{\textbf{Study 1: Manufacturing}} && \multicolumn{2}{c}{\textbf{Study 2: Medical}} && \multicolumn{2}{c}{\textbf{Study 3: Non-experts}}\\
\cmidrule(lr){2-3}\cmidrule(lr){5-6}\cmidrule(lr){8-9}
\textbf{Question} & \textbf{Human with black-box AI} & \textbf{Human with explainable AI} && \textbf{Human with black-box AI} & \textbf{Human with explainable AI} && \textbf{Human with black-box AI} & \textbf{Human with explainable AI} \\
\midrule
How strong would you consider your general IT skills? (1 = novice, 5 = expert) & 2.86 (0.94) & 3.08 (0.84) && 3.47 (1.01) & 3.77 (1.01) && 3.48 (0.99) & 3.48 (1.08) \\ 
\midrule
How often do you interact with Artificial Intelligence in your job? (1 = very little, 5 = very much) & 2.32 (1.13) & 2.23 (1.11) && \multicolumn{2}{c}{\emph{not part of preregistration}} && 3.18 (1.31) & 3.12 (1.32) \\ 
\midrule
How familiar do you feel with Artificial Intelligence in general? (1 = very little, 5 = very much) & 2.41 (1.10) & 2.88 (1.07) && 3.00 (1.07) & 3.07 (0.95) && 3.56 (0.98) & 3.48 (0.93) \\ 
\midrule
How well did the Artificial Intelligence System perform in comparison to your expectations? (1 = very poor, 7 = very good) & 5.00 (1.57) & 6.08 (1.06) && 4.49 (1.41) & 4.48 (1.47) && 5.72 (1.01) & 6.15 (0.85) \\ 
\midrule
How likely is the Artificial Intelligence System to make a bad estimate? (1 = very unlikely, 7 = very likely) & 3.55 (1.34) & 3.04 (1.08) && 3.93 (1.14) & 4.02 (1.13) && 3.25 (1.61) & 2.93 (1.68) \\ 
\midrule
Completing the quality inspections task has changed my opinion about Artificial Intelligence. (1 = strongly disagree, 7 = strongly agree) & 4.64 (1.36) & 4.08 (1.38) && \multicolumn{2}{c}{\emph{not part of preregistration}} && 4.68 (1.65) & 4.83 (1.61) \\ 
\midrule
How much did you rely on the Artificial Intelligence System? (1 = very little, 7 = very much) & 4.32 (1.46) & 4.77 (1.37) &&\multicolumn{2}{c}{\emph{not part of preregistration}} && 4.96 (1.43) & 5.54 (1.38) \\ 
\midrule
The Artificial Intelligence System provides clear explanations for its outputs. (1 = strongly disagree, 7 = strongly agree) & 4.59 (1.33) & 5.04 (1.00) && \multicolumn{2}{c}{\emph{not part of preregistration}} && 4.99 (1.54) & 5.71 (1.20) \\
\bottomrule
\end{tabular}
\end{center}
Notes: The table reports the average scores for previous experience and the perceived performance of the AI algorithm. Standard deviations are reported in parentheses. \\
\end{table}

\begin{table}[H]
\onehalfspacing
\tiny
\begin{center}
\caption{\textbf{Self-reported trust in AI algorithm}} 
\label{tab:trust}
\begin{tabular}{p{2.4cm} x{1.3cm} x{1.5cm} x{0cm} x{1.3cm} x{1.5cm} x{0cm} x{1.3cm} x{1.5cm}}
\toprule
& \multicolumn{2}{c}{\textbf{Study 1: Manufacturing}} && \multicolumn{2}{c}{\textbf{Study 2: Medical}} && \multicolumn{2}{c}{\textbf{Study 3: Non-experts}}\\
\cmidrule(lr){2-3}\cmidrule(lr){5-6}\cmidrule(lr){8-9}
\textbf{Question} & \textbf{Human with black-box AI} & \textbf{Human with explainable AI} && \textbf{Human with black-box AI} & \textbf{Human with explainable AI} && \textbf{Human with black-box AI} & \textbf{Human with explainable AI} \\
\midrule
The Artificial Intelligence System is deceptive. (1 = strongly disagree, 7 = strongly agree) & 3.27 (1.20) & 2.88 (1.18) && 3.31 (1.41) & 3.39 (1.28) && 3.79 (2.04) & 3.72 (2.16) \\ 
\midrule
I am suspicious of the Artificial Intelligence System’s intent, action, or outputs. (1 = strongly disagree, 7 = strongly agree) & 2.91 (1.27) & 2.62 (0.98) && 3.09 (1.69) & 2.98 (1.56) && 3.88 (2.04) & 3.69 (2.15) \\ 
\midrule
The Artificial Intelligence System’s actions will have a harmful outcome. (1 = strongly disagree, 7 = strongly agree) & 3.09 (1.41) & 2.65 (1.09) && 3.38 (1.51) & 3.34 (1.43) && 3.73 (2.04) & 3.48 (2.08) \\ 
\midrule
I am confident in the Artificial Intelligence System. (1 = very poor, 7 = very good) & 4.91 (1.06) & 4.96 (1.15) && 4.04 (1.31) & 4.34 (1.22) && 5.52 (1.20) & 5.89 (0.97) \\ 
\midrule
The Artificial Intelligence System is reliable. (1 = strongly disagree, 7 = strongly agree) & 4.95 (0.84) & 5.27 (1.04) && 4.07 (1.39) & 4.27 (1.09) && 5.63 (1.10) & 5.93 (0.85) \\ 
\midrule
I can trust the Artificial Intelligence System. (1 = strongly disagree, 7 = strongly agree) & 4.86 (0.77) & 5.04 (0.92) && 3.87 (1.39) & 4.07 (1.11) && 5.66 (1.06) & 5.80 (0.94) \\
\bottomrule
\end{tabular}
\end{center}
Notes: The table reports the average scores for the self-reported trust in the AI algorithm. Standard deviations are reported in parentheses. \\
\end{table}

\end{appendices}

\end{document}